\documentclass[a4paper,11pt]{article}
\pdfoutput=1
\usepackage{jcappub}

\usepackage{amsmath,amsfonts,amssymb,mathrsfs,graphicx,color,longtable,bm}
\usepackage{hyperref}
\usepackage{array}
\usepackage{cases} 
\usepackage{enumitem}
\usepackage[explicit]{titlesec}
\usepackage[utf8]{inputenc}
\usepackage[normalem]{ulem}
\usepackage{cleveref}
\usepackage{xcolor}

\hypersetup{
     colorlinks   = true,
     citecolor    = red,
     urlcolor     = blue,
     linkcolor    = blue
}

\newcommand{\dd}{\mathrm{d}}

\newcommand\smallO[1]{
  \mathchoice
    {{\scriptstyle\mathcal{O}}}
    {{\scriptstyle\mathcal{O}}}
    {{\scriptscriptstyle\mathcal{O}}}
    {\scalebox{.7}{$\scriptscriptstyle\mathcal{O}$}}
  {\left(#1\right)}}

\makeatletter
\gdef\@fpheader{}
\makeatother

\usepackage{physics}

\usepackage{lipsum}

\usepackage{subfigure}

\usepackage{verbatim}

\usepackage{cancel}

\usepackage{tikz}
\usetikzlibrary{mindmap, tikzmark, calc, decorations.pathreplacing}

\usepackage{tikz-feynman}
	\tikzfeynmanset{compat=1.1.0}

\newcommand{\noleft}{\left.\kern-\nulldelimiterspace}
\newcommand{\noright}{\right.\kern-\nulldelimiterspace}
\newcommand{\round}[1]{\left( #1 \right)}
\newcommand{\sround}[1]{\left[ #1 \right]}
\newcommand{\bround}[1]{\left\{ #1 \right\}}
\newcommand*\circled[1]{\tikz[baseline=(char.base)]{\node[shape=circle,draw,inner sep=2pt] (char) {#1};}}
\newcommand{\mybraket}[1]{\left\langle #1 \right\rangle}

\title{Refining the nonlinear modelling of primordial oscillatory features}

\author[a,b,c]{Mario Ballardini,}
\author[a,b]{Nicola Barbieri}

\affiliation[a]{Dipartimento di Fisica e Scienze della Terra, Universit\`a degli Studi di Ferrara, via Giuseppe Saragat 1, 44122 Ferrara, Italy}
\affiliation[b]{INFN, Sezione di Ferrara, via Giuseppe Saragat 1, 44122 Ferrara, Italy}
\affiliation[c]{INAF/OAS Bologna, via Piero Gobetti 101, 40129 Bologna, Italy}

\emailAdd{mario.ballardini@unife.it}
\emailAdd{nicola.barbieri@unife.it}

\abstract{Primordial oscillatory features in the power spectrum of curvature perturbations are sensitive probes of 
the dynamics of the early Universe and can provide insights beyond the standard inflationary scenario. While these 
features have been the focus of extensive studies using cosmic microwave background anisotropy data, large-scale 
structure surveys now provide competitive constraints with 
the opportunity to probe their effects at smaller scales with higher precision. 
In this paper, we present a complete description of the nonlinear model for primordial oscillatory features in the 
context of time-sliced perturbation theory extending the results already presented in the literature. 
We derive analytical expressions including novel contributions such as the mixed term between primordial oscillations 
and baryon acoustic oscillations, and we also calculate the corrections arising from the specific envelope of the 
oscillatory pattern, corresponding to a scale-dependent amplitude. These results are compared with N-body simulations 
using the COLA method and show consistent behaviour across different scales. Although the corrections are found to 
be small, they represent an important step 
to fully characterising the nonlinear imprints of primordial features 
on the matter power spectrum. Our results offer new 
calculations to be tested with future cosmological surveys that seek to detect 
these subtle signatures in the matter distribution.}

\begin{document}
\maketitle
\flushbottom

\section{Introduction} 
Features in the primordial power spectrum (PPS) of curvature perturbations provide a valuable source of information 
of the dynamics of the early Universe, extending our understanding beyond the simplest models of cosmic 
inflation~\cite{Starobinsky:1980te,Guth:1980zm,Linde:1981mu,Linde:1983gd,Mukhanov:1981xt}. 
These primordial features may arise from abrupt transitions in the inflationary 
potential~\cite{Starobinsky:1992ts,Adams:2001vc,Chen:2006xjb} or the presence of additional degrees of freedom during 
inflation~\cite{Achucarro:2010da,Chen:2011zf}, suggesting multi-field dynamics~\cite{Braglia:2020fms}, 
resonances~\cite{Chen:2008wn,Flauger:2009ab,Flauger:2010ja,Chen:2010bka} or even phase transitions in the early Universe. 
They could also indicate deviations from the standard Bunch-Davies vacuum for primordial 
perturbations~\cite{Martin:2000xs,Easther:2002xe,Armendariz-Picon:2003knj} and can be used to reveal the nature of the 
true scenario governing the early Universe~\cite{Chen:2011zf,Chen:2014joa,Chen:2018cgg}; all these effects 
lead to a breaking of scale invariance in the PPS, see 
Refs.~\cite{Chluba:2015bqa,Slosar:2019gvt,2022arXiv220308128A} for reviews. The identification and detailed characterisation of such 
signals is essential for a deeper understanding of the inflationary mechanism, and provides a unique opportunity to 
detect signs of new and high-energy physics in the early Universe. 

Despite numerous efforts, there is currently no statistically significant evidence for deviations from a power-law 
PPS, as predicted by the simplest single-field slow-roll inflationary models. The PPS has been constrained to within 
a few percent in amplitude over the wavenumber range $0.005 \lesssim k/(h\,{\rm Mpc}^{-1}) \lesssim 0.2$, as reported 
in Refs.~\cite{Akrami:2018odb,Handley:2019fll,Beutler:2019ojk,Ballardini:2022wzu,Mergulhao:2023ukp}.

While much of the search for primordial features has traditionally relied on measurements of the angular power spectra 
of the cosmic microwave background 
(CMB)~\cite{Wang:2000js,Adams:2001vc,Peiris:2003ff,Mukherjee:2003ag,Covi:2006ci,Hamann:2007pa,Meerburg:2011gd,Planck:2013jfk,Meerburg:2013dla,Benetti:2013cja,Miranda:2013wxa,Easther:2013kla,Chen:2014joa,Achucarro:2014msa,Hazra:2014goa,Hazra:2014jwa,Hu:2014hra,Ade:2015lrj,Gruppuso:2015zia,Gruppuso:2015xqa,Hazra:2016fkm,Torrado:2016sls,Akrami:2018odb,Zeng:2018ufm,Canas-Herrera:2020mme,Braglia:2021ckn,Braglia:2021sun,Braglia:2021rej,Naik:2022mxn,Hamann:2021eyw,Antony:2022ert,Antony:2024vrx}, 
and the CMB bispectra~\cite{Planck:2013wtn,Fergusson:2014tza,Planck:2015zfm,Meerburg:2015owa,Akrami:2018odb,Planck:2019izv}, 
recent attention has shifted towards large-scale structure (LSS) analyses~\cite{Beutler:2019ojk,Ballardini:2022wzu,Mergulhao:2023ukp}. 
In particular, the study of primordial oscillatory features with LSS promises to provide more precise measurements of 
the power spectrum at smaller scales, potentially increasing sensitivity to high-frequency signals 
compared to the constraints imposed by current CMB experiments
~\cite{Wang:1998gb,Zhan:2005rz,Huang:2012mr,Chen:2016vvw,Chen:2016zuu,Ballardini:2016hpi,Xu:2016kwz,Fard:2017oex,Palma:2017wxu,Ballardini:2017qwq,Zeng:2018ufm,Ballardini:2018noo,Slosar:2019gvt,Beutler:2019ojk,Ballardini:2019tuc,Debono:2020emh,Li:2021jvz,Euclid:2023shr}. However, this requires an accurate model to describe the galaxy power spectrum in the mildly nonlinear regime.
The nonlinear regime of structure formation, driven predominantly by gravitational interactions 
on quasi-linear scales, presents significant 
challenges that require the use of perturbative techniques.\footnote{Note that in general baryon physics is the main challenge to fully model the nonlinear regime, and that perturbative techniques only work in the quasi-linear scales.} The nonlinear modelling of oscillatory features has previously 
been explored within the framework of perturbation theory (PT) for LSS, extending the formalism originally developed to 
describe the damping of baryon acoustic oscillations (BAO) by the infrared (IR) resummation of large-scale bulk 
flows~\cite{Crocce:2007dt,Scoccimarro2011,Creminelli:2013poa,Baldauf:2015xfa,Blas:2015qsi,Senatore:2017pbn}. These methods have been already 
applied to both linear oscillations in Refs.~\cite{Blas:2016sfa,Beutler:2019ojk} and logarithmic oscillations in 
Refs.~\cite{Vasudevan:2019ewf,Beutler:2019ojk}. In parallel, several studies have focused on the numerical study of 
nonlinear dynamics in the presence of primordial features using N-body 
simulations~\cite{Vlah:2015zda,Ballardini:2019tuc,Chen:2020ckc,Li:2021jvz,Ballardini:2022vzh,Euclid:2023shr,2024arXiv240600103C}. These studies 
provide complementary insights into the effects of non-linearities on the power spectrum, particularly with respect to 
primordial features.

Building on the foundations laid in Ref.~\cite{Blas:2015qsi}, which introduced time-sliced perturbation theory (TSPT) 
as a novel approach to nonlinear perturbation theory, this paper applies the method to new scenarios and refines existing 
calculations by incorporating additional contributions that are typically neglected.
One of the key strengths of TSPT, as highlighted in Ref.~\cite{Blas:2016sfa}, is its ability to resum long wavelength 
perturbations that can cause large corrections to the power spectrum in the presence of oscillations. This IR resummation 
technique has been successfully applied to both biased tracers and redshift space distortions, yielding results in good 
agreement with numerical simulations; 
see Refs.~\cite{Ivanov:2018gjr,Ivanov:2019pdj,Ivanov:2019hqk,Nishimichi:2020tvu,Ivanov:2020ril,Chudaykin:2020ghx}. 
By applying these techniques to scenarios with primordial oscillatory features, we extend previous applications and 
calculate more accurate nonlinear templates, potentially useful 
in current and future survey of galaxies. This will be timely crucial to constrain primordial features with 
spectroscopic galaxy clustering analysis from stage-IV spectroscopic galaxy surveys such as DESI \cite{DESI:2024mwx} and {\em Euclid}~\cite{Euclid:2023shr,Euclid:2024yrr}.

We structure the paper as follows. After this introduction, in~\cref{sec:review_1} we review the basic technicalities 
and main results of the TSPT formalism. In~\cref{sec:osc_features}, we specify the main formalism to the addition of 
primordial oscillatory features, reproducing the results already obtained in the literature and extending them taking 
into account novel contributions. Finally in~\cref{sec:cola_sims} we compare our analytical results with N-body simulations 
using the COLA method. We draw our conclusions in~\cref{sec:conclusions}.

\section{Review of time-sliced perturbation theory} \label{sec:review_1}
In this section, we review the core elements of the TSPT formalism, along with its main features following the work presented 
in Refs.~\cite{Blas:2015qsi,Blas:2016sfa,Vasudevan:2019ewf}. For the sake of generality, we begin by considering a single field 
variable, denoted as $\Theta$. In cosmology, perturbative fields are dynamical variables. Therefore, without loss of generality, 
we assume that they satisfy deterministic equations of motion, given by
\begin{equation} \label{eq:eq_of_motion}
    \dot{\Theta}_\eta \left( \mathbf{k} \right) = \mathcal{I} \left[ \Theta_\eta ; \eta, \mathbf{k} \right] \,,
\end{equation}  
where $\eta$ is a generic time coordinate and the dot represents the derivative with respect to it. In the spirit of PT, the 
rhs (right hand side) is given as a power series in the fields 
\begin{equation}
    \mathcal{I} \left[ \Theta_\eta ; \eta, \mathbf{k} \right] = \sum_{n=1}^\infty \frac{1}{n!} \int \prod_{i=1}^n \dd^3q_i \; \Theta_\eta \left( \mathbf{q}_i \right) I_n \left( \eta; \mathbf{q}_1, \dots, \mathbf{q}_n \right) \delta^{\left(3\right)} \left( \mathbf{k} - \sum_{j=1}^n \mathbf{q}_j \right) \,.
\end{equation}
The $\delta$-function in this expression enforces the conservation of momentum, which we assume is satisfied by the system. 
We also assume that the latter respects parity, which is encoded by the following property 
\begin{equation}
    I_m \left( \eta; \mathbf{q}_1 , \dots , \mathbf{q}_m \right) = I_m \left( \eta ; - \mathbf{q}_1, \dots , - \mathbf{q}_m \right) \,.
\end{equation}
The second key element for constructing our framework is the statistical distribution at the initial time $\eta_0$, which is 
set deep within the linear regime. This choice allows us to recover the correlators of standard perturbation theory (SPT) within 
a path-integral formulation, using the following partition function~\cite{Carroll:2013oxa,Valageas:2003gm}
\begin{equation} \label{eq:spt_gen_functional}
    Z \left[J; \eta\right] = \mathcal{N}^{-1} \int \mathcal{D} \Theta_{\eta_0} \exp \left\{ -\frac{1}{2} \int \dd^3k \; \frac{\Theta_{\eta_0} \left( \mathbf{k} \right) \Theta_{\eta_0} \left( \mathbf{-k} \right)}{ P_{\eta_0} \left( k \right) } + \int \dd^3k \; \Theta_{\eta} \left( \mathbf{k} \right) J \left( -\mathbf{k} \right) \right\} \,,
\end{equation}
where $P_{\eta_0}$ is the power spectrum of the initial Gaussian distribution and  
\begin{equation}
    \mathcal{N} = \int \mathcal{D} \Theta_{\eta_0} \exp \left\{ -\frac{1}{2} \int \dd^3k \; \frac{\Theta_{\eta_0} \left( \mathbf{k} \right) \Theta_{\eta_0} \left( \mathbf{-k} \right)}{ P_{\eta_0} \left( k \right) } \right\} 
\end{equation}
is a normalization factor. 
It is important to note that the integration here is performed over the fields at the initial time, while the source $J$ couples 
to the final values of the field. At first glance, this may seem counterintuitive: since the second term in the exponential of~\cref{eq:spt_gen_functional} does not depend on the path integral variable, one might think that it could be factored out and 
included in the normalisation term. However, we must always remember that at any given time the field $\Theta_\eta$ is determined 
by a deterministic time evolution governed by~\cref{eq:eq_of_motion}, meaning it can always be expressed 
as a function of the initial conditions, $\Theta_\eta \left(\mathbf{k}\right) = \Theta_\eta \left[ \Theta_{\eta_0}; \mathbf{k} \right]$ 
(see section~4.2 of Ref.~\cite{Carroll:2013oxa}). We have explicitly imposed statistical homogeneity and isotropy by ensuring 
that, in the Gaussian weight, the momenta of the $\Theta$-fields sum to zero, and the power spectrum $P_{\eta_0} \left( k \right)$ 
depends only on the magnitude of the momentum. The equal-time correlation functions are then derived by differentiating 
$Z \left[J; \eta\right]$ with respect to the source and then setting $J=0$   
\begin{equation} \label{eq:correlators_def}
    \left\langle \Theta_\eta \left( \mathbf{k}_1 \right) \cdots \Theta_\eta \left( \mathbf{k}_n \right) \right\rangle = \left. \frac{\delta^n Z \left[J; \eta\right]}{\delta J \left( -\mathbf{k}_1 \right) \cdots \delta J \left( -\mathbf{k}_n \right)} \right\rvert_{J=0} \,.
\end{equation}
This fact follows directly from the expansion of the second factor at the exponential in the generating functional, \cref{eq:spt_gen_functional}, in power series 
\begin{align}
    Z \left[J; \eta\right] = \sum_{n=0}^{\infty} \frac{1}{n!} \int \prod_{i=1}^n \dd^3q_i \; J \left( - \mathbf{q}_i\right) \left\langle \Theta_\eta \left( \mathbf{q}_1 \right) \cdots \Theta_\eta \left( \mathbf{q}_n \right) \right\rangle \,,
\end{align}
which, noticing that 
\begin{equation}
    \left. \frac{\delta^n}{\delta J \left( -\mathbf{k}_1 \right) \cdots \delta J \left( -\mathbf{k}_n \right)} \int \prod_{i=1}^m \dd^3q_i \; J \left( - \mathbf{q} _i\right) \right\rvert_{J=0} = n! \, \delta_{mn} \int \prod_{i=1}^n \dd^3q_i \; \delta^{3} \left( \mathbf{k}_i - \mathbf{q}_i \right) \,,
\end{equation}
gives us back exactly~\cref{eq:correlators_def}. In particular, for the two-point function at the initial time we can explicitly calculate this formula. When 
$\eta = \eta_0$, the path-integral partition function of~\cref{eq:spt_gen_functional} can be easily computed as a Gaussian 
integral giving
\begin{align}
    Z \left[J; \eta_0\right] &= \mathcal{N}^{-1} \int \mathcal{D} \Theta_{\eta_0} \exp{ -\frac{1}{2} \int \dd^3k \left[ \frac{\Theta_{\eta_0} \left( \mathbf{k} \right) \Theta_{\eta_0} \left( \mathbf{-k} \right)}{ P_{\eta_0} \left( k \right) } + \Theta_{\eta_0} \left( \mathbf{k} \right) J \left( -\mathbf{k} \right) \right] } \,,\notag \\
    &= \exp{ \frac{1}{2} \int \dd^3k \; J \left( \mathbf{k} \right) J \left( -\mathbf{k} \right) P_{\eta_0} \left(k\right) } \,.
\end{align}
It is then easy to see that the two-point correlation function is 
\begin{equation}
    \left\langle \Theta_{\eta_0} \round{\mathbf{k}_1} \Theta_{\eta_0} \round{\mathbf{k}_2} \right\rangle \equiv \left. \frac{\delta^2 Z \left[ J; \eta_0 \right]}{\delta J \left( -\mathbf{k}_1 \right) \delta J \left( -\mathbf{k}_2 \right)} \right\rvert_{J=0} = \delta^{\left(3\right)} \left( \mathbf{k}_1 + \mathbf{k}_2 \right) P_{\eta_0} \left(k\right) \,,
\end{equation}
which correspond to the initial power spectrum, as expected. 

In SPT, it is customary to solve the equations of motion, \cref{eq:eq_of_motion}, iteratively and express the field at final 
time, $\Theta_\eta \left(\mathbf{k}\right)$, as a Taylor series in powers of the initial configuration, $\Theta_{\eta_0} 
\left(\mathbf{k}\right)$. The key distinction between SPT and TSPT lies in the way we choose to approach this problem. 
Since we are generally interested in correlation functions at a specific time, say $\eta$, it is natural to use the field 
evaluated at that moment as the primary variable in our formalism. To implement this, we replace the integration variable 
in~\cref{eq:spt_gen_functional} with the field at time $\eta$, which then defines a time-dependent probability density function 
(PDF), $\mathcal{P} \left[ \Theta_\eta; \eta \right]$, as follows
\begin{equation} \label{eq:tspt_gen_functional}
    Z \left[ J; \eta \right] = \int \mathcal{D} \Theta_\eta \; \mathcal{P} \left[ \Theta_\eta; \eta \right] \exp \left\{ \int \dd^3k \; \Theta_\eta \left(\mathbf{k}\right) J \left(-\mathbf{k}\right) \right\} \,.
\end{equation}
Note that $\mathcal{P} \left[ \Theta_\eta; \eta \right]$ not only generalises~\cref{eq:spt_gen_functional}, which defines 
a time-dependent PDF, but also takes into account deviations from the Gaussian one, if any. The equation that determines the 
time evolution of $\mathcal{P}$ is nothing more than the classical Liouville equation, where the value of the field at a given 
$\mathbf{k}$ is considered as a stochastic variable. It can be derived by performing a substitution of the integration variables 
in~\cref{eq:tspt_gen_functional} in terms of the fields at time $\eta + \delta \eta$, namely 
\begin{equation} \label{eq:change_variables}
    \Theta_{\eta + \delta \eta} \left(\mathbf{k}\right) = \Theta_{\eta} \left(\mathbf{k}\right) + \delta \eta \dot{\Theta}_{\eta} \left(\mathbf{k}\right) \stackrel{\left(\ref{eq:eq_of_motion}\right)}{=} \Theta_{\eta} \left(\mathbf{k}\right) + \delta \eta \mathcal{I} \left[ \Theta_\eta ; \eta , \mathbf{k} \right] \,.
\end{equation}
This change of variables affects all the factors appearing in~\cref{eq:tspt_gen_functional}. Then, asking for the invariance of $Z \left[ J; \eta \right]$ one 
finally obtain the following time evolution equation 
\begin{equation} \label{eq:liouville_eq}
    \frac{\partial}{\partial \eta} \mathcal{P} \left[ \Theta_{\eta} ; \eta \right] + \int \dd^3p \; \frac{\delta}{\delta \Theta_\eta \left( \mathbf{p} \right)} \left( \mathcal{I} \left[ \Theta_\eta; \mathbf{p} , \eta \right] \mathcal{P} \left[ \Theta_\eta ; \eta \right] \right) = 0 \,,
\end{equation}
which, as expected, is the functional version of Liouville's equation for the conservation of probability. We note that once 
$\mathcal{P}$ is found from~\cref{eq:liouville_eq}, it is possible to compute all the different correlation functions 
using~\cref{eq:tspt_gen_functional}. In this way, TSPT allows to disentangle the time evolution of the fields from their 
statistical averaging. From now on we will omit the subindex $\eta$ on the field $\Theta$ whenever it appears as an argument of 
the distribution function.
    
In the spirit of a perturbative expansion, in analogy of what it is usually done in field theories (FTs) it is convenient 
to search for solution of~\cref{eq:liouville_eq} in the form 
\begin{equation}
    \mathcal{P} \sround{\Theta ; \eta} = \mathcal{N}^{-1} \exp \left\{ - \sum_{n=1}^{\infty} \frac{1}{n!} \int \prod_{i=1}^n \dd^3 k_i \; \Theta \round{\mathbf{k}_i} \Gamma_n^\text{tot} \round{\eta; \mathbf{k}_1, \cdots, \mathbf{k}_n } \right\} \,,
\end{equation}
where $\Gamma_n^\text{tot}$ is reminiscent of the power expansion of the 1-particle irreducible (1PI) FT effective action. This 
expression allow us to translate the conservation of probability in a hierarchy of equations which replace the dynamical equation 
of SPT, 
\begin{align}
    & \dot{\Gamma}_n^\text{tot} \round{ \eta ; \mathbf{k}_1, \dots, \mathbf{k}_n } + \sum_{m=1}^n \frac{1}{m!\round{n-m}!} \notag \\
    & \quad \; \times \sum_\sigma I_m \round{ \eta ; \mathbf{k}_{\sigma \round{1}}, \dots, \mathbf{k}_{\sigma \round{m}}} \Gamma^\text{tot}_{n-m+1} \round{\eta; \sum_{l=1}^m \mathbf{k}_{\sigma \round{l}} , \mathbf{k}_{\sigma \round{m+1}}, \dots, \mathbf{k}_{\sigma \round{n}} } = \notag \\
    & \qquad \;\; = \delta^{\round{3}} \round{ \sum_{i=1}^n \mathbf{k}_i } \int \dd^3 p \; I_{n+1} \round{\eta; \mathbf{p}, \mathbf{k}_1 , \dots , \mathbf{k}_n } \,, \label{eq:gamman_rec}
\end{align}
where in the second term on the lhs (left hand side) the sum runs over all permutations $\sigma$ of $n$ indices.

It is useful to decompose the solutions of this equation as
\begin{equation}
    \Gamma_n^\text{tot} = \Gamma_n + C_n \,,
\end{equation}
where $\Gamma_n$ are the solutions of the homogeneous equations with initial conditions matching the statistical distribution of 
the primordial fluctuations, while $C_n$ are the solutions of the non-homogeneous equations with vanishing initial conditions. The 
$\Gamma_n$ vertices have the physical meaning of the 1PI contributions to the tree-level correlators, and $C_n$ are counterterms 
whose role is to cancel UV divergences in the loop corrections.\footnote{To avoid confusion, let us stress that the $C_n$ differ 
from the counterterms of FTs in that their values cannot be adjusted at will: they are uniquely fixed by the solution of 
the Liouville equation. They arise to cancel certain UV divergences that arise from the (singular) Jacobian describing the change 
in functional measure when going from~\cref{eq:spt_gen_functional} to~\cref{eq:tspt_gen_functional}. A proper UV renormalisation 
of the theory is likely to require additional counterterms to capture the true physical effects of the short modes.} For the 
purposes of this paper, the expressions for $C_n$ are not needed.

\subsection{Featureless perturbation theory}
In order to solve the regular part of~\cref{eq:gamman_rec}, we assume to be in an Einstein-de Sitter (EdS) universe. This is a 
good approximation for the study of LSS (see section~2 of Ref.~\cite{Blas:2016sfa}) and leads a great simplification of equations. 
For a more realistic $\Lambda$CDM cosmology, the deviations from a EdS regime can be taken into account perturbatively. These 
assumptions allow us to assume that the evolution kernels $I_n$ are time-independent. By resorting to the statistical homogeneity 
of the fields, $\left \langle \Theta \right \rangle = 0 $, it is possible to set $\Gamma_1 = 0$, while for $n \geq 2$ we use the 
ansatz
\begin{equation}
    \Gamma_n \round{\eta , \mathbf{k}_1 , \dots, \mathbf{k}_n } = \sum_{l=2}^n e^{-l\eta} \Gamma^{\round{l}}_n \round{\mathbf{k}_1 , \dots, \mathbf{k}_n } \,,
\end{equation}
which separates time and momentum dependence. This, together with~\cref{eq:gamman_rec}, gives the relations for 
$\Gamma_n^{\round{l}}$ with $2 < l < n$, 
\begin{align}
    & \Gamma_n^{\round{l}} \round{ \mathbf{k}_1, \dots, \mathbf{k}_n } = - \frac{1}{n-l} \sum_{m=2}^{n-l+1} \frac{1}{m!\round{n-m}!} \notag \\
    & \quad \, \times \sum_{\sigma} I_m \round{ \mathbf{k}_{\sigma \round{1}}, \dots, \mathbf{k}_{\sigma \round{m}}}  \Gamma^{\round{l}}_{n-m+1} \round{\sum_{i=1}^m \mathbf{k}_{\sigma \round{i}} , \mathbf{k}_{\sigma \round{m+1}}, \dots, \mathbf{k}_{\sigma \round{n}} } \,. \label{eq:recursive_rel}
\end{align}
The remaining vertices, $\Gamma_n^{\round{n}}$, should be fixed from initial conditions.    

In case of Gaussian initial conditions, the problem is simplified: the PDF $\mathcal{P} \sround{\Theta, \eta}$ must reduce to 
a Gaussian distribution at early times after having rescaled the fields with the linear growth factor 
$D \round{\eta} = e^\eta$.\footnote{As the discussion now focuses on the specific problem of cosmological perturbation evolution, 
$\eta$ is no longer just a generic time coordinate; rather, it is that which allows perturbation equations to be written without 
any explicit time dependence, as is customary in the study of LSS evolution. It is defined as $\eta \equiv \ln D \round{\tau}$, 
where $\tau$ is the conformal time.} Basically, one can require
\begin{equation} \label{eq:gauss_ic}
    \lim_{\eta \rightarrow -\infty} \mathcal{P} \sround{ D \round{\eta} \Tilde{\Theta}, \eta } = \mathcal{N} \exp \bround{ -\int \dd^3 k \; \frac{\Tilde{\Theta}_\mathbf{k} \Tilde{\Theta}_{-\mathbf{k}}}{2 P_L \round{k}} } \,,
\end{equation}
where $P_L \round{k}$ is the linear power spectrum.
Under these assumptions one finds that the $\Gamma_n$ vertices have a universal dependence on time 
\begin{equation}
    \Gamma_n \round{\eta ; \mathbf{k}_1, \dots, \mathbf{k}_n} = \frac{1}{g^2 \round{\eta}} \bar{\Gamma}_n \round{\mathbf{k}_1, \dots, \mathbf{k}_n} \,,
\end{equation}
where $g \round{\eta} \equiv D \round{\eta}$ plays the role of TSPT coupling constant. In particular, all the vertices are sourced 
by the Gaussian initial weight inherited by $P_L \round{k}$ and propagated from the simplest vertex $\Gamma^{\round{2}}_2$ to all 
the other through recursion relations like those of~\cref{eq:recursive_rel}. In TSPT it is possible to build a perturbative structure 
upon expanding the generating functional of~\cref{eq:tspt_gen_functional} around the Gaussian part of $\mathcal{P}$, which is 
equivalent to an expansion in the coupling constant. At this step the analogy with the FTs is complete and it is possible to 
represent this calculation as a sum of Feynman diagrams, whose first elements (for the regular vertices) are given by 
\begin{equation}
    \begin{tikzpicture}[baseline=(b)]
          \begin{feynman}[inline=(b.base)]
              \vertex (a);
            \vertex [right=1.5cm of a] (b);						
              \diagram*{
                (a) -- [edge label=$\mathbf{k}$] (b),
            };
          \end{feynman}
    \end{tikzpicture} 
    \quad = \quad g^2 \round{\eta} P_L \round{k} \,,
    \hspace{.3cm}
    \begin{tikzpicture}[baseline=(b)]
        \begin{feynman}[inline=(b.base)]
            \vertex (a);
            \vertex [right=1.2cm of a] (b);		
            \vertex [above left=1.2cm of a] (c);
            \vertex [below left=1.2cm of a] (d);			
            \diagram*{
              (b) -- [edge label=$\mathbf{k}_3$] (a),
              (c) -- [edge label=$\mathbf{k}_2$] (a),
              (d) -- [edge label=$\mathbf{k}_1$] (a),
          };
        \end{feynman}
        \draw [fill=black] (a) circle(2pt);
  \end{tikzpicture} 
  \quad = \quad -\frac{1}{g^2 \round{\eta}} \frac{1}{3!} \bar{\Gamma}_3 \round{\mathbf{k}_1, \mathbf{k}_2, \mathbf{k}_3} \,.
\end{equation}
Let us notice that, in contrast to SPT, time does not flow along the diagrammatic elements, but is taken care of by the time 
dependence of the coupling constant. To calculate an $n$-point correlation function of the field $\Theta$ one needs to draw all 
diagrams with n external legs. It is straightforward to see that diagrams with higher number of loops are proportional to higher 
powers of $g\round{\eta}$. One should also include vertices corresponding to counterterms $C_n$, with $n \geq 1$, in order to 
subtract certain UV divergences in loop diagrams. In this sense counterterms appear quite naturally in TSPT.

\subsection{Application to baryon acoustic oscillations}
The key objects for describing LSS are the matter density contrast, $\delta$, and the velocity divergence, $\Theta$, fields. 
The nonlinear evolution is captured by the cosmological perturbation theory, which uses as seeds the linear 
fields $\delta_L$ and $\Theta_L$ evolved up to the present epoch as if the perturbations were always in the linear regime. In 
the case of adiabatic initial conditions corresponding to a growing mode, the two linear fields are identical, and they are related 
to the curvature perturbation, $\zeta$, by a transfer function $T \left(\eta_0, \mathbf{k}\right)$ which encodes the evolution of 
perturbations eventually from cosmic inflation, through recombination, up to a given time $\eta_0$. So we write 
\begin{equation}
    \Theta_L \round{\mathbf{k}} = \delta_L \round{k} \equiv T \left(\eta_0, \mathbf{k}\right) \zeta \round{\mathbf{k}} \,.
\end{equation}
The statistical properties of the field $\delta_L$ are captured by the power spectrum, defined as the Fourier transform of the 
two-point correlation function, given by 
\begin{equation}
    \mybraket{\delta_L \round{\mathbf{k}} \delta_L \round{\mathbf{k} '}} \equiv \round{2\pi}^3 \delta^{\round{3}} \round{\mathbf{k} + \mathbf{k}'} P_L \round{k} \,.
\end{equation}
In this way we can relate the linear matter power spectrum to the primordial power spectrum of curvature perturbations 
as\footnote{From now on we will omit the time dependence of the matter power spectrum for simplicity.}
\begin{equation} \label{eq:lin_evol}
    P_L \left(k\right) = \left[ T \left(\eta_0, \mathbf{k}\right) \right]^2 P_\zeta \left( k \right) \,.
\end{equation}

To start with the TSPT formalism, it is convenient to decompose the linear power spectrum into a smooth (non-wiggly, nw) and 
an oscillatory (wiggly, w) part, which describes the imprint of the BAO and any other oscillations, as
\begin{equation} \label{eq:matt_ps_split}
    P_L \left( k\right) = P_L^\mathrm{nw} \left( k \right) + P_L^\mathrm{w} \left( k \right) \,.
\end{equation}

The vertices (limiting ourselves to the case of Gaussian fields), $\bar{\Gamma}_n$, are functionals of the linear power spectrum. 
It is therefore natural to expand them to the first order in $P_L^\mathrm{w}$ to separate them into non-wiggly and wiggly parts 
\begin{equation}
    \bar{\Gamma}_n \round{\mathbf{k}_1, \dots, \mathbf{k}_n} = \bar{\Gamma}_n^\mathrm{nw} \round{\mathbf{k}_1, \dots, \mathbf{k}_n} + \bar{\Gamma}_n^\mathrm{w} \round{\mathbf{k}_1, \dots, \mathbf{k}_n} \,. 
\end{equation}
As in Ref.~\cite{Blas:2016sfa}, let us split the momenta into hard $\left\{ \mathbf{k}_i \right\}$ and soft 
$\left\{ \mathbf{q}_j \right\}$ and analyze the structure of the wiggly vertices in the limit $\varepsilon \sim q/k \ll 1$. 
Consider first the cubic vertex. The leading contribution in the soft limit is 
\begin{equation}
    \Gamma_3'{}^\text{w} \round{\mathbf{k}, -\mathbf{k} -\mathbf{q}, \mathbf{q}} \simeq \frac{\round{\mathbf{q} \cdot \mathbf{k}}}{q^2} \frac{P_L^\mathrm{w} \round{\abs{\mathbf{k} + \mathbf{q}}} - P_L^\mathrm{w} \round{k}}{\sround{P_L^\mathrm{nw} \round{k}}^2} \,.
\end{equation}
The pole of the first factor cancels the numerator of the second factor at $q \rightarrow 0$. However, the cancellation does 
not occur if $q$ is greater than the period of the oscillations, $q \gtrsim k/ \omega$, 
where $\omega$ refers to the frequency of the BAO,\footnote{Note that in presence of different oscillatory pattern, this frequency $\omega$ must be replaced with the quantity which describes the modulation of feature signal.} 
in which case the contribution is enhanced by $1/\varepsilon$. To keep track of 
this enhancement, we introduce the linear operator $\mathcal{D}_\mathbf{q}$ acting on the wiggly power spectrum, 
\begin{align} 
    \mathcal{D}_\mathbf{q} P_L^\mathrm{w} \round{\mathbf{k}} &\equiv \frac{\round{\mathbf{q} \cdot \mathbf{k}}}{q^2} \sround{P_L^\mathrm{w} \round{\abs{\mathbf{k} + \mathbf{q}}} - P_L^\mathrm{w} \round{k}} \notag\\
    &= \left. \frac{\round{\mathbf{q} \cdot \mathbf{k}}}{q^2} \sround{\exp \round{\mathbf{q} \cdot \nabla_{\mathbf{k'}}} -1} P_L^\mathrm{w} \round{k'} \right\rvert_{\mathbf{k}' = \mathbf{k}} \,. \label{eq:d_op_def}
\end{align}
It is useful here to introduce a \textit{translation operator}, 
$\mathcal{T}_\mathbf{q} \equiv \exp \round{\mathbf{q} \cdot \nabla_{\mathbf{k}}}$, in analogy to quantum mechanics, which, acting 
on the linear power spectrum, shifts its evaluation point in momentum space by $\mathbf{q}$. Then, the second equality 
in~\cref{eq:d_op_def} follows by the Taylor series expansion of the power spectrum 
\begin{align}
    P_L^\mathrm{w} \left( \abs{\mathbf{k} + \mathbf{q}} \right) &= P_L^\mathrm{w} \left(k\right) + \mathbf{q} \cdot \nabla_\mathbf{k} P_L^\mathrm{w} \left(k\right) + q_i q_j \partial_{k_i} \partial_{k_j} P_L^\mathrm{w} \left(k\right) + \cdots \notag \\
    &= \exp\left(\mathbf{q} \cdot \nabla_{\mathbf{k}} \right) P_L^\mathrm{w} \left(k\right) = \mathcal{T}_\mathbf{q} P_L^\mathrm{w} \left(k\right) \,.
\end{align}
Each insertion of this operator is treated as a quantity of order $1/\varepsilon$. Next, it can be shown that the leading 
contribution in an $n$-point vertex with $n-2$ soft momenta is of order $\round{1/\varepsilon}^{n-2}$ and has the form 
\begin{equation} \label{eq:n_soft_vertex}
    \bar{\Gamma}_n'{}^\mathrm{w} \round{\mathbf{k}, -\mathbf{k} - \sum_{i=1}^{n-2} \mathbf{q}_i, \mathbf{q}_1, \dots, \mathbf{q}_{n-2}} \simeq \frac{\round{-1}^{n-1}}{\sround{P_L^\mathrm{nw} \round{k}}^2} \prod_{j=1}^{n-2} \mathcal{D}_{\mathbf{q}_j} P_L^\mathrm{w} \round{k} \,.
\end{equation}
The proof of this formula can be found in appendix C of Ref.~\cite{Vasudevan:2019ewf}. 

The most IR-enhanced diagrams among the loop corrections to the wiggly power spectrum are the so-called \textit{daisy} diagrams, 
which are obtained by dressing the wiggly vertices with soft loops. Thus, at the leading IR order, we have 
\begin{equation}
    P^{\rm w,\,LO}_{\delta\delta}(k) \quad = \quad 
    \begin{tikzpicture}[baseline=(b)]
          \begin{feynman}[inline=(b.base)]
              \vertex (a);
            \vertex [right=1.5cm of a] (b);						
              \diagram*{
                (a) -- [photon] (b),
            };
          \end{feynman}
    \end{tikzpicture} 
    \quad + \quad 
    \begin{tikzpicture}[baseline=(b)]
          \begin{feynman}[inline=(b.base)]
              \vertex [blob] (b) {};
            \vertex [above=1cm of b] (b1);
            \vertex [left=1cm of b] (a);
            \vertex [right=1cm of b] (c);
              \diagram*{
                (a) -- (b) -- (c),
                (b) -- [out=45,in=0, min distance=0.5cm] (b1),
                (b) -- [out=135,in=180, min distance=0.5cm] (b1),
            };
          \end{feynman}
    \end{tikzpicture}
    \quad + \quad 
    \begin{tikzpicture}[baseline=(b)]
          \begin{feynman}[inline=(b.base)]
              \vertex [blob] (b) {};
            \vertex [above=1cm of b] (b1);
            \vertex [below=1cm of b] (b2);
            \vertex [left=1cm of b] (a);
            \vertex [right=1cm of b] (c);
              \diagram*{
                (a) -- (b) -- (c),
                (b) -- [out=45,in=0, min distance=0.5cm] (b1),
                (b) -- [out=135,in=180, min distance=0.5cm] (b1),
                (b) -- [out=-45,in=0, min distance=0.5cm] (b2),
                (b) -- [out=-135,in=180, min distance=0.5cm] (b2),
            };
          \end{feynman}
    \end{tikzpicture}
    \quad + \quad \dots \,,
\end{equation}
where we have used a wavy line and shaded circles to denote the wiggly linear power spectrum and the wiggly vertices respectively. 
The smooth lines represent the non-wiggly propagator $g^2 P_L^\mathrm{nw} \round{k}$. The term with $\ell$ loops in this expression 
is of the order $g^2 \round{g^2/ \varepsilon^2}^\ell$. We see that the loop suppression represented by $g^2$ is partially compensated 
by the IR enhancement, so these contributions must be resummed. The leading part of the diagram with $\ell$ loops reads  
\begin{align}
    &P_{\delta \delta}^{\mathrm{w},\, {\rm LO},\, \ell-\text{loop}} \round{\eta, k} = - \frac{1}{\round{2 \ell + 2}!} \cdot \round{2\ell +2} \cdot \round{2\ell +1} \cdot \round{2\ell -1}!! \cdot g^4 \round{\eta} \sround{P_L^\mathrm{nw} \round{k}}^2 \notag \\
    & \quad\quad \,\, \times \prod_{i=1}^{\ell} \sround{\int \dd^{3} q_i \, g^2 \round{\eta} P_L^\mathrm{nw} \round{q_i}} g^{-2} \round{\eta} \bar{\Gamma}'{}_{2\ell +2}^\mathrm{w} \round{\mathbf{k}, -\mathbf{k}, \mathbf{q}_1, -\mathbf{q}_1, \dots, \mathbf{q}_\ell, -\mathbf{q}_\ell} \notag \\
    & \quad \, \simeq \frac{g^{2\ell +2} \round{\eta}}{2^\ell \ell!} \prod_{i=1}^\ell \sround{ \int_{q_i \leq k_{\rm S}} \dd^{3} q_i \, P_L^\mathrm{nw} \round{q_i} \mathcal{D}_{\mathbf{q}_i} \mathcal{D}_{-\mathbf{q}_i}} P_L^\mathrm{w} \round{k} \label{eq_l_loop_spectrum} \,, 
\end{align} 
where we have used~\cref{eq:n_soft_vertex} to go to the second line. We have also restricted the loop integrals to the IR domain 
$g \leq k_{\rm S}$. The separation scale $k_{\rm S}$ defining this domain must be in the range $k/\omega < k_{\rm S} < k$. Otherwise 
its choice is arbitrary and represents an intrinsic freedom in the resummation scheme. The sensitivity of the final result to the 
precise value of $k_{\rm S}$ provides an estimate of the theoretical uncertainty due to considering only the IR part of the loops 
and omitting the integrals over the hard momenta $q>k_{\rm S}$. This sensitivity decreases when higher orders in the hard loops are 
included in the calculation; see Refs.~\cite{Vasudevan:2019ewf,Blas:2016sfa}. 

Adding up the contributions~\cref{eq_l_loop_spectrum} with all $\ell$ we obtain the leading-order IR-resummed wiggly power spectrum 
in a closed form, 
\begin{equation}
    P_{\delta \delta}^{\mathrm{w}, \text{LO}} \round{\eta , k} = g^2 \round{\eta} e^{-g^2 \round{\eta} \mathcal{S}} P_L^\mathrm{w} \round{k} \,,
\end{equation}
where we introduced a new operator 
\begin{equation} \label{eq:s_operator}
    \mathcal{S} \equiv - \frac{1}{2} \int_{q \leq k_S} \dd^3 q \, P_L^\mathrm{nw} \round{q} \mathcal{D}_\mathbf{q} \mathcal{D}_{-\mathbf{q}} \,.
\end{equation}
It should be emphasised that this operator encodes the damping of the oscillations by IR resummation of the large-scale bulk flows. 
The total IR-resumed power spectrum at the leading order is obtained by adding the non-wiggly contribution, which remains unchanged, 
\begin{equation} \label{eq:lo_power_spec}
    P_{\delta\delta}^{\text{LO}} \round{\eta , k} = g^2 \round{\eta} \sround{ P_L^\mathrm{nw} \round{k} + e^{-g^2 \round{\eta} \mathcal{S}} P_L^\mathrm{w} \round{k}} \,.
\end{equation}
The resummation procedure can be extended to include the corrections from hard loops. We will not repeat the derivation here and 
refer the interested reader to Ref.~\cite{Blas:2016sfa}. For the next-to-leading order (NLO) in the hard loops, the result is 
\begin{align} \label{eq:nlo_power_spec}
    P_{\delta \delta}^{\text{NLO}} \round{\eta , k} &= g^2 \round{\eta} \sround{ P_L^\mathrm{nw} \round{k} + \round{1 + g^2\round{\eta} \mathcal{S} } e^{-g^2 \round{\eta} \mathcal{S} } P_L^\mathrm{w} \round{k} } \notag \\
    & \quad \, + g^4\round{\eta} P_{\delta \delta}^{1-\text{loop}} \sround{P_L^\mathrm{nw} \round{k} + e^{-g^2 \round{\eta} \mathcal{S}} P_L^\mathrm{w} \round{k}} \,,
\end{align}
where the last term is the usual 1-loop correction to the power spectrum evaluated using the linear spectrum with damped 
oscillations corresponding to the LO result.

\section{Inclusion of primordial oscillatory features} \label{sec:osc_features}
In this section, we will first outline the general procedure for incorporating primordial oscillatory features into a TSPT-like 
approach to perturbation theory. This procedure is completely general in the sense that it does not depend on the particular 
shape of the wiggly part of the power spectrum.
We then focus on two cases of interest: linear and logarithmic primordial oscillations. This discussion is mainly based on 
Refs.~\cite{Blas:2016sfa,Vasudevan:2019ewf} where we refer the interested reader for further details. 

Before we look at the details of IR resummation with oscillatory features, we want to say a few words about its theoretical 
modelling, since we have tried to stay as general as possible so far. In the following, we will assume that the PPS can be written as 
    \begin{equation} \label{eq:prim_spectrum_split}
    P_\zeta \round{k} = P_{\zeta,0} \round{k} \sround{ 1 + \delta P^X_\zeta \round{k} } \,,
\end{equation}
where 
\begin{equation}
    P_{\zeta, 0} \equiv \frac{2 \pi^2}{k^3} A_\text{s} \round{\frac{k}{k_*}}^{n_\text{s}-1} \,,
\end{equation}
is the standard almost scale-invariant PPS predicted by the simplest inflationary models while $\delta P_\zeta \round{k}$ encodes 
any deviations from the scale invariance, the primordial feature. Here, $A_\mathrm{s}$ and $n_\mathrm{s}$ are the scalar amplitude 
and spectral index at the pivot scale $k_*$, which is assumed to be $k_* = 0.05 \, \mathrm{Mpc}^{-1}$. We consider primordial 
features with linearly- and logarithmically-spaced oscillations with $X = \bround{\mathrm{lin}, \mathrm{log}}$.

Within this work, we consider the following generic parameterisation
\begin{equation} \label{eqn:Pk_template}
    \delta P_\zeta^\text{X} \round{k} = A \round{k} \sin \round{\omega_\text{X} \Xi_\text{X} + \varphi_\text{X}} \,,
\end{equation}
where $\Xi_\text{X} = \bround{ k/k_*, \ln \round{k/k_*} }$, which we refer to as \textit{linear features} (LIN) and 
\textit{logarithmic features} (LOG), respectively. On top of the sinusoidal component, one can additionally consider the presence 
of a scale-dependent amplitude. We consider three different cases: one with a constant amplitude and two with a scale-dependent 
amplitude modulated by a Gaussian function (to which we refer to as \textit{wave packet}, WP) or by a power law (PL). We parameterise 
the generic amplitude appearing in~\cref{eqn:Pk_template} as
\begin{subnumcases}{A \round{k} =}
    A_\text{X} \label{eq:amp} \\ 
    A^\text{WP}_\text{X} \exp \sround{-\round{k-\mu}^2 / 2\Delta^2 } \label{eq:gauss_amp} \,.\\ 
    A^\text{PL}_\text{X} \round{k/k_\mathrm{PL}}^n \label{eq:pl_amp} 
\end{subnumcases} 
  
The templates described by~\cref{eqn:Pk_template} cover a wide range of primordial oscillatory signals. Linear oscillations in 
the PPS of density perturbations can be generated by abrupt changes in the background parameters, also known as {\em sharp features}, 
at certain moments during the evolution of the inflationary dynamics. These sudden shifts give rise to transient oscillatory 
features both in single-field inflationary 
models~\cite{Starobinsky:1992ts,Adams:2001vc,Chen:2006xjb,Chen:2008wn,Bean:2008na,Miranda:2012rm} and in multi-field 
models~\cite{Achucarro:2010da,Chen:2011zf,Braglia:2020fms}. 
Logarithmic oscillations, also known as {\em resonant features}, can arise from periodic modulations in the inflationary potential, as 
seen in Refs.~\cite{Chen:2008wn,Flauger:2009ab,Chen:2011zf,Flauger:2010ja,Chen:2010bka}. The amplitude of these oscillatory signals 
is highly model dependent, influenced by the nature of the primordial feature for both case with linear and logarithmic 
oscillations~\cite{Starobinsky:1992ts,Adams:2001vc,Armendariz-Picon:2003knj,Achucarro:2010da,Chen:2011zf}. 
Starting from templates that assume oscillatory features with a constant amplitude~\eqref{eq:amp}, we study two different cases 
where the envelope evolves as a Gaussian~\eqref{eq:gauss_amp} or as a power-law~\eqref{eq:pl_amp}. In addition to these behaviours, 
some models, such as the {\em classical primordial standard clock} models, predict a complex superposition of both types of linear 
and logarithmic features~\cite{Chen:2011zf,Braglia:2022ftm} and other models like 
{\em axion monodromy inflation}~\cite{McAllister:2008hb,Silverstein:2008sg} predict a running frequency, further extending 
the testable phenomenology.
    
The information in the PPS is transferred to the matter power spectrum by the evolution of~\cref{eq:lin_evol}. Given the linearity 
property of the transfer function, it is natural to assume a split like that of~\cref{eq:prim_spectrum_split} also in the matter power 
spectrum. This is in perfect analogy to what has been done for the description of BAO in~\cref{eq:matt_ps_split}, so it is natural 
to constrain the feature models as contributions to the BAO spectrum. In presence of primordial features, it is useful to further 
separate the wiggly component of the matter power spectrum as 
\begin{equation} \label{eq:wiggle_spectrum}
    P^\text{w} \left(k \right) \equiv P_\text{BAO}^\text{w} \left(k\right) + P_X^\text{w} \left(k\right) + P_\text{BAO}^\text{w} \left(k\right) \delta P^X_\zeta \left(k\right) \,,
\end{equation}
where $P_\mathrm{BAO}^\mathrm{w} \round{k}$ is the standard BAO features in a flat $\Lambda$CDM cosmology, which can be 
parameterised as~\cite{Seo:2007ns}
\begin{equation}
    P^\mathrm{w}_\mathrm{BAO} \round{k} = P^\mathrm{nw} \round{k} \delta P^\mathrm{BAO} \round{k} \simeq A_\mathrm{BAO} \round{k} \sin \round{ \tilde{\omega}_\text{BAO} k } \, .
\end{equation}
Here, $\tilde{\omega}_\text{BAO} \simeq 110 \,h^{-1}\,\mathrm{Mpc}$ is the frequency of the BAO~\cite{Planck:2018vyg}, 
$A_\mathrm{BAO} \round{k}$ is the amplitude of the BAO, where the $k$ dependence stems for the modulation originated by the 
diffusion damping, which start to be efficient around the Silk damping scale~\cite{Silk:1967kq,Hu:1995en,Eisenstein:1997ik}. 
The component proportional to the primordial features is given by 
\begin{equation}
    P_X^\mathrm{w} \round{k} = P^{\mathrm{nw}} \round{k} \delta P^X_\zeta \round{k} \,.
\end{equation}
Finally, the third term in~\cref{eq:wiggle_spectrum} is the cross-correlation term between the BAO and the primordial features. 
Since it is proportional to $A_\mathrm{BAO} \times A_X$, where $A_\mathrm{BAO} \approx 0.07$ and $A_X < 1$, it is subdominant and therefore it has been usually neglected in the literature.

\subsection{Linear oscillations} \label{sec:lin_osc}
The key ingredient to compute the damping of oscillatory features by IR resummation is to evaluate the action of the $\mathcal{S}$ 
operator, defined in~\cref{eq:s_operator}, on the analytic expression of the wiggly part of the power spectrum. As already discussed 
in~\cref{sec:osc_features}, the first step is the evaluation of the translation operator, which we recall here to be defined as  
\begin{equation} \label{eq:t_op_sum}
    \mathcal{T}_\mathbf{q} \sround{ \; . \;} = \sum_{n=0}^\infty \frac{1}{n!} \left( \mathbf{q} \cdot \nabla_{\mathbf{k}} \right)^n \,. 
\end{equation}
From this equation it is clear how the problem moves to the evaluation of the derivative operator 
$\left( \mathbf{q} \cdot \nabla_{\mathbf{k}} \right)$ and its integer powers. Considering the case of linear oscillations with 
constant amplitude, as introduced in~\cref{eqn:Pk_template}, we have to evaluate the following quantity
\begin{equation}
    \mathbf{q} \cdot \nabla_{\mathbf{k}} \sround{ A_\text{lin} \sin \round{\omega_\text{lin} \frac{k}{k_*} + \varphi_\text{lin}} } \,.
\end{equation}
The next steps become simpler by considering separately the complex exponential functions building the sine function, on which 
the derivative operator of our interest act as 
\begin{equation}
    \mathbf{q} \cdot \nabla_{\mathbf{k}} e^{\pm i \omega_\text{lin} k/k_*} = \pm i \frac{\omega_\text{lin}}{k_*} \frac{\left(\mathbf{q} \cdot \mathbf{k}\right)}{k} e^{\pm i \omega_\text{lin} k/k_*} \,.
\end{equation}
Thanks to this relation we can now start looking to the action of the translation operator on the exponential components of our parameterisation. By explicitly carrying out the first two steps of the series we find 
\begin{align} \label{eq:damping_approx_lin}
    &\mathcal{T}_\mathbf{q} \sround{e^{\pm i \omega_\text{lin} k/k_*}} = \sum_{n=0}^{\infty} \left( \mathbf{q} \cdot \nabla_{\mathbf{k}} \right)^{\left( n-1 \right)} \left[ \pm i \frac{\omega_\text{lin}}{k_*} \frac{\left(\mathbf{q} \cdot \mathbf{k}\right)}{k} e^{\pm i \omega_\text{lin} k/k_*} \right] \notag \\
    & \quad \, = \sum_{n=0}^{\infty} \round{\mathbf{q} \cdot \nabla_\mathbf{k}}^{\round{n-2}} \bround{\pm i \frac{\omega_\mathrm{lin}}{k_*} \sround{ \pm i \frac{\omega_\mathrm{lin}}{k_*} \frac{\round{\mathbf{q} \cdot \mathbf{k}}^2}{k^2} + \frac{q^2}{k} - \frac{\round{\mathbf{q} \cdot \mathbf{k}}^2}{k^3}} e^{\pm i \omega_\mathrm{lin} k/k_*} } \,.
\end{align}
The terms in the squared brackets are the second derivative of the exponential factor (analogous to the previous step) plus the 
derivatives of the polynomial factor in front of the exponential in the first line of~\cref{eq:damping_approx_lin}. We notice that it 
is not possible to write this sum with a closed-form because of the problem of the $n$th derivatives of a fraction and therefore 
it is not possible to write the action of the translation operator for our linear feature. 
The common practice here is to recall the hierarchy between soft and hard momenta, and noticing that all the corrections coming 
from subsequent derivatives of $\round{\mathbf{q} \cdot \mathbf{k}}/k$ are of order $\mathcal{O}\round{\varepsilon}$, and thus 
negligible in the limit of soft momenta corrections. Under this approximation, we can finally write the action of the translation 
operator as 
\begin{equation} \label{eq:t_op_lin}
    \mathcal{T}_\mathbf{q} \sround{e^{\pm i \omega_\text{lin} k/k_*}}  = e^{\pm i \frac{\omega_\mathrm{lin}}{k_*} \frac{\round{\mathbf{q} \cdot \mathbf{k}}}{k} } e^{\pm i \omega_\text{lin} k/k_*} \,. 
\end{equation}
As we are going to show, the possibility to build this kind of relationship is fundamental to compute in an analytic way the damping 
due to IR resummation of the features. 

To compute the 1-loop corrections to the matter power spectrum we need now to evaluate the derivative operator defined 
in~\cref{eq:s_operator} on the wiggly part of the spectrum, corresponding to~\cref{eqn:Pk_template}. We start by noticing that 
using~\cref{eq:d_op_def}, we can cast~\cref{eq:s_operator} in the following way 
\begin{align} 
    \mathcal{S} P_L^\mathrm{w} \left(k\right) &= + \left. \frac{1}{2} \int_{q \leq k_\mathrm{S}} \frac{\dd^3 q}{\left(2 \pi\right)^3} \; P_L^\mathrm{nw} \left(q\right) \frac{\left(\mathbf{q} \cdot \mathbf{k}'\right)^2}{q^4} \left( \mathcal{T}_\mathbf{q} - 1 \right) \left( \mathcal{T}_{-\mathbf{q}} - 1 \right) P_L^\mathrm{w} \left(k'\right) \right\rvert_{\mathbf{k}'= \mathbf{k}} \notag \\
    &= \left. \int_{q \leq k_\mathrm{S}} \frac{\dd^3 q}{\left(2 \pi\right)^3} \; P_L^\mathrm{nw} \left(q\right) \frac{\left(\mathbf{q} \cdot \mathbf{k}'\right)^2}{q^4} \left[ 1 - \cosh \left( \mathbf{q} \cdot \nabla_{\mathbf{k'}} \right) \right] P_L^\mathrm{w} \left(k'\right) \right\rvert_{\mathbf{k}'= \mathbf{k}} \,. \label{eq:s_op_start}
\end{align} 
It is now easy to show that 
\begin{equation}
    \left[ 1 - \cosh \left( \mathbf{q} \cdot \nabla_{\mathbf{k'}} \right) \right] P_L^\mathrm{w} \left(k'\right) \rvert_{\mathbf{k}'= \mathbf{k}} = \left[ 1- \cos \left( \frac{\omega_\text{lin}}{k_*} \frac{ \mathbf{q} \cdot \mathbf{k} }{k} \right) \right] P_L^\mathrm{w} \left(k\right) \,, \label{eq:operator_action}
\end{equation}
which is true every time that the translation operator has exponential ``eigenvalues'' like in~\cref{eq:t_op_lin}. Thanks to 
this result it is possible to calculate the angular part of the integral in~\cref{eq:s_op_start} leaving us with 
\begin{align} 
    \mathcal{S} P_L^\mathrm{w} \left(k\right) &= P_L^\mathrm{w} \left( k \right) \frac{k^2}{6 \pi^2} \int_0^{k_{\rm S}} \dd q \; P_L^\mathrm{nw} \left( q \right) \left[ 1 - j_0 \left( \frac{\omega_\text{lin}}{k_*} q \right) + 2 j_2 \left( \frac{\omega_\text{lin}}{k_*} q \right) \right] \notag  \\ 
    &\equiv k^2 \Sigma_\text{lin} \left(k \right) P_L^\mathrm{w} \left( k \right) \,. \label{eq:s_lin_osc}
\end{align}
This formula for the damping of linear oscillations coincides with that of the BAO case (up to modifications in the frequency 
from $\omega_{\rm lin}/k_*$ to $\tilde{\omega}_\mathrm{BAO}$), already present in the literature in Ref.~\cite{Blas:2016sfa}. This result 
already appeared in the treatment of primordial linear oscillations in Ref.~\cite{Beutler:2019ojk} (albeit in a slightly different 
formalism) and can be seen here as a test of our methodology and a summary of its main steps. Finally, let us recall that this 
is all we need for the calculation of the LO matter power spectrum, using~\cref{eq:lo_power_spec}, but also for the calculation 
of the NLO in the hard loops, as shown in~\cref{eq:nlo_power_spec}.

\subsubsection{Mixed term between baryon and primordial linear oscillations}
The goal of this section is to further improve the calculation of the nonlinear matter power spectrum with linear oscillatory 
features by the addition of the primordial features to BAO mixed term in~\cref{eq:wiggle_spectrum}. Thanks to some basic trigonometric 
relations,\footnote{From addition and subtraction relations one finds 
\begin{equation}
        \sin \alpha \sin \beta = \frac{1}{2} \left[ \cos \left(\alpha + \beta\right) - \cos \left(\alpha -\beta\right) \right] \,.
\end{equation}} 
we can write 
\begin{align}
    &P_\text{BAO}^\text{w} \left(k\right) \delta P^\text{lin}_\zeta \left(k\right) = \notag \\
    & \quad \, = P_L^\text{nw} \round{k} A_\text{BAO} \round{k} \sin \round{ \tilde{\omega}_\text{BAO} k } A_\text{lin} \sin \left(\omega_\text{lin} \frac{k}{k_*} + \varphi_\text{lin}\right) \notag \\
    & \quad \, = \frac{1}{2} A_\text{BAO} \round{k} P_L^\text{nw} \left(k\right) A_\text{lin} \notag \\
    & \qquad \,\, \times \left\{ \cos \left[ \left( \omega_\text{lin} - k_* \tilde{\omega}_\text{BAO} \right) \frac{k}{k_*} +\varphi_\text{lin} \right] - \cos \left[ \left( \omega_\text{lin} + k_* \tilde{\omega}_\text{BAO} \right) \frac{k}{k_*} +\varphi_\text{lin} \right] \right\} \notag \\
    & \quad \, \equiv \frac{1}{2} A_\text{BAO} \round{k}  \left[ \left. P_\text{lin}^\text{w} \left(k\right) \right\rvert_{\omega_\text{lin} - k_* \tilde{\omega}_\text{BAO}}^{\varphi_\text{lin} +\pi/2 }  - \left. P_\text{lin}^\text{w} \left(k\right) \right\rvert_{\omega_\text{lin} + k_* \tilde{\omega}_\text{BAO}}^{\varphi_\text{lin} +\pi/2 } \right] \,. \label{eq:cross_linear}
\end{align} 
From now on, the vertical line on the side of factors within an equation indicates that all factors to its left should be evaluated using the values to its right. For example, in~\cref{eq:LO_cross_lin},
\begin{equation}
    \noleft P_\text{lin}^\mathrm{w} \left(k\right) \right\rvert_{ \omega_\text{lin} - k_* \tilde{\omega}_\text{BAO}}^{ \varphi_\text{lin} + \pi/2} 
\end{equation}
means that both the damping and the wiggly power spectrum must be evaluated with frequency $\omega_\mathrm{lin} - k_* \tilde{\omega}_\mathrm{BAO}$ and phase $\varphi_\mathrm{lin} + \pi/2$.

From the previous equation we notice that it is possible to describe the mixed term as interference between linear waves of 
different frequencies. Let us notice that the shift in the phase has been introduced to recover the initial expression of the 
wiggly power spectrum.

Given that all operators needed to compute the damping are linear operators and that~\cref{eq:cross_linear} is linear as well, 
we can easily compute the damping sourced by the mixed term using the results of~\cref{sec:lin_osc}. In particular, thanks 
to~\cref{eq:s_lin_osc}, we can write 
\begin{align} 
    & \mathcal{S} \left[ P_\text{BAO}^\mathrm{w} \left(k\right) \delta P^\text{lin}_\zeta \left(k\right) \right] = \notag \\
    & \quad \, = \frac{k^2}{2} A_\text{BAO} \round{k} \left[ \left. \Sigma_\text{lin} \left(k \right) P_\text{lin}^\mathrm{w} \left(k\right) \right\rvert_{\omega_\text{lin} - k_* \tilde{\omega}_\text{BAO}}^{\varphi_\text{lin} + \pi/2} - \left. \Sigma_\text{lin} \left(k \right) P_\text{lin}^\mathrm{w} \left(k\right) \right\rvert_{ \omega_\text{lin} + k_* \tilde{\omega}_\text{BAO}}^{ \varphi_\text{lin} + \pi/2} \right] \,. \label{eq:s_op_cross_lin}
\end{align}
In this case, it is not possible to define a single damping factor for mixed term (in our terminology we could say that it is not an ``eigenfunction'' of the $\mathcal{S}$ operator). However, to proceed with our evaluations we just need to know what has been reported in~\cref{eq:s_op_cross_lin}. The approximations underlying this result are exactly the same used in the case of a single component of linear oscillations. In addition, we neglected corrections coming from the derivatives of the BAO amplitude, which shows a weak dependence on $k$.
Thanks to this result, and those of the previous section, we can finally write the LO matter power spectrum with linear oscillatory features as 
\begin{align}
    &P_{\delta \delta}^\text{LO} \round{\eta, k} = g^2 \round{\eta} \left\{  \vphantom{\frac{1}{2} A_\text{BAO} \round{k} } P_L^\text{nw} \round{k}  +  e^{-g^2 \round{\eta} k^2 \Sigma_\text{BAO} \round{k}} P_\text{BAO}^\text{w} \round{k}  +  e^{-g^2 \round{\eta} k^2 \Sigma_\text{lin} \round{k}} P_\text{lin}^\text{w} \round{k}    \noright \notag \\
    & \quad \, + \frac{1}{2} A_\text{BAO} \round{k} \left[  \noleft e^{ -g^2 \round{\eta} k^2 \Sigma_\text{lin} \left(k \right)} P_\text{lin}^\mathrm{w} \left(k\right) \right\rvert_{ \omega_\text{lin} - k_* \tilde{\omega}_\text{BAO}}^{ \varphi_\text{lin} + \pi/2} 
    - \noleft   \noleft e^{ -g^2 \round{\eta} k^2 \Sigma_\text{lin} \left(k \right)} P_\text{lin}^\mathrm{w} \left(k\right) \right\rvert_{\omega_\text{lin} + k_* \tilde{\omega}_\text{BAO}}^{ \varphi_\text{lin} + \pi/2}   \right] \right\} \,, \label{eq:LO_cross_lin}
\end{align}
where $\Sigma_\mathrm{lin}$ is defined in~\cref{eq:s_lin_osc} while $\Sigma_\mathrm{BAO}$ can be obtained from the latter by substitution of $\omega_\text{lin}/k_*$ with $\tilde{\omega}_\text{BAO}$. 
Using the same ingredients, it is possible to compute also the NLO matter power spectrum, that is 
\begin{align} \label{eq:NLO_cross_lin}
    P_{\delta \delta}^\text{NLO} \round{\eta, k} &= g^2 \round{\eta} \left\{ \vphantom{\noleft e^{ - g^2 \round{\eta} \Sigma_P^\text{lin} \left(k , k_S\right)} P_\text{lin}^w \left(k\right) \right\rvert_{\omega \rightarrow \omega - \tilde{\omega}_\text{BAO}}^{\varphi \rightarrow \varphi + \pi/2}} P_L^\text{nw} \round{k} + g^2 \round{\eta} P^\text{1-loop}_{\delta \delta} \sround{ g^{-2} \round{\eta} P_{\delta \delta}^\text{LO} \round{\eta, k}} \noright \notag \\
    & \quad \, + \round{1 + g^2 \round{\eta} k^2 \Sigma_\text{BAO} \round{k} } e^{-g^2 \round{\eta} k^2 \Sigma_\text{BAO} \round{k}} P_\text{BAO}^\text{w} \round{k} \notag \\
    & \quad \, + \round{1 + g^2 \round{\eta} k^2 \Sigma_\text{lin} \round{k} } e^{-g^2 \round{\eta} k^2 \Sigma_\text{lin} \round{k}} P_\text{lin}^\text{w} \round{k} \notag \\
    & \quad \, + \frac{1}{2} A_\text{BAO} \round{k} \left[ \round{1 + g^2 \round{\eta} k^2 \Sigma_\text{lin} \left(k \right)} \noleft e^{ - g^2 \round{\eta} k^2 \Sigma_\text{lin} \left(k \right)} P_\text{lin}^\mathrm{w} \left(k\right) \right\rvert_{\omega_\text{lin} - k_* \tilde{\omega}_\text{BAO}}^{ \varphi_\text{lin} + \pi/2} \noright \notag \\
    & \qquad \,\, \noleft \noleft - \round{1 + g^2 \round{\eta} k^2 \Sigma_\text{lin} \left(k\right)} \noleft e^{ - g^2 \round{\eta} k^2 \Sigma_\text{lin} \left(k\right)} P_\text{lin}^\mathrm{w} \left(k\right) \right\rvert_{\omega_\text{lin} + k_* \tilde{\omega}_\text{BAO}}^{ \varphi_\text{lin} + \pi/2} \right] \right\} \,. \notag \\
\end{align}
This formula, together with~\cref{eq:LO_cross_lin}, shows how the addition of the mixed term between baryon and primordial 
linear oscillations affects the calculation of the matter power spectrum, which has never been explicitly calculated.

\subsection{Logarithmic oscillations}   \label{sec:log_osc}
The case of logarithmic oscillations, besides some subtleties that we are going to discuss later, is analogous to that of linear 
oscillations. Again, the key ingredient is the evaluation of the translation operator, which pass through the computation of the 
integer powers of $\round{\mathbf{q} \cdot \nabla_{\mathbf{k}}}$ on the wiggly power spectrum of~\cref{eqn:Pk_template}, namely 
\begin{equation}
    \mathbf{q} \cdot \nabla_{\mathbf{k}} \sround{ A_\text{log} \sin \round{\omega_\text{log} \ln \frac{k}{k_*} + \varphi_\text{log}} } \,.
\end{equation}
As before, we consider separately the exponential components of the sine function, on which the derivative operator acts as 
\begin{equation} \label{eq:master_relation}
    \mathbf{q} \cdot \nabla_{\mathbf{k}} e^{\pm i \omega_\text{log} \ln \left( k/k_* \right)} = \pm i \omega_\text{log} \frac{\left(\mathbf{q} \cdot \mathbf{k}\right)}{k^2} e^{\pm i \omega_\text{log} \ln \left( k/k_* \right)} \,.
\end{equation} 
To find a suitable form for the eigenvalues of the translation operator is necessary to operate some approximations. The natural 
choice, in analogy to what we have done in the previous section, is to neglect derivatives of the factor 
$\round{\mathbf{q} \cdot \mathbf{k}}/k^2$ in front of the oscillating exponential; as it was done in 
Refs.~\cite{Vasudevan:2019ewf,Beutler:2019ojk}.
For logarithmic oscillations, this approximation is stronger than for the linear ones, since we are neglect the derivatives of an 
additional $1/k$ factor in the first term. We will refer to this prescription, which only applies to the logarithmic case, as 
\textit{strong approximation}. Under this assumption, we thus have
\begin{equation} \label{eq:exp_form}
    \mathcal{T}_\mathbf{q} \sround{e^{\pm i \omega_\text{log} \ln \left( k/k_* \right)}} = e^{\pm i \omega_\text{log} \frac{\left( \mathbf{q} \cdot\mathbf{k} \right)}{k^2}} e^{\pm i \omega_\text{log} \ln \left( k/k_* \right)} \,.
\end{equation}
The effect of this approximation can be assessed with the same argument used in the calculation of~\cref{eq:damping_approx_lin} 
for linear oscillations. In the present case we obtain
\begin{align}
    & \mathcal{T}_\mathbf{q} \sround{ e^{\pm i \omega_\text{log} \ln \left( k/k_* \right)} } = \sum_{n=0}^\infty \left( \mathbf{q} \cdot \nabla_{\mathbf{k}} \right)^{\left( n-1 \right)} \left[ \pm i \omega_\text{log} \frac{\left( \mathbf{q} \cdot \mathbf{k} \right)}{k^2} e^{\pm i \omega_\text{log} \ln \left( k/k_* \right)} \right] \notag \\
    & \quad \, = \sum_{n=0}^\infty \left( \mathbf{q} \cdot \nabla_{\mathbf{k}} \right)^{\left( n-2 \right)} \left\{ \pm i \omega_\text{log} \sround{ \pm i \omega_\text{log} \frac{\round{\mathbf{q} \cdot \mathbf{k}}^2}{k^4} + \frac{q^2}{k^2} - 2 \frac{\round{\mathbf{q} \cdot \mathbf{k}}^2}{k^4} } e^{\pm i \omega_\text{log} \ln \left( k/k_* \right)} \right\} \,, \label{eq:nth_exp_form}
\end{align}
from which it is easy to see that in this case the corrections (those coming from the derivatives of the rational factor) are of 
the same order in $\varepsilon$ as the first term (those coming from the subsequent derivatives of the exponential). However, we 
avoid inconsistency in our approximation because, by choosing sufficiently large feature frequencies, we can distinguish these 
oscillations from the broadband shape of the power spectrum, such as the BAO. This choice improves the first term 
of~\cref{eq:nth_exp_form} with respect to the derivative corrections. To complete the analogy with the linear case, we conclude the 
evaluation of the damping within this approximation, leaving further comments on possible strategies to overcome this strong 
approximation in~\cref{app:relaxation}. 

Following the steps shown in~\cref{eq:s_op_start,eq:operator_action}, we end up with the following result for the calculation of 
the 1-loop corrected matter power spectrum
\begin{align} 
    \mathcal{S} P_L^\mathrm{w} \left(k\right) &= P_L^\mathrm{w} \left( k \right) \frac{k^2}{6 \pi^2} \int_0^{k_{\rm S}} \dd q \; P_L^\mathrm{nw} \left( q \right) \left[ 1 - j_0 \left( \frac{\omega_\text{log}}{k} q \right) + 2 j_2 \left( \frac{\omega_\text{log}}{k} q \right) \right] \notag \\
    & \equiv k^2 \Sigma_\text{log} \left(k\right) P_L^\mathrm{w} \left( k \right) \,.  \label{eq:s_operator_explicit}
\end{align}
This result is similar to the linear case except replacing $\omega_\text{log}/k$ with $\omega_\text{lin}/k_*$.\footnote{This is 
valid only under the hypothesis $q/k \ll 1$. In fact, as shown in~\cref{app:relaxation}, when this hypothesis is not strictly 
satisfied, one must start to take into account corrections like the one in~\cref{eq:operator_action_gen}, which make impossible 
to map the linear case into the logarithmic one with a simple redefinition of the frequency.}
This result already appeared explicitly in the framework of TSPT in Ref.~\cite{Vasudevan:2019ewf} and, with a slightly different 
formalism, in Ref.~\cite{Beutler:2019ojk}. Finally, it should be emphasised that, for the purpose of calculating the LO and NLO 
matter power spectra, it is sufficient to replace this result by~\cref{eq:lo_power_spec,eq:nlo_power_spec}, respectively.

\subsubsection{Mixed term between baryon and primordial logarithmic oscillations}
The inclusion of the mixed term for the case of logarithmic oscillation is more subtle than for the linear case because it cannot 
be seen as a superposition of linear waves due to the additional $k$ dependence arising from the logarithmic term. It is useful to 
exploit the same idea of the linear case and write the mixed term as 
\begin{align}
    P_\text{BAO}^\text{w} \left(k\right) \delta P^\text{log}_\zeta \left(k\right) &= P_L^\text{nw} \left(k\right) A_\text{BAO} \round{k} \sin \round{\tilde{\omega}_\text{BAO} k} A_\text{log} \sin \left(\omega_\text{log} \ln \frac{k}{k_*} + \varphi_\text{log}\right) \notag \\
    & = \frac{1}{2} P_L^\text{nw} \left(k\right) A_\text{BAO} \round{k} A_\text{log} \left\{ \cos \left( \omega_\text{log} \ln \frac{k}{k_*} - \tilde{\omega}_\text{BAO} k + \varphi_\text{log} \right) \noright \notag \\
    & \quad \, - \noleft \cos \left( \omega_\text{log} \ln \frac{k}{k_*} + \tilde{\omega}_\text{BAO} k + \varphi_\text{log} \right) \right\} \,.
\end{align} 
This expression does not represent a limitation to our purposes, it just leads to more cumbersome calculations. Indeed, also for 
this exotic oscillatory term is possible to prove that (under the same hypotheses assumed in the case of logarithmic oscillations)
\begin{equation}
    \mathbf{q} \cdot \nabla_\mathbf{k} \left[ e^{i \left( \omega_\text{log} \ln \round{k/k_*} \pm \tilde{\omega}_\text{BAO} k \right)} \right] =  i \frac{\left(\mathbf{q} \cdot \mathbf{k}\right)}{k^2} \left( \omega_\text{log} \pm \tilde{\omega}_\text{BAO}k \right) e^{i \left( \omega_\text{log} \ln \round{k/k_*} \pm \tilde{\omega}_\text{BAO} k \right)} \,.
\end{equation}
To build the translation operator here, we need to specify a power counting. As a first approximation we can neglect all the $k$ 
dependence that is not coming from the exponential term, thus finding 
\begin{equation} \label{eq:top_log_cross}
    \mathcal{T}_\mathbf{q} \left[ e^{i \left( \omega_\text{log} \ln \round{k/k_*} \pm \tilde{\omega}_\text{BAO} k \right)} \right] =  e^{i \frac{\left(\mathbf{q} \cdot \mathbf{k}\right)}{k^2} \left( \omega_\text{log} \pm \tilde{\omega}_\text{BAO}k \right)} e^{i \left( \omega_\text{log} \ln \round{k/k_*} \pm \tilde{\omega}_\text{BAO} k \right)} \,.
\end{equation}
However, differently from the pure logarithmic case, we do not have only the polynomial factor in from of the exponential, but also 
an additional $k$ factor in the frequency term coming from the differences between linear and logarithmic oscillations in the mixed 
term. 

Thanks to~\cref{eq:top_log_cross}, we can compute also in this case the damping coming from the IR resummation, obtaining 
\begin{align}
    & \mathcal{S} \left[ P_\text{BAO}^\text{w} \left(k\right) \delta P^\text{log}_\zeta \left(k\right) \right] = \frac{k^2}{2} A_\text{BAO} \round{k} P_L^\text{nw} \left(k\right) A_\text{log} \notag\\
    & \qquad \times\left[ \left. \Sigma_\text{log} \left( k\right) \right\rvert_{ \omega_\text{log} - \tilde{\omega}_\text{BAO} k } \cos \left( \omega_\text{log} \ln \frac{k}{k_*} - \tilde{\omega}_\text{BAO} k + \varphi_\text{log} \right) \right. \notag \\
    & \qquad \left. - \left. \Sigma_\text{log} \left( k\right) \right\rvert_{ \omega_\text{log} + \tilde{\omega}_\text{BAO} k } \cos \left( \omega_\text{log} \ln \frac{k}{k_*} + \tilde{\omega}_\text{BAO} k + \varphi_\text{log} \right)  \right] \,.  \label{eq:S_mix_osc_1}
\end{align} 
Then, from~\cref{eq:lo_power_spec}, we obtain at LO 
\begin{align} \label{eqn:LO_cross_log}
    &P_{\delta \delta}^\text{LO} \round{\eta, k} = g^2 \round{\eta} \left\{ 
    P_L^\text{nw} \round{k} + e^{-g^2 \round{\eta} k^2 \Sigma_\text{BAO} \round{k}} P_\text{BAO}^\text{w} \round{k} \right.\notag\\
    &\qquad\left. +\, e^{-g^2 \round{\eta} k^2 \Sigma_\text{log} \round{k}} P_\text{log}^\text{w} \round{k} 
    + \frac{1}{2} A_\text{BAO} \round{k} P_L^\text{nw} \round{k} A_\text{log} \right.\notag\\
    &\qquad \left.\left.\times\left[e^{-g^2 \round{\eta} k^2 \Sigma_\text{log} \round{k }} \right\rvert_{ \omega_\text{log} - k \tilde{\omega}_\text{BAO} } \cos \left( \omega_\text{log} \ln \frac{k}{k_*} - \tilde{\omega}_\text{BAO}k + \varphi_\text{log} \right) \right.\right. \notag \\
    &\qquad \left.\left.\left. -\, e^{-g^2 \round{\eta} k^2 \Sigma_\text{log} \round{k }} \right\rvert_{ \omega_\text{log} + k \tilde{\omega}_\text{BAO} } \cos \left( \omega_\text{log} \ln \frac{k}{k_*} + \tilde{\omega}_\text{BAO}k + \varphi_\text{log} \right) \right] \right\}  \,,
\end{align}
where $\Sigma_\mathrm{log}$ is defined in~\cref{eq:s_operator_explicit}. 
With the same ingredients, it is possible to compute also the NLO matter power spectrum, namely
\begin{align} \label{eqn:NLO_cross_log}
    & P_{\delta \delta}^\text{NLO} \round{\eta, k} = g^2 \round{\eta} \left\{ \vphantom{\cos \left( \omega_\text{log} \ln \frac{k}{k_*} + \tilde{\omega}_\text{BAO}k + \varphi_\text{log} \right)} P_L^\text{nw} \round{k} + P^\text{1-loop}_{\delta \delta} \sround{ g^{-2} \round{\eta} P_{\delta \delta}^\text{LO} \round{\eta, k}} \right.\notag\\
    &\left. \qquad + \round{1 + g^2 \round{\eta} k^2 \Sigma_\text{BAO} \round{k} e^{-g^2 \round{\eta} k^2 \Sigma_\text{BAO} \round{k}} P_\text{BAO}^\text{w} \round{k} } \right.\notag\\
    &\left. \qquad  + \round{1 + g^2 \round{\eta} k^2 \Sigma_\text{log} \round{k } e^{-g^2 \round{\eta} k^2 \Sigma_\text{log} \round{k }} P_\text{log}^\text{w} \round{k} } \right.\notag\\
    &\left. \qquad +\, \frac{1}{2} A_\text{BAO} \round{k} P_L^\text{nw} \round{k} A_\text{log} \right.\notag\\
    &\left.\left. \qquad\qquad\qquad\qquad \times \left[ \round{1 + g^2 \round{\eta} k^2 \Sigma_\text{log} \left(k\right)} e^{-g^2 \round{\eta} k^2 \Sigma_\text{log} \round{k}} \right\rvert_{\omega_\text{log} - k \tilde{\omega}_\text{BAO} } \right.\right.\notag\\
    &\left.\left. \qquad\qquad\qquad\qquad\qquad \times \cos \left( \omega_\text{log} \ln \frac{k}{k_*} - \tilde{\omega}_\text{BAO}k + \varphi_\text{log} \right) \right.\right. \notag \\
    &\left.\left.\left. \qquad\qquad\qquad\qquad \quad - \round{1 + g^2 \round{\eta} k^2 \Sigma_\text{log} \left(k\right)} e^{-g^2 \round{\eta} k^2 \Sigma_\text{log} \round{k}} \right\rvert_{\omega_\text{log} + k \tilde{\omega}_\text{BAO} } \right.\right.\notag\\
    &\left.\left. \qquad\qquad\qquad\qquad\qquad \times \cos \left( \omega_\text{log} \ln \frac{k}{k_*} + \tilde{\omega}_\text{BAO}k + \varphi_\text{log} \right) \right] \right\} \,.
\end{align}
\Cref{eqn:LO_cross_log,eqn:NLO_cross_log} show the calculation of the matter power spectrum at LO and NLO, 
respectively, including the mixed term between baryon and primordial logarithmic oscillations, in analogy 
to~\cref{eq:LO_cross_lin,eq:NLO_cross_lin}. 
We note that it is still possible to extend this result by partially relaxing the $q\ll k$ assumption used in the derivation of~\cref{eq:exp_form}, 
as shown for the non-mixed term only in Ref.~\cite{Beutler:2019ojk}. 
In~\cref{app:relaxation} we present the calculation of the mixed term that relaxes the $q\ll k$ assumption. Note that the contribution from relaxing this assumption is consistently subdominant compared to the effect of including the mixed term discussed in this section. Although the magnitude of the $q\ll k$ contribution increases as the frequency is lowered, we have checked that the cross term remains larger across the range of scales of interest for all frequencies considered. In particular, we find that the cross term is an order of magnitude larger than the $q\ll k$ corrections for $\omega \sim \smallO{100}$ and about a factor of $3$ for $\omega \sim \smallO{1}$.

\subsection{Gaussian wave packets}
In this subsection, we examine both linear and logarithmic features modulated by a Gaussian distribution, using the same 
methodology as in the previous sections. These templates are particularly useful for assessing the effects of a scale-dependent 
amplitude on the description of primordial oscillatory features. This is usually the case in realistic models. 
The perturbations to the primordial power spectrum can be expressed as in~\cref{eqn:Pk_template}, namely
\begin{equation}
    \delta P^\mathrm{WP, lin}_\zeta \round{k} = A^\mathrm{WP}_\mathrm{lin} \exp\sround{ - \frac{\round{k - \mu}^2}{2 \Delta^2}} \sin \round{\omega_\mathrm{lin} \frac{k}{k_*} + \varphi_\mathrm{lin}} \,,
\end{equation} 
and 
\begin{equation}
    \delta P^\mathrm{WP, log}_\zeta \round{k} = A^\mathrm{WP}_\mathrm{log} \exp\sround{ - \frac{\round{k - \mu}^2}{2 \Delta^2}} \sin \round{\omega_\mathrm{log} \ln \frac{k}{k_*} + \varphi_\mathrm{log}} \,,
\end{equation}
for linear and logarithmic oscillations, respectively. Let us briefly comment on the structure of these equations: the first 
factor is the Gaussian weight, while the second is the standard oscillatory term introduced in~\cref{eqn:Pk_template} for linear 
and logarithmic oscillations, respectively. Here, $\mu$ is the centre of the Gaussian envelope of the oscillations and $\Delta$ 
is its width.

As with other parameterisations, the initial step is to examine the effect of the translation operator, $\mathcal{T}_\mathbf{q}$, 
on the oscillatory component of the power spectrum. For a single application of the operator $\mathbf{q} \cdot \nabla_\mathbf{k}$, 
we have 
\begin{equation}
    \mathbf{q} \cdot \nabla_\mathbf{k} \sround{e^{ - \frac{\round{k - \mu}^2}{2 \Delta^2}} e^{\pm i \omega_\mathrm{lin} k/k_* } } = i \sround{ \pm \frac{\omega_\mathrm{lin}}{k_*} + \frac{i}{\Delta^2} \round{k - \mu} } \frac{\round{\mathbf{q} \cdot \mathbf{k}}}{k} e^{ - \frac{\round{k - \mu}^2}{2 \Delta^2}} e^{\pm i \omega_\mathrm{lin} k/k_* } \,, \label{eq:deriv_ass_1}
\end{equation}
and
\begin{equation}
    \mathbf{q} \cdot \nabla_\mathbf{k} \sround{e^{ - \frac{\round{k-\mu}^2}{2 \Delta^2}} e^{\pm i \omega_\mathrm{log} \ln \round{ k/k_*  } } } = i \sround{\pm \omega_\mathrm{log} + \frac{i k}{\Delta^2} \round{k-\mu} } \frac{ \round{\mathbf{q} \cdot \mathbf{k}}}{k^2} e^{ - \frac{\round{k-\mu}^2}{2 \Delta^2}} e^{\pm i \omega_\mathrm{log} \ln \round{ k/k_*  } } \,. \label{eq:deriv_ass_2}
\end{equation}
In order to extend these relationships to $n$-evaluations of the operator and thereby construct the translation operator, it is necessary to establish a power counting scheme between successive applications of the derivative operator. It should be noted that while it is always possible to compute the action of $n$-copies of the derivative operator exactly, it is not possible to obtain a closed-form expression for their sum, as found previously for linear and logarithmic oscillations. An alternative method for evaluating the impact of this approximation would be to truncate the sum after a specified number of terms and compare it to the approximated version. This approach is analogous to that employed in~\cref{sec:lin_osc,sec:log_osc}, with the key distinction being that, in this case, the power counting is determined not only by the oscillation frequency but also by the amplitude of the Gaussian weights.
Indeed, depending on the amplitude of the Gaussian, these corrections may render the perturbative approach to the problem inapplicable, as they could exceed the damping effects caused by the oscillatory component. Therefore, this method is valid only within a specific range of Gaussian weight parameters. Within this approximation, we can express the translation operator as
\begin{equation} \label{eq:t_wp_lin}
    \mathcal{T}_\mathbf{q} \sround{e^{ - \frac{\round{k - \mu}^2}{2 \Delta^2}} e^{\pm i \omega_\mathrm{lin} k/k_* } } = e^{i \sround{ \pm \frac{\omega_\mathrm{lin}}{k_*} + \frac{i}{\Delta^2} \round{k - \mu} } \frac{\round{\mathbf{q} \cdot \mathbf{k}}}{k}} e^{ - \frac{\round{k - \mu}^2}{2 \Delta^2}} e^{\pm i \omega_\mathrm{lin} k/k_* } \,,
\end{equation} 
and 
\begin{equation}
    \mathcal{T}_\mathbf{q} \sround{e^{ - \frac{\round{k - \mu}^2}{2 \Delta^2}} e^{\pm i \omega_\mathrm{log} \ln \round{ k/k_*  } } } = e^{i \sround{\pm \omega_\mathrm{log} + \frac{ik}{\Delta^2} \round{k - \mu} } \frac{ \round{\mathbf{q} \cdot \mathbf{k}}}{k^2}} e^{ - \frac{\round{k - \mu}^2}{2 \Delta^2}} e^{\pm i \omega_\mathrm{log} \ln \round{ k/k_*  } } \,. \label{eq:t_wp_log}
\end{equation}
In order to complete the computation of the 1-loop corrections to the matter power spectrum, it is sufficient to follow the steps 
outlined in~\cref{eq:s_op_start,eq:operator_action}. Here, however, the presence of an additional imaginary factor in the frequency 
results in a significant alteration to the computation. The subsequent steps are illustrated for the linear case only; the logarithmic 
case differs solely in terms containing the frequency, which can be reinterpreted from the linear one. In this context, the analogue 
of~\cref{eq:operator_action} is
\begin{align}
    &\sround{1 - \cosh \round{\mathbf{q} \cdot \nabla_{\mathbf{k}}} } P^\mathrm{w}_\mathrm{WP, lin} \round{k} = \notag \\
    & \quad \, = \frac{1}{2} \left\lbrace 1 - \cos \sround{ \round{ \frac{\omega_\mathrm{lin}}{k_*} + \frac{ik}{\Delta^2} \round{k-\mu}} \frac{\round{\mathbf{q} \cdot \mathbf{k} }}{k}} \sround{ \sin \round{\omega_\mathrm{lin} \frac{k}{k_*}} - i \cos \round{\omega_\mathrm{lin} \frac{k}{k_*}} } \noright \notag \\
    & \qquad \,\, - \noleft \cos \sround{ \round{\frac{\omega_\mathrm{lin}}{k_*} - \frac{ik}{\Delta^2} \round{k-\mu}} \frac{\round{\mathbf{q} \cdot \mathbf{k} }}{k} } \sround{ \sin \round{\omega_\mathrm{lin} \frac{k}{k_*}} + i \cos \round{\omega_\mathrm{lin} \frac{k}{k_*}} } \right\rbrace e^{ - \frac{\round{k - \mu}^2}{2 \Delta^2}} \,,
\end{align}
which, by means of relationships between trigonometric and hyperbolic functions can be casted in a manifestly real expression, namely
\begin{align}
    &\sround{1 - \cosh \round{\mathbf{q} \cdot \nabla_{\mathbf{k}}} }  P^\mathrm{w}_\mathrm{WP, lin} \round{k} = \notag \\
    & \quad \, = \bround{ 1 - \cos \sround{ \frac{\omega_\mathrm{lin}}{k_*} \frac{\round{\mathbf{q} \cdot \mathbf{k}}}{k} } \cosh \sround{\frac{\round{k-\mu}}{\Delta^2} \frac{\round{\mathbf{q} \cdot \mathbf{k}}}{k}} } P^\mathrm{w}_\mathrm{WP, lin}\round{k} \notag \\ 
    & \qquad \,\, + \sin \sround{ \frac{\omega_\mathrm{lin}}{k_*} \frac{\round{\mathbf{q} \cdot \mathbf{k}}}{k}} \sinh \sround{\frac{\round{k-\mu}}{\Delta^2} \frac{\round{\mathbf{q} \cdot \mathbf{k}}}{k}} \noleft  P^\mathrm{w}_\mathrm{WP, lin}\round{k} \right\rvert_{\varphi_\text{lin}+\pi/2} \,. \label{eq:cosh_wp_lin}
\end{align} 
It is evident from this expression that the inclusion of a scale-dependent amplitude in the computation of the 
damping factor introduces an out-of-phase contribution, as seen in~\cref{eq:cosh_wp_lin}. This result is to be 
expected: derivative operators induce a phase shift when acting on oscillatory functions. However, when acting 
on the scale-dependent amplitude, the derivative does not introduce any phase shift in the oscillatory component. 
This results in the emergence of two oscillatory factors that are out of phase.
It is crucial to highlight that in order to compute the oscillatory component of the matter power spectrum at 
LO, the operator defined in~\cref{eq:cosh_wp_lin} must be applied multiple times. In light of the out-of-phase 
contribution, the application of~\cref{eq:cosh_wp_lin} alone is not enough. Consequently, the computation must be 
repeated in order to determine in which the operator acts upon the phase-shifted power spectrum. 
It can be demonstrated that
\begin{align}
    &\sround{1 - \cosh \round{\mathbf{q} \cdot \nabla_{\mathbf{k}}} } \noleft  P^\mathrm{w}_\mathrm{WP, lin} \round{k} \right\rvert_{\varphi+\pi/2} = \notag \\
    & \quad \, = \bround{ 1 - \cos \sround{ \frac{\omega_\mathrm{lin}}{k_*} \frac{\round{\mathbf{q} \cdot \mathbf{k}}}{k} } \cosh \sround{\frac{\round{k-\mu}}{\Delta^2} \frac{\round{\mathbf{q} \cdot \mathbf{k}}}{k}} } \noleft P^\mathrm{w}_\mathrm{WP, lin}\round{k} \right\rvert_{\varphi_\text{lin}+\pi/2} \notag \\ 
    & \qquad \,\, - \sin \sround{ \frac{\omega_\mathrm{lin}}{k_*} \frac{\round{\mathbf{q} \cdot \mathbf{k}}}{k}} \sinh \sround{\frac{\round{k-\mu}}{\Delta^2} \frac{\round{\mathbf{q} \cdot \mathbf{k}}}{k}} P^\mathrm{w}_\mathrm{WP, lin}\round{k} \,. \label{eq:cosh_wp_lin2}
\end{align}
Now, following what has been shown in appendix A.2 of Ref.~\cite{Beutler:2019ojk}, we find the $\ell$-loop contribution to the 
matter power spectrum to be 
\begin{align}
    &P^\mathrm{w}_{\ell-\mathrm{loop}, \mathrm{LO}}\round{\eta, k} = \frac{\sround{ik g \round{\eta}}^{2\ell}}{\ell!} \left\{ \frac{1}{2} \left[ \round{\Sigma^{2}_{\rm lin}(k) + i \hat{\Sigma}^{2}_{\rm lin}(k)}^{\ell} + \round{\Sigma^{2}_{\rm lin}(k) - i \hat{\Sigma}^{2}_{\rm lin}(k)}^{\ell} \right] P_{\rm WP,lin}^\mathrm{w}(k) \right. \notag \\
    & \quad \, + \left. \frac{1}{2i} \left[ \round{\Sigma^{2}_{\rm lin}(k) + i \hat{\Sigma}^{2}_{\rm lin}(k)}^{\ell} - \round{\Sigma^{2}_{\rm lin}(k) - i \hat{\Sigma}^{2}_{\rm lin}(k)}^{\ell} \right] \noleft P^\mathrm{w}_\mathrm{WP, lin}\round{k} \right\rvert_{\varphi_\text{lin}+\pi/2} \right\} \,.  \label{eq:l_loop_spectrum_2}
\end{align}
Adding together the contributions from~\cref{eq:l_loop_spectrum_2} for all $\ell$, we obtain the leading-order IR-resummed wiggly 
power spectrum
\begin{align}
    P_{\delta \delta}^\mathrm{w,LO} \round{\eta, k} &= g^2 \round{\eta} \left\{ e^{- g^2 \round{\eta} k^2 \Sigma_{\rm lin}^{\rm WP} \round{k} } \cos \sround{g^2 \round{\eta} k^2 \hat{\Sigma}_{\rm lin}^{\rm WP} \round{k}} P_\mathrm{WP,lin}^\mathrm{w} \round{k} \noright \notag \\
    & \noleft \quad \, - e^{- g^2 \round{\eta} k^2 \Sigma_{\rm lin}^{\rm WP} \round{k} } \sin \sround{g^2 \round{\eta} k^2 \hat{\Sigma}_{\rm lin}^{\rm WP} \round{k}} \noleft P_\mathrm{WP,lin}^\mathrm{w} \round{k} \right\rvert_{\varphi_\text{lin}+\pi/2} \right\} \,, \label{eq:LO_power_packets}
\end{align}
where
\begin{align} 
    &\Sigma^\text{WP}_\text{lin} \left(k\right) \equiv \frac{1}{6\pi^2} \int_0^{k_\mathrm{S}} \dd q \; P_\mathrm{nw} \round{q} \biggl\{ 1 - \frac{3\Delta^2}{q^3 \round{\round{k-\mu}^2 + \Delta^4 \tilde{\omega}_{\rm lin}^2}^3} \notag \\
    & \quad\, \times \left[ \Delta^2 \cosh\round{\frac{q \round{k - \mu}}{\Delta^2}} \left( \vphantom{\round{\mu^2 + \Delta^4 \tilde{\omega}_{\rm lin}^2}^2} -2 q \cos\round{\tilde{\omega}_{\rm lin} q} \round{ \round{k-\mu}^4 - \Delta^8 \tilde{\omega}_{\rm lin}^4} \noright \noright \notag \\
    & \qquad\quad\, \,\, + \tilde{\omega}_{\rm lin} \sin \round{\tilde{\omega}_{\rm lin} q} \left( \vphantom{\round{\mu^2 + \Delta^4 \tilde{\omega}_{\rm lin}^2}^2}  
    q^2 \round{k - \mu}^4 + 6 \round{k - \mu}^2 \Delta^4 + q^2 \Delta^8 \tilde{\omega}_{\rm lin}^4 \noright \notag  \\
    & \qquad\quad\quad\, \,\,\, \noleft \noleft 
    -2 \Delta^4 \tilde{\omega}_{\rm lin}^2 \round{\Delta^4 -q^2 \round{k-\mu}^2}  \vphantom{\round{\mu^2 + \Delta^4 \tilde{\omega}_{\rm lin}^2}^2} \right) \right) \notag \\
    & \quad\quad\, \, + \round{k-\mu} \sinh\round{\frac{q \round{k - \mu}}{\Delta^2}} \left( \vphantom{\round{\mu^2 + \Delta^4 \tilde{\omega}_{\rm lin}^2}^2} -4 q \Delta^4 \tilde{\omega}_{\rm lin} \sin\round{q \tilde{\omega}_{\rm lin}} \round{ \round{k -\mu}^2 + \Delta^4 \tilde{\omega}_{\rm lin}^2} \noright \notag \\
    & \qquad\quad\, \,\, + \cos \round{\tilde{\omega}_{\rm lin} q} \left(  
    q^2 \round{k - \mu}^4 + 2 \round{k - \mu}^2 \Delta^4 + q^2 \Delta^8 \tilde{\omega}_{\rm lin}^4 \noright \notag \\
    & \qquad\quad\quad\, \,\,\, \noleft \noleft \noleft 
    - 2 \Delta^4 \tilde{\omega}_{\rm lin}^2 \round{3\Delta^4 - q^2 \round{k - \mu}^2} \right) \right) \vphantom{\sinh\sround{\frac{q \round{k - \mu}}{\Delta^2}}}\right]
    \biggr\} \,,
\end{align} 
and  
\begin{align}
    &\hat{\Sigma}^\text{WP}_\text{lin} \left(k\right) \equiv - \frac{1}{6\pi^2} \int_0^{k_\mathrm{S}} \dd q \; P_\mathrm{nw} \round{q} \frac{3\Delta^2}{q^3 \round{\round{k-\mu}^2 + \Delta^4 \tilde{\omega}_{\rm lin}^2}^3} \notag \\
    & \quad\, \times \left\{ \round{k-\mu} \cosh\sround{\frac{q \round{k - \mu}}{\Delta^2}} \left[ \vphantom{\round{\mu^2 + \Delta^4 \tilde{\omega}_{\rm lin}^2}^2} 4 q \Delta^4 \tilde{\omega}_{\rm lin} \cos\round{q \tilde{\omega}_{\rm lin}} \round{ \round{k - \mu}^2 + \Delta^4 \tilde{\omega}_{\rm lin}^2} \noright \noright \notag \\
    & \qquad\quad\, \,\, + \sin \round{\tilde{\omega}_{\rm lin} q} \left( \vphantom{\round{\mu^2 + \Delta^4 \tilde{\omega}_{\rm lin}^2}^2}     q^2 \round{k - \mu}^4 + 2 \round{k - \mu}^2 \Delta^4 + q^2 \Delta^8 \tilde{\omega}_{\rm lin}^4     \noright \notag  \\
    & \qquad\quad\quad\, \,\,\, \noleft \noleft     -2 \Delta^4 \tilde{\omega}_{\rm lin}^2 \round{3\Delta^4 -q^2 \round{k-\mu}^2}  \vphantom{\round{\mu^2 + \Delta^4 \tilde{\omega}_{\rm lin}^2}^2}     \right) \right] \notag \\
    & \quad\quad\, \, - \Delta^2 \sinh\sround{\frac{q \round{k - \mu}}{\Delta^2}} \left[ \vphantom{\round{\mu^2 + \Delta^4 \tilde{\omega}_{\rm lin}^2}^2} 2q \sin\round{\tilde{\omega}_{\rm lin} q} \round{\round{k -\mu}^4 - \Delta^8 \tilde{\omega}_{\rm lin}^4} \noright \notag \\
    & \qquad\quad\, \,\, + \tilde{\omega}_{\rm lin} \cos \round{\tilde{\omega}_{\rm lin} q} \left( \vphantom{\round{\mu^2 + \Delta^4 \tilde{\omega}_{\rm lin}^2}^2}    q^2 \round{k - \mu}^4 + 6 \round{k - \mu}^2 \Delta^4 + q^2 \Delta^8 \tilde{\omega}_{\rm lin}^4      \noright \notag \\
    & \qquad\quad\quad\, \,\,\, \noleft \noleft \noleft    -2 \Delta^4 \tilde{\omega}_{\rm lin}^2 \round{\Delta^4 -q^2 \round{k-\mu}^2} \right) \right] \vphantom{\sinh\sround{\frac{q \round{k - \mu}}{\Delta^2}}}\right\} \,,
\end{align}
are the two contributions to the damping. Here we used for simplicity $\tilde{\omega}_{\rm lin} \equiv \omega_{\rm lin} / k_*$. 
Finally, in this case, we can include the NLO corrections in the hard loops by analogy with~\cref{eq:nlo_power_spec}. With minor adjustments to the procedure used in the previous case, we now find
\begin{align} \label{eq:NLO_power_packets}
    P_{\delta \delta}^\mathrm{w,NLO} \round{\eta, k} &= g^2 \round{\eta} e^{- g^2 \round{\eta} k^2 \Sigma_{\rm lin}^{\rm WP} \round{k} } \left\{ \vphantom{\noleft P_\mathrm{WP,lin}^\mathrm{w} \round{k} \right\rvert_{\varphi+\pi/2}} \round{1 + k^2 \Sigma_{\rm lin}^{\rm WP} \round{k}} \noright \notag \\
    & \quad \, \times \left[  \cos \round{g^2 \round{\eta} k^2 \hat{\Sigma}_{\rm lin}^{\rm WP} \round{k}} P_\mathrm{WP,lin}^\mathrm{w} \round{k} \noright \notag \\
    & \noleft \qquad \,\, -  \sin \round{g^2 \round{\eta} k^2 \hat{\Sigma}_{\rm lin}^{\rm WP} \round{k}} \noleft P_\mathrm{WP,lin}^\mathrm{w} \round{k} \right\rvert_{\varphi_\text{lin}+\pi/2} \right] \notag \\
    & \quad\, + k^2 \hat{\Sigma}_{\rm lin}^{\rm WP} \round{k} \left[ \sin \round{g^2 \round{\eta} k^2 \hat{\Sigma}_{\rm lin}^{\rm WP} \round{k}} P_\mathrm{WP,lin}^\mathrm{w} \round{k} \noright \notag \\
    & \noleft\noleft \qquad \,\, + \cos \round{g^2 \round{\eta} k^2 \hat{\Sigma}_{\rm lin}^{\rm WP} \round{k}} \noleft P_\mathrm{WP,lin}^\mathrm{w} \round{k} \right\rvert_{\varphi_\text{lin}+\pi/2} \right] \right\} \notag \\
    & \quad\, + g^4\round{\eta} P_{\delta \delta}^{1-\text{loop}} \sround{ g^{-2} \round{\eta} P_{\delta \delta}^\mathrm{LO} \round{k} } \,.
\end{align}

Eq.~\eqref{eq:LO_power_packets} and~\cref{eq:NLO_power_packets} hold for the LO and NLO wave packets with logarithmic 
oscillations, respectively. In this case the two damping factors are given by 
\begin{align} 
    &\Sigma^\text{WP}_\text{log} \left(k \right) \equiv \frac{1}{6\pi^2} \int_0^{k_\mathrm{S}} \dd q \; P_\mathrm{nw} \round{q} \biggl\{ 1 - \frac{3k \Delta^2}{q^3 \round{k^2\round{k-\mu}^2 + \Delta^4 \omega_\mathrm{log}^2}^3} \notag \\
    & \quad\, \times \left[ \Delta^2 \cosh\round{\frac{q \round{k - \mu}}{\Delta^2}} \left( \vphantom{\round{\mu^2 + \Delta^4 \omega_\mathrm{log}^2}^2} -2 k q \cos\round{\omega_\mathrm{log} \frac{q}{k}} \round{ k^4 \round{k-\mu}^4 - \Delta^8 \omega_\mathrm{log}^4} \noright \noright \notag \\
    & \qquad\quad\, \,\, + \omega_\mathrm{log} \sin \round{\omega_\mathrm{log} \frac{q}{k}} \left( \vphantom{\round{\mu^2 + \Delta^4 \omega_\mathrm{log}^2}^2}   k^4 \round{k - \mu}^2 \round{q^2 \round{k - \mu}^2 + 6 \Delta^4} + q^2 \Delta^8 \omega_\mathrm{log}^4      \noright \notag  \\
    & \qquad\quad\quad\, \,\,\, \noleft \noleft     - 2 k^2 \Delta^4 \omega_\mathrm{log}^2 \round{\Delta^4 - q^2 \round{k - \mu}^2}      \vphantom{\round{\mu^2 + \Delta^4 \omega_\mathrm{log}^2}^2} \right) \right) \notag \\
    & \quad\quad\, \, + k \round{k-\mu} \sinh\round{\frac{q \round{k - \mu}}{\Delta^2}} \left( \vphantom{\round{\mu^2 + \Delta^4 \omega_\mathrm{log}^2}^2} -4 k q \Delta^4 \omega_\mathrm{log} \sin\round{\omega_\mathrm{log} \frac{q}{k}} \round{ k^2 \round{k -\mu}^2 + \Delta^4 \omega_\mathrm{log}^2} \noright \notag \\
    & \qquad\quad\, \,\, + \cos \round{\omega_\mathrm{log} \frac{q}{k}} \left(     \vphantom{\round{\mu^2 + \Delta^4 \omega_\mathrm{log}^2}^2}   k^4 \round{k - \mu}^2 \round{q^2 \round{k - \mu}^2 + 2 \Delta^4} + q^2 \Delta^8 \omega_\mathrm{log}^4       \noright \notag \\
    & \qquad\quad\quad\, \,\,\, \noleft \noleft \noleft    - 2 k^2 \Delta^4 \omega_\mathrm{log}^2 \round{3 \Delta^4 - q^2 \round{k - \mu}^2} \right) \right) \vphantom{\sinh\sround{\frac{q \round{k - \mu}}{\Delta^2}}}\right] \biggl\} \,,
\end{align}
and 
\begin{align}
    &\hat{\Sigma}^\text{WP}_\text{log} \left(k \right) \equiv - \frac{1}{6\pi^2} \int_0^{k_\mathrm{S}} \dd q \; P_\mathrm{nw} \round{q} \frac{3k \Delta^2}{q^3 \round{k^2\round{k-\mu}^2 + \Delta^4 \omega_\mathrm{log}^2}^3} \notag \\
    & \quad\, \times \left\{ k \round{k-\mu} \cosh\sround{\frac{q \round{k - \mu}}{\Delta^2}} \left[ \vphantom{\round{\mu^2 + \Delta^4 \omega_\mathrm{log}^2}^2} 4 k q \Delta^4 \omega_\mathrm{log} \cos\round{\omega_\mathrm{log} \frac{q}{k}} \round{ k^2 \round{k-\mu}^2 + \Delta^4 \omega_\mathrm{log}^2} \noright \noright \notag \\
    & \qquad\quad\, \,\, + \sin \round{\omega_\mathrm{log} \frac{q}{k}} \left( \vphantom{\round{\mu^2 + \Delta^4 \omega_\mathrm{log}^2}^2}   k^4 \round{k - \mu}^2 \round{q^2 \round{k - \mu}^2 + 2 \Delta^4} + q^2 \Delta^8 \omega_\mathrm{log}^4      \noright \notag  \\
    & \qquad\quad\quad\, \,\,\, \noleft \noleft     - 2 k^2 \Delta^4 \omega_\mathrm{log}^2 \round{3 \Delta^4 - q^2 \round{k - \mu}^2}      \vphantom{\round{\mu^2 + \Delta^4 \omega_\mathrm{log}^2}^2} \right) \right] \notag \\
    & \quad\quad\, \, - \Delta^2 \sinh\sround{\frac{q \round{k - \mu}}{\Delta^2}} \left[ \vphantom{\round{\mu^2 + \Delta^4 \omega_\mathrm{log}^2}^2} 2 k q \sin\round{\omega_\mathrm{log} \frac{q}{k}} \round{ k^4 \round{k -\mu}^4 - \Delta^8 \omega_\mathrm{log}^4} \noright \notag \\
    & \qquad\quad\, \,\, + \omega_\mathrm{log} \cos \round{\omega_\mathrm{log} \frac{q}{k}} \left(     \vphantom{\round{\mu^2 + \Delta^4 \omega_\mathrm{log}^2}^2}   k^4 \round{k - \mu}^2 \round{q^2 \round{k - \mu}^2 + 6 \Delta^4} + q^2 \Delta^8 \omega_\mathrm{log}^4       \noright \notag \\
    & \qquad\quad\quad\, \,\,\, \noleft \noleft \noleft    - 2 k^2 \Delta^4 \omega_\mathrm{log}^2 \round{ \Delta^4 - q^2 \round{k - \mu}^2} \right) \right] \vphantom{\sinh\sround{\frac{q \round{k - \mu}}{\Delta^2}}}\right\}  \,.
\end{align}
These formulas represent the main result of this section and provide a generalisation 
of~\cref{eq:s_lin_osc,eq:s_operator_explicit} when a scale-dependent Gaussian amplitude is introduced. 
In the limit of $\Delta \to \infty$, corresponding to oscillations with constant amplitude, we recover  
\begin{equation}
    \lim_{\Delta \rightarrow\infty} \Sigma^\mathrm{WP}_\mathrm{lin} \round{k} = \Sigma_\mathrm{lin} \round{k} \qquad \text{and} \qquad \lim_{\Delta \rightarrow\infty} \hat{\Sigma}^\mathrm{WP}_\mathrm{lin} \round{k} = 0  \,,
\end{equation}
from which it is straightforward to see how~\cref{eq:LO_power_packets,eq:NLO_power_packets} reduce 
to~\cref{eq:lo_power_spec,eq:nlo_power_spec}. 
The same behaviour occurs for logarithmic oscillations. 
The key parameter governing the size of the corrections to the damping, arising from the scale-dependent amplitude, 
is $\Delta$. A Taylor expansion around $\Delta \rightarrow \infty$ can be used to assess the impact of these corrections on linear and logarithmic oscillations. In particular, we find
\begin{align}
    \Sigma^\mathrm{WP}_\mathrm{lin} \round{k} &= \Sigma_\mathrm{lin} \round{k} + \frac{1}{\Delta^4} \delta \Sigma^{\round{4}}_\mathrm{lin} \round{k} + \mathcal{O} \round{\frac{1}{\Delta^8}} \,,  \label{eq:sigma_corr_1} \\
    \hat{\Sigma}^\mathrm{WP}_\mathrm{lin} \round{k} &=\frac{1}{\Delta^2} \delta \hat{\Sigma}^{\round{2}}_\mathrm{lin} \round{k} + \mathcal{O} \round{\frac{1}{\Delta^6}} \,. \label{eq:sigma_corr_2}
\end{align}
The largest corrections come from the out-of-phase spectrum term. It would be useful to establish a criterion to 
assess the impact of corrections and to define the region of the parameter space where our methodology is valid. 
As emphasised throughout this work, our analysis relies on certain approximations. When these break down, the 
results are no longer reliable. This means that we can ignore the derivatives in~\cref{eq:deriv_ass_1} if we 
assume that $\Delta$ is large enough. If $\Delta$ is too small, the assumption fails and the corrections are 
unreliable. We need to find a safe range for $\Delta$ where our formulas still work. It is difficult to establish 
this range based on the damping terms. It would be easier to check if this is right by looking at the LO wiggly 
power spectrum.

By propagating the corrections from~\cref{eq:sigma_corr_1,eq:sigma_corr_2} into~\cref{eq:LO_power_packets}, 
we obtain a series expansion in $\Delta$, with the first three terms given by
\begin{align}
    \Delta^0 \quad &\to \quad e^{-k^2 \Sigma_\mathrm{lin} \round{k}} P_L^\mathrm{w} \round{k} \\[5pt]
    \Delta^{-2} \quad &\to \quad k^2 e^{-k^2 \Sigma_\mathrm{lin} \round{k}} \delta \hat{\Sigma}_\mathrm{lin}^{\round{2}} \round{k} \noleft P_L^\mathrm{w} \round{k} \right\rvert_{\varphi_\text{lin}+\pi/2} \\
    \Delta^{-4} \quad &\to \quad -k^2 e^{-k^2 \Sigma_\mathrm{lin} \round{k}} \sround{ \frac{k^2}{2} \delta \hat{\Sigma}_\mathrm{lin}^{\round{2}} \round{k} + \delta \Sigma_\mathrm{lin}^{\round{4}} \round{k}  } P_L^\mathrm{w} \round{k} \,.
\end{align}
Finally, we can compute the relative impact of the corrections to the LO power spectrum as 
\begin{equation} \label{eq:error_formula}
    \frac{P_{\delta\delta}^\mathrm{w,LO} \round{k} - \noleft P_{\delta\delta}^\mathrm{w,LO} \round{k} \right\rvert_{\Delta\rightarrow\infty}}{\noleft P_{\delta\delta}^\mathrm{w,LO} \round{k} \right\rvert_{\Delta\rightarrow\infty}} = \frac{k^2}{\Delta^2} \delta \hat{\Sigma}_\mathrm{lin}^{\round{2}} \round{k} \frac{\noleft P_L^\mathrm{w} \round{k} \right\rvert_{\varphi_\text{lin}+\pi/2}}{P_L^\mathrm{w} \round{k}} \,,
\end{equation} 
where 
\begin{align}
    \delta \hat{\Sigma}_\mathrm{lin}^{\round{2}} \round{k} &\equiv \frac{1}{6\pi^2} \int_0^{k_\mathrm{S}} \dd q \; P_\mathrm{nw} \round{q} \frac{3 \round{k - \mu}}{q^3 \tilde{\omega}_\mathrm{lin}^4 } \notag \\
    & \quad \, \times \left[ q \tilde{\omega}_\mathrm{lin} \round{-6 + q^2 \tilde{\omega}_\mathrm{lin}^2} \cos\round{q \tilde{\omega}_\mathrm{lin}} - 3 \round{-2 + q^2 \tilde{\omega}_\mathrm{lin}^2} \sin\round{q \tilde{\omega}_\mathrm{lin}} \right] \,.
\end{align}

This formula can be used to determine the range of $\Delta$ where our method is valid, given other fixed parameters. Analogous considerations hold for the case with logarithmic oscillations, with the terms arising from the corrections to the damping given by
\begin{align}
    \delta \hat{\Sigma}_\mathrm{log}^{\round{2}} \round{k} &\equiv \frac{1}{6\pi^2} \int_0^{k_\mathrm{S}} \dd q \; P_\mathrm{nw} \round{q} \frac{3k(k - \mu)}{q^3 \omega_\mathrm{log}^4} \notag \\
    & \times \left[ q \omega_\mathrm{log} \left( -6k^2 + q^2 \omega_\mathrm{log}^2 \right) \cos\left( \frac{q \omega_\mathrm{log}}{k} \right) + 3k \left( 2k^2 - q^2 \omega_\mathrm{log}^2 \right) \sin\left( \frac{q \omega_\mathrm{log}}{k} \right) \right] \,.
\end{align}

While these equations suggest that an analytical determination could be attempted, we find that empirical validation through simulation provides the most reliable approach. We have verified that our approximations hold in all cases analysed, and extending this analysis to more general scenarios would require a dedicated study beyond the scope of this work. 
We find that for the Gaussian wave packet the calculations work well for $\Delta > 0.1$, while for lower values such as $\Delta \sim 0.01$ they only work for high frequencies such as $\omega > 10$ and $\omega > 100$ for linear and logarithmic oscillations, respectively.

\subsection{Power-law amplitude}
In this section, we study models of linear and logarithmic oscillations modulated by a power-law scale-dependent amplitude. 
The motivation for this, beyond providing a more comprehensive analysis of feature models, is that these templates allow a 
broader assessment of the impact of scale-dependent amplitudes, complementing the Gaussian models studied in the previous 
section. The two templates we intend to explore can be written as
\begin{equation} \label{eq:pl_1}
    \delta P^\mathrm{PL, lin}_\zeta \round{k} = A^\mathrm{PL}_\mathrm{lin} \round{\frac{k}{k_\mathrm{PL}}}^n \sin \round{\omega_\mathrm{lin} \frac{k}{k_*} + \varphi_\mathrm{lin}} \,,
\end{equation} 
and 
\begin{equation}  \label{eq:pl_2}
    \delta P^\mathrm{PL, log}_\zeta \round{k} = A^\mathrm{PL}_\mathrm{log} \round{\frac{k}{k_\mathrm{PL}}}^n \sin \round{\omega_\mathrm{log} \ln \frac{k}{k_*} + \varphi_\mathrm{log}} \; ,
\end{equation}
for linear and logarithmic oscillations, respectively. Here, the scale-dependent amplitude is regulated by a new scale $k_\mathrm{PL}$ and an index $n$.
Note the presence of two distinct pivot scales in \cref{eq:pl_1,eq:pl_2}. The scale in the amplitude can be chosen independently 
of the other since, as we will see, it never enters explicitly into the computation of the damping factors. These templates are 
conceptually analogous to the wave packets discussed in the previous section, differing only in the function used to represent the 
scale-dependent amplitude. Therefore, we will not detail every step of the computation. Instead, we will proceed by drawing analogies 
with the main equations and focus more on general insights.

The evaluation of the translation operator can be carried out under the same assumptions discussed above~\cref{eq:t_wp_lin,eq:t_wp_log}. 
In particular, any model with a scale-dependent amplitude represented by a regular function will show a similar issue: applying the 
derivative operator within the translation operator introduces an additional scale dependence that cannot be fully addressed 
analytically. Typically, this dependence is weaker than that of the oscillatory factor and can be neglected within a certain range 
of the parameters that model the envelope. This confines the effect of the amplitude's shape to a modification in the eigenvalues 
of the translation operator. Specifically, for the model of interest in this section, we find
\begin{equation}
    \mathcal{T}_\mathbf{q} \sround{ e^{n\ln \round{k/k_*} \pm i \omega_\mathrm{lin} k/k_*} } = e^{\round{\frac{n}{k} \pm i \frac{\omega_\mathrm{lin}}{k_*}} \frac{\round{\mathbf{q} \cdot \mathbf{k}} }{k}}  e^{n\ln \round{k/k_*} \pm i \omega_\mathrm{lin} k/k_*} \,,
\end{equation}
and 
\begin{equation}
    \mathcal{T}_\mathbf{q} \sround{e^{ \round{n\pm i \omega_\mathrm{log}} \ln \round{k/k_*}}} = e^{\round{n \pm i \omega_\mathrm{log}} \frac{\round{\mathbf{q} \cdot \mathbf{k}} }{k^2}} e^{ \round{n\pm i \omega_\mathrm{log}} \ln \round{k/k_*}} \,.
\end{equation}

This change will, of course, impact the computation of the damping factors. Whenever a non-oscillatory envelope is used, it will introduce the out-of-phase contribution discussed in the previous section due to the separation of derivatives. In such a case we 
need to resort again to~\cref{eq:LO_power_packets} for the computation of the damping. In this particular case we do not show all 
the passages leading to the final expression which can be recovered from the previous section through the substitution 
\begin{equation}
    \frac{\round{k-\mu}}{\Delta^2} \quad \rightarrow \quad \frac{n}{k}
\end{equation}
in all the arguments of hyperbolic functions. We finally give the expression for the damping factors which are 

\begin{align}
     &\Sigma^\text{PL}_\text{lin} \left(k\right) \equiv \frac{1}{6\pi^2} \int_0^{k_\mathrm{S}} \dd q \; P_\mathrm{nw} \round{q} \left\{ 1 - \frac{3 k}{q^3 \round{n^2 + k^2 \tilde{\omega}_{\rm lin}^2}^3} \noright \notag \\
     & \quad \, \times \left[ k \cosh\round{\frac{n q}{k}} \Bigl( -2 q \cos\round{q \tilde{\omega}_{\rm lin}} \round{n^4 - k^4 \tilde{\omega}_{\rm lin}^4} \noright \notag \\
     & \qquad\quad \,\,\, + \tilde{\omega}_{\rm lin} \sin\round{q \tilde{\omega}_{\rm lin}} \left( n^4 q^2 + k^4 \tilde{\omega}_{\rm lin}^2 \round{-2 + q^2 \tilde{\omega}_{\rm lin}^2} + 2 k^2 n^2 \round{3 + q^2 \tilde{\omega}_{\rm lin}^2} \right) \Bigr) \notag \\
     & \qquad \,\, + n\sinh \round{\frac{nq}{k}} \Bigl( -4 k^2 q \tilde{\omega}_{\rm lin} \sin\round{q \tilde{\omega}_{\rm lin}} \round{n^2 + k^2 \tilde{\omega}_{\rm lin}^2} \notag \\
     & \qquad\quad \,\,\, \noleft \noleft + \cos\round{q \tilde{\omega}_{\rm lin}} \round{n^4 q^2 + k^4 \tilde{\omega}_{\rm lin}^2 \round{-6 + q^2 \tilde{\omega}_{\rm lin}^2} + 2 k^2 n^2 \round{1 + q^2 \tilde{\omega}_{\rm lin}^2}} \Bigr) \right] \vphantom{\frac{3 k}{q^3 \round{n^2 + k^2 \tilde{\omega}_{\rm lin}^2}^3}} \right\} \,,
\end{align}
and
\begin{align}
     &\hat{\Sigma}^\text{PL}_\text{lin} \left(k\right) \equiv - \frac{1}{6\pi^2} \int_0^{k_\mathrm{S}} \dd q \; P_\mathrm{nw} \round{q} \frac{3 k}{q^3 \round{n^2 + k^2 \tilde{\omega}_{\rm lin}^2}^3} \notag \\
     & \quad \, \times \Bigl\{ n \cosh\round{\frac{n q}{k}} \Bigl[ 4 k^2 q \tilde{\omega}_{\rm lin} \cos\round{q \tilde{\omega}_{\rm lin}} \round{n^2 + k^2 \tilde{\omega}_{\rm lin}^2}  \notag \\
     & \qquad\quad \,\,\, \left. + \sin\round{q \tilde{\omega}_{\rm lin}} \round{n^4 q^2 + k^4 \tilde{\omega}_{\rm lin}^2 \round{-6 + q^2 \tilde{\omega}_{\rm lin}^2} + 2 k^2 n^2 \round{1 + q^2 \tilde{\omega}_{\rm lin}^2}}  \Bigr] \right. \notag \\
     & \qquad \,\, \left.- k \sinh\round{\frac{n q}{k}} \Bigl[ 2 q \sin\round{q \tilde{\omega}_{\rm lin}} \round{n^4 - k^4 \tilde{\omega}_{\rm lin}^4} \right. \notag \\
     & \qquad\quad \,\,\,  + \tilde{\omega}_{\rm lin} \cos\round{q \tilde{\omega}_{\rm lin}} \left( n^4 q^2 + k^4 \tilde{\omega}_{\rm lin}^2 \round{-2 + q^2 \tilde{\omega}_{\rm lin}^2} + 2 k^2 n^2 \round{3 + q^2 \tilde{\omega}_{\rm lin}^2} \right) \Bigr] \Bigr\} \,,
\end{align}
for the linear oscillations model. Also here we used $\tilde{\omega}_{\rm lin} \equiv \omega_{\rm lin} / k_*$. Instead, for the logarithmic case we have
\begin{align}
     &\Sigma^\text{PL}_\text{log} \left(k \right) \equiv \frac{1}{6\pi^2} \int_0^{k_\mathrm{S}} \dd q \; P_\mathrm{nw} \round{q} \Bigl\{ 1 - \frac{3 k}{q^3 \round{n^2 + \omega_\mathrm{log}^2}^3} \notag \\
     & \quad \, \times \left[ \cosh\round{\frac{n q}{k}} \Bigl( -2 k q \cos\round{\omega_\mathrm{log} \frac{q}{k}} \round{n^4 - \omega_\mathrm{log}^4} \noright \notag \\
     & \qquad\quad \,\,\, + \omega_\mathrm{log} \sin\round{\omega_\mathrm{log} \frac{q}{k}} \left( k^2 \round{6 n^2 -2 \omega^2_\mathrm{log}} + q^2 \round{n^2 + \omega^2_\mathrm{log}}^2 \right) \Bigr) \notag \\
     & \qquad \,\, + n\sinh \round{\frac{nq}{k}} \Bigl( -4 k q \omega_\mathrm{log} \sin\round{\omega_\mathrm{log} \frac{q}{k}} \round{n^2 + \omega_\mathrm{log}^2} \notag \\
     & \qquad\quad \,\,\, \noleft + \cos\round{\omega_\mathrm{log} \frac{q}{k}} \round{2k^2 \round{n^2 - 3 \omega^2_\mathrm{log}} +q^2 \round{n^2 + \omega^2_\mathrm{log}}^2 } \Bigr) \right] \vphantom{\frac{3 k}{q^3 \round{n^2 + k^2 \omega_\mathrm{lin}^2}^3}} \Bigr\} \,,
\end{align}
and
\begin{align}
     &\hat{\Sigma}^\text{PL}_\text{log} \left(k \right) \equiv - \frac{1}{6\pi^2} \int_0^{k_\mathrm{S}} \dd q \; P_\mathrm{nw} \round{q} \frac{3 k}{q^3 \round{n^2 + \omega_\mathrm{log}^2}^3} \notag \\
     & \quad \, \times \left[ n \cosh\round{\frac{n q}{k}} \Bigl( 4 k q \omega_\mathrm{log} \cos\round{ \omega_\mathrm{log} \frac{q}{k}} \round{n^2 + \omega_\mathrm{log}^2}  \right. \notag \\
     & \qquad\quad \,\,\,  \left.+ \sin\round{\omega_\mathrm{log} \frac{q}{k}} \round{ 2k^2 \round{n^2 -3 \omega^2_\mathrm{log}} + q^2 \round{n^2 + \omega^2_\mathrm{log}}^2 }  \Bigr) \right. \notag \\
     & \qquad \,\, \left.- \sinh\round{\frac{n q}{k}} \Bigl( 2 k q \sin\round{\omega_\mathrm{log} \frac{q}{k}} \round{n^4 - \omega_\mathrm{log}^4}  \right.\notag \\
     & \qquad\quad \,\,\, \left. + \omega_\mathrm{log} \cos\round{ \omega_\mathrm{log} \frac{q}{k}} \left(  k^2 \round{6n^2 - 2 \omega_{\mathrm{log}}} +q^2 \round{n^2 + \omega^2_\mathrm{log}}^2  \right) \Bigr) \right] \,.
\end{align}
Finally, these new sets of damping factors, evaluated for the case of the power-law envelope, can be used in combination 
to~\cref{eq:LO_power_packets,eq:NLO_power_packets} to obtain the nonlinear matter power spectra at LO and NLO, respectively.

\section{Cosmological simulations} \label{sec:cola_sims}
To quantify the accuracy of the analytic templates based on perturbation theory of the dark matter clustering in the nonlinear 
regime, we produce sets of large-scale N-body simulations. In particular, we explore the following primordial features models:
\begin{description}
    \item[LIN] with $A_{\rm lin} = 0.1$, $\log_{10} \round{\omega_{\rm lin}} = [0.6, 1.0, 1.4, 1.8]$, and $\varphi_{\rm lin} = 0$;
    \item[LOG] with $A_{\rm log} = 0.1$, $\log_{10} \round{\omega_{\rm log}} = [0.6, 1.0, 1.4, 1.8]$, and $\varphi_{\rm log} = 0$;
    \item[WPLIN] with $A_{\rm lin} = 0.1$, $\mu = [0.1, 0.3]$, $\Delta = [0.01, 0.1]$, $\log_{10} \round{\omega_{\rm lin}} = [0.8, 1.4]$, and $\varphi_{\rm lin} = 0$;
    \item[WPLOG] with $A_{\rm log} = 0.1$, $\mu = [0.1, 0.3]$, $\Delta = [0.01, 0.1]$, $\log_{10} \round{\omega_{\rm log}} = [0.8, 1.4]$, and $\varphi_{\rm log} = 0$;
    \item[PLLIN] with $A_{\rm lin} = 0.1$, $k_\mathrm{PL} = 0.05$, $n = [-0.5, 0.5]$, $\log_{10} \round{\omega_{\rm lin}} = [0.8, 1.4]$, and $\varphi_{\rm lin} = 0$;
    \item[PLLOG] with $A_{\rm log} = 0.1$, $k_\mathrm{PL} = 0.1$, $n = 0.5$, $\log_{10} \round{\omega_{\rm log}} = [0.8, 1.4]$, and $\varphi_{\rm log} = 0$.
\end{description}
Here $\mu$ and $\Delta$ have the dimension of $h\,{\rm Mpc}^{-1}$, $k_\mathrm{PL}$ has the dimension of $\mathrm{Mpc}^{-1}$, and $n$ is a dimensionless number. We consider an amplitude 
for the primordial features $A_{\rm X} = 0.1$ larger compared to the one allowed by current constraints for most of the 
template considered here. Bounds obtained from current CMB and LSS surveys correspond to $A_{\rm X} \lesssim 0.02-0.03$ at 95\% confidence level for constant-amplitude primordial oscillations \cite{Akrami:2018odb,Beutler:2019ojk,Ballardini:2022vzh,Mergulhao:2023ukp} 
while it could be larger for localised primordial oscillations \cite{Akrami:2018odb,Ballardini:2022wzu,Mergulhao:2023ukp} and specific best fits. The other feature parameters have been chosen to have an impact in the range of interest for LSS surveys. We fix the values of the standard cosmological parameters to~\cite{Planck:2018vyg}
\begin{equation}
    \Omega_{\rm m} = 0.31377,\, \Omega_{\rm b} = 0.04930,\, h = 0.6736,\, \sigma_8 = 0.8107,\, n_{\rm s} = 0.9649
\end{equation}
the value of $\sigma_8$ corresponds to the value for a flat $\Lambda$CDM cosmological model with one massive neutrinos with 
minimal mass of $60\, {\rm meV}$ and $\ln \left(10^{10} A_{\rm s}\right) = 3.044$. Note that high frequency oscillations are 
averaged to zero and do not modify the value of $\sigma_8$. Vice versa, cases with a low frequency modify $\sigma_8$ with respect 
the $\Lambda$CDM model. We keep it fixed to make the comparison on small scales clearer. 

The simulations produced in this paper make use of the {\tt COLA} (COmoving Lagrangian Acceleration) method 
\cite{Tassev:2013pn,Tassev:2015mia,Winther:2017jof,Wright:2017dkw}, an approximate N-body method alternative to full N-body 
simulations that combines second-order Lagrangian Perturbation Theory (2LPT) with a Particle-Mesh (PM) algorithm. The main 
advantage of this hybrid approach is its ability to efficiently evolve large-scale and intermediate-scale structures using 2LPT, 
while the PM algorithm handles the small-scale evolution. This combination allows for a significant reduction in the number of 
time-steps required compared to conventional N-body simulations, without compromising the accuracy of the large-scale dynamics. 
        
In traditional N-body simulations, particle positions and velocities are updated using the equations of motion
\begin{subequations} 
    \begin{align} 
        \frac{\mathrm{d} \mathbf{x}}{\mathrm{d} t} &= \mathbf{v} \,, \\ 
        \frac{\mathrm{d} \mathbf{v}}{\mathrm{d} t} &= -\nabla \Phi \,,
    \end{align}
\end{subequations}
where $\mathbf{x}$ and $\mathbf{v}$ represent the position and velocity of a particle, and $\Phi$ is the gravitational potential.
In the COLA framework, the particle displacement is expressed as $\mathbf{x}_{\rm COLA} = \mathbf{x} - \mathbf{x}_{\rm LPT}$, where 
$\mathbf{x}_{\rm LPT}$ denotes the trajectory determined by 2LPT. Consequently, the equations of motion are modified to
\begin{subequations} 
    \begin{align} 
        \frac{\mathrm{d} \mathbf{x}}{\mathrm{d} t} &= \mathbf{v}_{\text{COLA}} + \frac{\mathrm{d} \mathbf{x}_{\text{LPT}}}{\mathrm{d} t} \,, \\ 
        \frac{\mathrm{d} \mathbf{v}_{\text{COLA}}}{\mathrm{d} t} &= -\nabla \Phi - \frac{\mathrm{d}^2 \mathbf{x}_{\text{LPT}}}{\mathrm{d} t^2} \,, 
    \end{align}
\end{subequations}
where the variables $\mathbf{x}$ and $\mathbf{v}_{\rm COLA}$ are evolved over time. At the start of the simulation 
$\mathbf{v}_{\rm COLA} = 0$, ensuring that the particle trajectories closely follow the LPT evolution on large scales. This 
method effectively functions as a PM N-body code in the COLA frame, binding particles to LPT trajectories at large 
scales.

COLA-based simulations as well as full N-body simulations, and the agreement between the two approaches, have already been used to study the nonlinear dynamics in the presence of 
primordial features in Refs.~\cite{Ballardini:2019tuc,Chen:2020ckc,Li:2021jvz,Ballardini:2022vzh,Euclid:2023shr}.

\subsection{Simulation settings}
We executed COLA simulations using a modified version of the publicly available code {\tt L-PICOLA}~\cite{Howlett:2015hfa}\footnote{\url{https://github.com/CullanHowlett/l-picola}, \url{https://github.com/HAWinther/MG-PICOLA-PUBLIC}} evolving $2048$ particles in a 
simulation box of $(2048\,h^{-1}\,{\rm Mpc})^3$ and a mesh grid of $N_{\rm mesh} = 1024$.
        
All simulations commenced from an initial redshift of $z_{\rm ini} = 9$ using 2LPT to generate the initial displacement 
fields, ensuring that transient effects from initial conditions are minimised. The simulations were performed with 50 
time-steps across five redshift intervals, each with a time resolution of $\Delta a \approx 0.02$, with 10 time-steps for 
each of the following interval $9 \geq z > 2$, $2 \geq z > 1.5$, $1.5 \geq z > 1$, $1 \geq z > 0.5$, and $0.5 \geq z > 0$.

To suppress the undersamples large modes, due to the finite volume of the simulations, we use spectra averaged over pairs 
of simulations with the same initial seeds and inverted initial conditions, and with amplitude fixing in order to minimise 
the cosmic variance~\cite{Viel:2010bn,Villaescusa-Navarro:2018bpd}. Finally, we generate 5 realisations with different 
initial seed for each pairs of cosmological model studied.

\subsection{Comparing predictions of perturbation theory with N-body simulations}
\begin{figure}
    \centering
    \includegraphics[width=\linewidth]{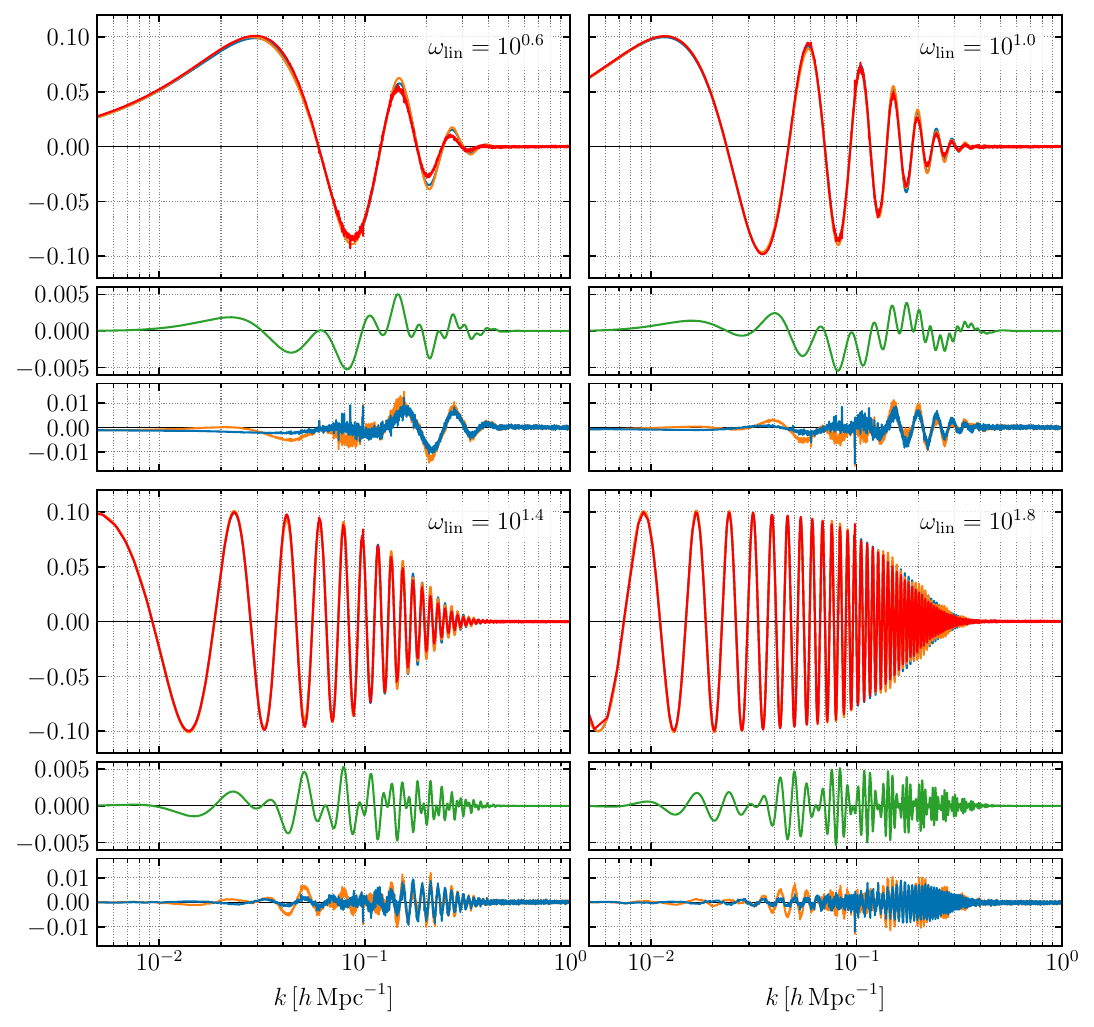}
    \caption{In each of the four panels we show: the relative differences between the matter power spectra computed 
    for a model with primordial linear oscillations and the one with power-law PPS divided by the non-wiggly power 
    spectrum with (orange line) and without (blue line) the mixed term at NLO and the one obtained from the N-body 
    simulations (red line), the relative differences of the NLO matter power spectrum with mixed term to that without 
    (green line), and the differences between the theoretical spectra at NLO, with (orange line) and without (blue line) 
    the mixed term, and the N-body simulations. All the matter power spectra are computed at redshift $z = 0$.}
    \label{fig:LIN_panel}
\end{figure}
In this section, we present a comparison between the results from TSPT and N-body simulations for the nonlinear matter 
power spectrum at redshift $z=0$. We analyse all six different types of primordial feature modelled, with linear and 
logarithmic oscillations and with constant and scale-dependent amplitudes.\footnote{We used {\tt FastPT} 
\cite{2016JCAP...09..015M,2017JCAP...02..030F} to compute the 1-loop power spectrum; \url{https://github.com/JoeMcEwen/FAST-PT}.}
\begin{figure}
    \centering
    \includegraphics[width=\linewidth]{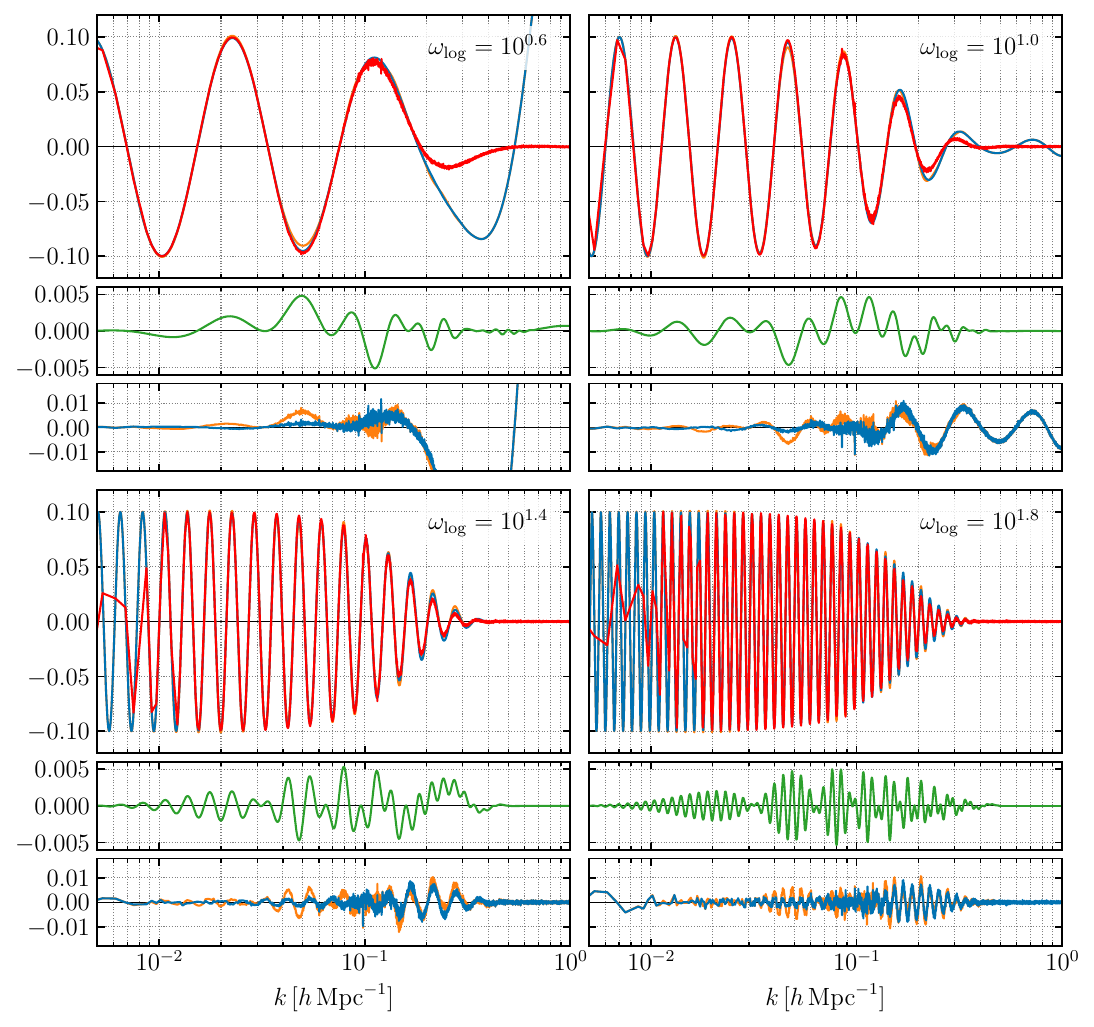}
    \caption{As in~\cref{fig:LIN_panel} for the primordial logarithmic oscillations.}
    \label{fig:LOG_panel}
\end{figure}

In~\cref{fig:LIN_panel} we show the comparison for primordial linear oscillations with constant amplitude, focusing on 
the results at NLO. At LO, differences between TSPT spectra and N-body simulations are reduced and closer to the NLO 
results by choosing the separation scale $k_{\rm S} = 0.2\,h\, {\rm Mpc}^{-1}$; see in Refs.~\cite{Ballardini:2019tuc,Euclid:2023shr} 
for a comparison between LO and NLO for a wider range of values for $k_{\rm S}$. At NLO, however, the comparison improves 
and the dependence on the separation scale $k_{\rm S}$ is minimal. This suggests that higher-order corrections are critical 
to accurately capture the nonlinear behaviour, especially at smaller scales where nonlinear damping becomes important.
A comparable trend is observed when logarithmic oscillations are taken into account, as shown in~\cref{fig:LOG_panel}. 
However, for both LO and NLO, discrepancies with N-body simulations are more pronounced, particularly for low primordial 
frequencies such as $\log_{10} \round{\omega_{\rm log}} = 0.6$, as already pointed out in 
Refs.~\cite{Vasudevan:2019ewf,Beutler:2019ojk,Ballardini:2019tuc,Euclid:2023shr}. This is due to the different correlation 
length in real space between linear and logarithmic oscillations, from $\omega_{\rm lin}/k_*$ to $\omega_{\rm log}/k$.
The TSPT results at NLO are in agreement with the findings of the simulations for both linear and logarithmic oscillations 
with constant amplitude, with discrepancies of less than 1\% observed, with the exception of the logarithmic 
model at the lower frequency.\footnote{For effective frequencies lower than the BAO one the validity of the perturbation theory results are expected to worsted as previously discussed and shown in Refs.~\cite{Vasudevan:2019ewf,Beutler:2019ojk,Ballardini:2019tuc,Euclid:2023shr}.} 

We explore the addition of nonlinear modelling of the mixed term between baryon and primordial oscillations, neglected in 
previous studies, to further refine the TSPT predictions. Given the relative small amplitude of this term, with the amplitude 
of the BAO feature accounting for approximately 7\% of the total matter power spectrum and the amplitude of the primordial 
features expected to be less than 10\% depending on the specific model, the effect is almost undetectable when compared to 
the simulation results; see~\cref{fig:LIN_panel,fig:LOG_panel}.

\begin{figure}
    \centering
    \includegraphics[width=\linewidth]{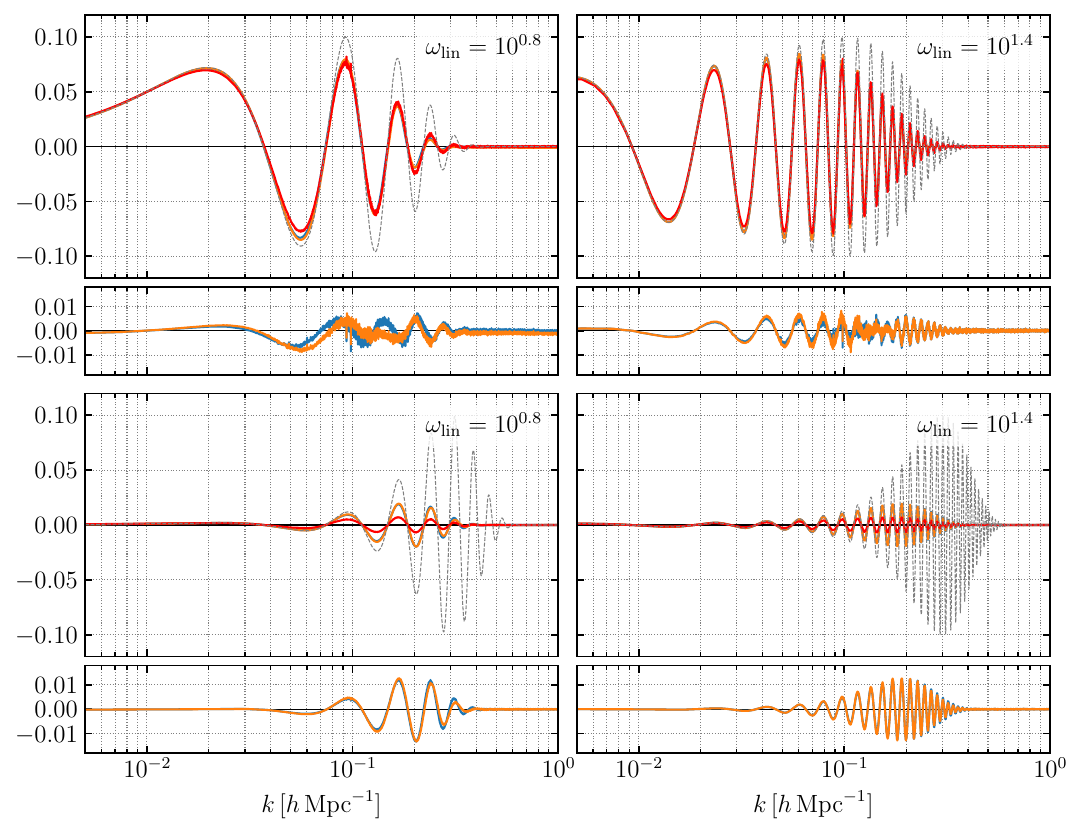}
    \caption{In each of the four panels we show: in the upper subpanel the relative differences between the matter power 
    spectra computed for a model with primordial linear oscillations, with Gaussian amplitude, and the one with power-law PPS 
    divided by the non-wiggly power spectrum at NLO (orange line) and LO (blue line), for the N-body simulations (red line) and for 
    the linear theory (dashed grey line), and in the lower subpanel the relative differences between the theoretical spectra, 
    at NLO (orange line) and LO (blue line), and the N-body simulations. The two upper panels correspond to 
    $\mu = 0.1\, h\, {\rm Mpc}^{-1}$ while the lower ones to $\mu = 0.3\, h\, {\rm Mpc}^{-1}$; here we always assume  
    $\Delta = 0.1\, h\, {\rm Mpc}^{-1}$. All the matter power spectra are computed at redshift $z = 0$.}
    \label{fig:GAUSSLIN_panel}
\end{figure}

\begin{figure}
    \centering
    \includegraphics[width=\linewidth]{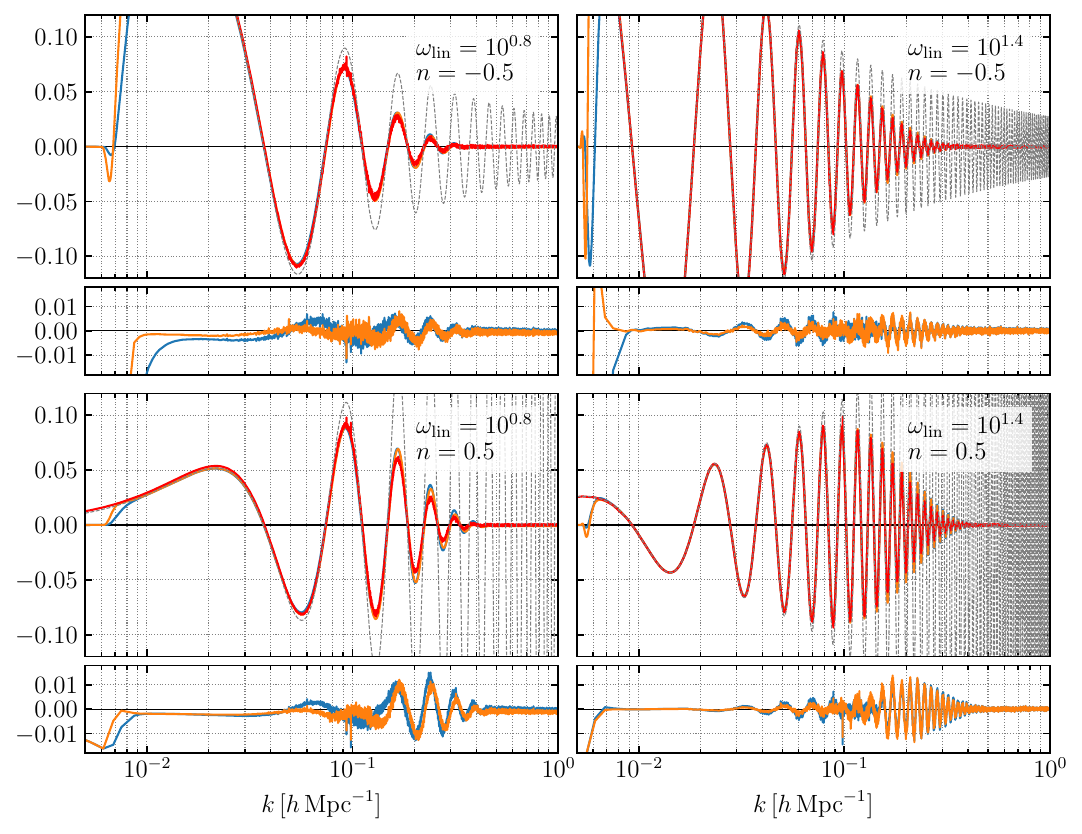}
    \caption{As in~\cref{fig:GAUSSLIN_panel} for the primordial linear oscillations with power-law amplitude 
    with $n=\pm 0.5$, $k_\mathrm{PL} = 0.05\, {\rm Mpc}^{-1}$, and $\log_{10} \round{\omega_{\rm log}} = 0.8,\,1.4$.}
    \label{fig:PLLIN_panel}
\end{figure}

\begin{figure}[t]
    \centering
    \includegraphics[width=\linewidth]{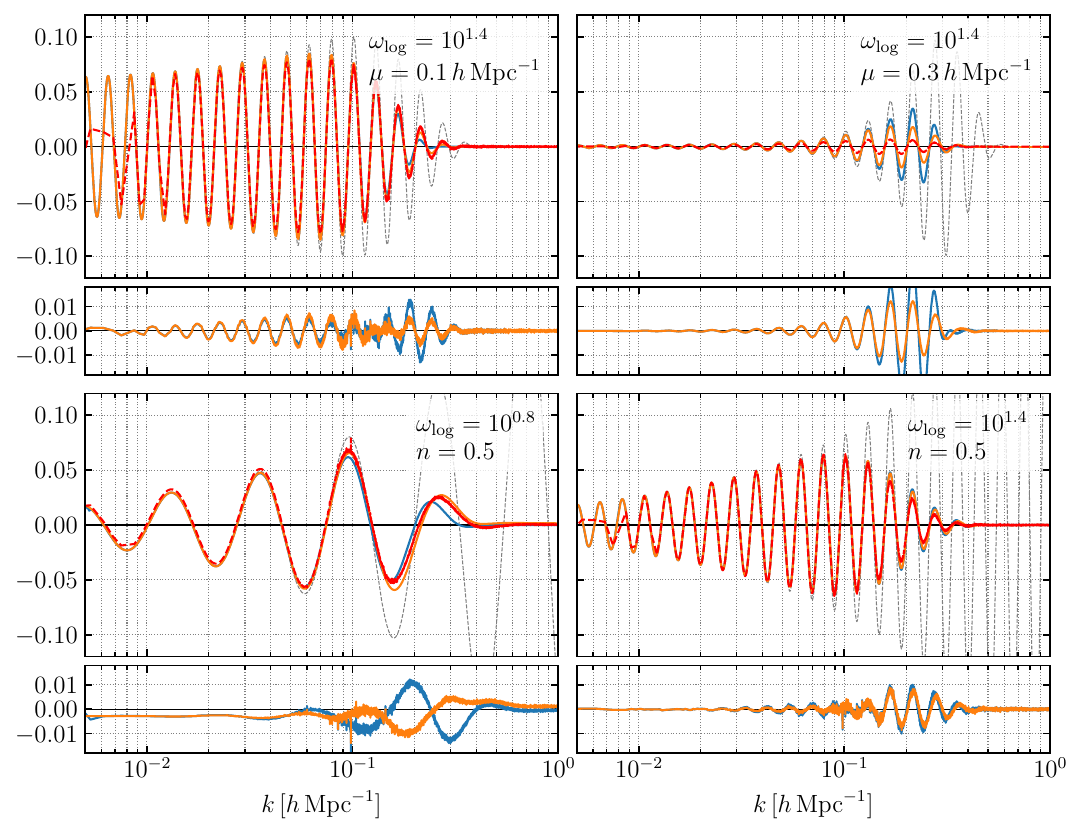}
    \caption{In each of the four panels we show: in the upper subpanel the relative differences between the matter power 
    spectra computed for a model with primordial logarithmic oscillations, with scale-dependent amplitude, and the one with power-law 
    PPS divided by the non-wiggly power spectrum at NLO (orange line) and LO (blue line), for the N-body simulations (red line) and 
    for the linear theory (dashed grey line), and in the lower subpanel the differences between the theoretical spectra, at NLO 
    (orange line) and LO (blue line), and the N-body simulations. 
    The two upper panels correspond to the case with Gaussian wave-packet amplitude with $\log_{10} \round{\omega_{\rm log}} = 1.4$, 
    $\Delta = 0.1\, h\, {\rm Mpc}^{-1}$, and $\mu = 0.1\, (0.3)\, h\, {\rm Mpc}^{-1}$ on the left (right).
    The two lower panels correspond to the power-law amplitude with $n=0.5$, $k_\mathrm{PL} = 0.1\, {\rm Mpc}^{-1}$, and 
    $\log_{10} \round{\omega_{\rm log}} = 0.8$ (1.4) on the left (right).
    All the matter power spectra are computed at redshift $z = 0$.}
    \label{fig:MIXLOG_panel}
\end{figure}

Furthermore, in addition to the oscillatory models with constant amplitude, we present the comparison with N-body simulations 
for the wave-packet and power-law models, see~\cref{fig:GAUSSLIN_panel,fig:PLLIN_panel,fig:MIXLOG_panel}. These scale-dependent 
amplitudes, which are closer to the expected primordial features produced by exact models, result in distinctive signatures 
within the matter power spectrum. In this case, the TSPT predictions show reasonable agreement with simulations at larger 
scales, but the inclusion of NLO corrections is crucial to improve accuracy over a wider range of scales. The results demonstrate 
the efficacy of TSPT in capturing the physics of primordial features, particularly when higher-order terms are considered.

For all models considered, the power spectrum at $z=0$ shows that non-linearities on small scales tend to wash out the 
primordial oscillations, especially at wavenumbers beyond $k = 0.3\, h\, {\rm Mpc}^{-1}$. 
In summary, the comparison between TSPT and N-body simulations shows that NLO corrections significantly improve the agreement 
for both linear and logarithmic oscillations, as well as for the scale-dependent amplitude case. This highlights the importance 
of going beyond the LO to capture the complexity of nonlinear structure formation, especially when studying primordial features 
in the matter power spectrum, in order to reduce the theoretical uncertainties arising from the modelling of nonlinear damping.

It should be noted that the discrepancies observed at large scales, particularly for logarithmically-spaced high frequencies, 
are numerical artefacts resulting from the finite number of samples from the nonlinear matter power spectrum of the N-body 
simulations. These discrepancies have no impact on the conclusions drawn in this study.

\section{Conclusions} \label{sec:conclusions}
In this work, we have presented a comprehensive analysis of the impact of primordial oscillatory features on the matter 
power spectrum within the framework of TSPT. 
We derive new contributions to the nonlinear matter power spectrum, namely 
the mixed term between primordial 
oscillations and BAO, as well as corrections arising from a scale-dependent envelope modulating the oscillation patterns. 
These corrections, although small, represent an important refinement of current theoretical models, bringing us closer to 
a full characterisation of the nonlinear imprints of primordial features.

The inclusion of IR resummation techniques, as implemented in TSPT, has proved essential to handle long-wavelength perturbations 
that significantly affect the power spectrum, especially in the presence of oscillatory features. The good agreement between our 
analytical results and N-body simulations underlines the robustness of the formalism and highlights the importance of using such 
perturbative techniques for future LSS analyses of primordial features.

It has been shown that in both the cases of linear and logarithmic oscillatory features any contribution is generated from small-scale inhomogeneities in the nonlinear region of the power spectrum~\cite{Beutler:2019ojk,2017JCAP...11..007B}. Corrections due to the mixed term do not spoil the argument used to prove this statement, which still holds without additional issues. Differently, when considering scale-dependent amplitudes additional scales enter the discussion (like that coming from the amplitude of the Gaussian shape, for our GAUSSLIN and GAUSSLOG models). We found that when these scales become too small, the power counting developed between derivatives of the oscillatory template breaks down making the computation unstable and unreliable. These additional constraints on the allowed parameter space can be derived through~\cref{eq:error_formula} or analogous.

Our results are relevant for upcoming LSS surveys, such as DESI~\cite{DESI:2024mwx} and {\em Euclid}~\cite{Euclid:2023shr,Euclid:2024yrr}, which are expected to provide more precise measurements of the matter power spectrum 
at small scales, potentially revealing subtle signatures of new physics in the early Universe. By considering the nonlinear 
regime of structure formation, our analysis provides an improved theoretical framework for interpreting these future data.

With the increasing precision of cosmological surveys, the detection of primordial features remains a key goal in probing the 
physics of the early Universe. The tools developed in this study provide a solid foundation for these efforts and offer new 
avenues for exploration beyond the standard cosmological model. 
In the future, it will be crucial to further refine these models and extend them to other potential sources of primordial 
features and more complex scenarios. In addition, primordial features are expected to leave correlated signatures on higher 
order correlators~\cite{Chen:2006xjb,Chen:2008wn,Cyr-Racine:2011bjz,2014PhRvD..89f3540G,2015JCAP...10..062M}, 
such as the primordial bispectrum, as well as small but characteristic imprints in the galaxy bias expansion~\cite{Cabass:2018roz}, 
which can be modelled within the TSPT approach.

\acknowledgments
We thank Xingang Chen and Hans Winther for useful comments.
MB and NB acknowledge financial support from the INFN InDark initiative. MB also acknowledges financial support from the COSMOS network (\url{www.cosmosnet.it}) through the ASI (Italian Space Agency) Grants 2016-24-H.0, 2016-24-H.1-2018, 2020-9-HH.0 (participation in LiteBIRD phase A).

\appendix
    \section{Partial relaxation of $q\ll k$ limit} \label{app:relaxation}
        In this appendix, following the approach of Ref.~\cite{Beutler:2019ojk}, we aim to demonstrate how it is possible to include some of the terms neglected in~\cref{eq:exp_form} due to the assumption of working in the limit $q \ll k$. 
        Note beforehand that the intermediate results and steps in this subsection differ from those in Ref.~\cite{Beutler:2019ojk}, 
        while the final result for the damping factor matches when integrated via~\cref{eq:s_operator_explicit}.
        In contrast to the strong approximation used in the main text, we will refer to this approach as the \textit{weak approximation}. To avoid overloading the paper with unnecessary derivation steps, we assume a phase $\varphi_\mathrm{log} = 0$, and will provide comments on this assumption at the end of the calculation.

        \subsection{Logarithmic oscillations}
        The crucial point is writing the $2n$-th element in the sum defining the translation operator, see~\cref{eq:t_op_sum}, 
        stripped of the momenta $\mathbf{q}$ and evaluated on the oscillating function in the wiggle part of the spectrum as 
        \begin{align}
            &\partial_{k_{i_n}} \cdots \partial_{k_{i_{2n}}} \sin \round{\omega_\text{log} \ln \frac{k}{k_*}} = \notag \\
            & \quad\, = \frac{1}{2i} \frac{ \hat{k}_{i_1} \cdots \hat{k}_{i_{2n}}}{k^{2n}} \sround{f_n \round{i \omega_\text{log}} \round{\frac{k}{k_*}}^{i \omega_\text{log}} - f_n \round{-i\omega_\text{log}} \round{\frac{k}{k_*}}^{-i \omega_\text{log}} } \notag \\
            & \quad\, = \frac{1}{2} \frac{ \hat{k}_{i_1} \cdots \hat{k}_{i_{2n}}}{k^{2n}} \left\{ \sround{ f_n \round{i \omega_\text{log}} + f_n \round{-i \omega_\text{log}} } \sin \round{\omega_\text{log} \ln \frac{k}{k_*} } \noright \notag \\
            & \noleft \quad\, \quad\, -i \sround{ f_n \round{i \omega_\text{log}} - f_n \round{-i \omega_\text{log}} } \cos \round{\omega_\text{log} \ln \frac{k}{k_*} } \right\} \,, \label{eq:nth_derivative}
       \end{align}
       where we employed the following decomposition of the sine  
       \begin{equation}
           \sin \round{y \ln x} = \frac{1}{2i} \round{ e^{iy \ln x} - e^{-iy \ln x} } = \frac{1}{2i} \round{x^{iy} - x^{-iy}} \,,
       \end{equation}
       and we defined the function 
       \begin{equation}
           f_n \round{i\omega_\text{log}} \equiv \frac{\round{i \omega_\text{log}}!}{\round{i \omega_\text{log} -2n }!} \,.
       \end{equation}
       The action of the operator appearing in~\cref{eq:operator_action} on these term is 
       \begin{align}
           &\sround{1 - \cosh \round{\mathbf{q} \cdot \nabla_{\mathbf{k}}}} P^\mathrm{w}_\text{log} \round{k} = \notag \\
           & \quad\, = \sround{ 1- \sum_{n=0}^{\infty} \frac{1}{\round{2n}!} \round{\mathbf{q} \cdot \nabla_{\mathbf{k}} }^{2n} } P^\mathrm{nw} \round{k} A_\text{log} \sin \round{ \omega_\text{log} \ln{\frac{k}{k_*}} } \notag \\
           & \quad \, = P^\mathrm{nw} \round{k} A_\text{log} \left\{ \sin \round{ \omega_\text{log} \ln{\frac{k}{k_*}}} - \frac{1}{2} \sum_{n=0}^{\infty} \frac{1}{\round{2n}!} q_{i_1} \cdots q_{i_{2n}} \frac{ \hat{k}_{i_1} \cdots \hat{k}_{i_{2n}}}{k^{2n}} \noright \notag \\
           & \qquad \,\, \times \left[ \round{ f_n \round{i \omega_\text{log}} + f_n \round{-i \omega_\text{log}} } \sin \round{\omega_\text{log} \ln \frac{k}{k_*} } \noright \notag \\
           & \qquad \quad \,\,\, \noleft \noleft -i \round{ f_n \round{i \omega_\text{log}} - f_n \round{-i \omega_\text{log}} } \cos \round{\omega_\text{log} \ln \frac{k}{k_*} } \right] \vphantom{\frac{ \hat{k}_{i_1}}{k^{2n}}} \right\}  \,. \label{eq:operator_action_2}
       \end{align} 
        As anticipated, the computation was performed with a phase of $\varphi_\mathrm{log} \neq 0$. While this would have 
        resulted in a more extensive calculation, it would not have been significantly more challenging or different from 
        what we have done. In that case, one simply needs to express the oscillating function in~\cref{eq:nth_derivative} as
        \begin{equation} 
            \sin \round{\omega_\text{log} \ln \frac{k}{k_*} + \varphi_\mathrm{log}} = \cos\round{\varphi_\mathrm{log} }\sin \round{\omega_\text{log} \ln \frac{k}{k_*}} + \sin\round{\varphi_\mathrm{log} }\cos \round{\omega_\text{log} \ln \frac{k}{k_*}} \,.  \label{eq:phase_separation}
        \end{equation}
        The first term on the right hand side is identical to that which was previously studied, with the exception of the multiplication by the constant 
        $\cos\round{\varphi_\mathrm{log}}$. The second term is analogous to the first, differing only in the sign. Given that 
        the operator in~\cref{eq:operator_action_2} is linear, it is feasible to compute the two terms separately and subsequently 
        add the results together. By summing the two results, weighted by the corresponding phase functions, and combining the trigonometric functions using~\cref{eq:phase_separation}, it is possible to recast the result in the same form as~\cref{eq:operator_action_2}, with the phase appearing in the argument of the sine and cosine functions in the second line, showing that this result is unaffected by the presence of a phase.

        For the next step of the computation, let us break~\cref{eq:operator_action_2} into two parts. We begin by analyzing the first term, namely
        \begin{equation}
            \circled{1} \equiv \frac{1}{2} \sum_{n=0}^{\infty} \frac{1}{\round{2n}!} q_{i_1} \cdots q_{i_{2n}} \frac{ \hat{k}_{i_1} \cdots \hat{k}_{i_{2n}}}{k^{2n}} \sround{ f_n \round{i \omega_\text{log}} + f_n \round{-i \omega_\text{log}} } \,. \label{eq:term_1}
        \end{equation}
        The $2n$ scalar products in~\cref{eq:term_1} can be rewritten as 
        \begin{equation} \label{eq:hint_1}
            \frac{(\mathbf{q} \cdot \hat{\mathbf{k}})}{k} =  \frac{q}{k} (\hat{\mathbf{q}} \cdot \hat{\mathbf{k}}) = \frac{q \mu }{k} \,,
        \end{equation}
        and we can also notice that 
        \begin{equation} \label{eq:hint_2}
            \frac{1}{\round{2n}!} f_n \round{i \omega_\text{log}} = \binom{i \omega_\text{log}}{i\omega_\text{log} - 2n} = \binom{i \omega_\text{log}}{2n} \,.
        \end{equation}
        Using~\cref{eq:hint_1,eq:hint_2} then, our sum can be rewritten as 
        \begin{equation}
            \circled{1} = \frac{1}{2} \sum_{n=0}^\infty \left\{ \binom{i \omega_\text{log}}{2n} \round{\frac{q\mu}{k}}^{2n} + \binom{-i \omega_\text{log}}{2n}  \round{\frac{q\mu}{k}}^{2n} \right\} \,.
        \end{equation}
        Here it is easy to recognize that each terms has the same form of a Newton's series with only even powers. In particular, it is possible to build such a series by means of the following trick 
        \begin{align}
            (1+x)^\alpha + (1-x)^\alpha &= \sum_{n=0}^{+\infty} \binom{\alpha}{n} \left[ 1 + \round{-1}^n \right] x^n = 2 \sum_{n=0}^{+\infty} \binom{\alpha}{2n} x^{2n} \,,
        \end{align}
        where, the second equality comes from the fact that all the elements corresponding to odd indices in the sum are identically vanishing. Thanks to this observations, the first term in~\cref{eq:operator_action_2} can be finally written as
        \begin{align}
            \circled{1} &= \frac{1}{4} \left\{ \round{1+ \frac{q\mu}{k}}^{i\omega_\text{log}} + \round{1 - \frac{q\mu}{k}}^{i\omega_\text{log}} + \round{1+ \frac{q\mu}{k}}^{-i\omega_\text{log}} + \round{1- \frac{q\mu}{k}}^{-i\omega_\text{log}} \right\}  \notag \\
            &= \frac{1}{2} \left\{ \cos \left[ \omega_\text{log} \ln \round{1 + \frac{q\mu}{k}} \right] + \cos \left[ \omega_\text{log} \ln \round{1 - \frac{q\mu}{k}} \right] \right\} \,.
        \end{align}
        Concerning the second term of~\cref{eq:operator_action_2}, we have 
        \begin{equation}
            \circled{2} \equiv \frac{i}{2} \sum_{n=0}^{+\infty} \frac{1}{\round{2n}!} q_{i_1} \cdots q_{i_{2n}} \frac{ \hat{k}_{i_1} \cdots \hat{k}_{i_{2n}}}{k^{2n}} \sround{ f_n \round{i \omega_\text{log}} - f_n \round{-i \omega_\text{log}} } \,.
        \end{equation}
        The analysis is completely analogous to that of $\circled{1}$, following what has been done before, we can write
        \begin{equation}
            \circled{2} = \frac{i}{2} \sum_{n=0}^\infty \left\{ \binom{i \omega_\text{log}}{2n} \round{\frac{q\mu}{k}}^{2n} - \binom{-i \omega_\text{log}}{2n}  \round{\frac{q\mu}{k}}^{2n} \right\} \,,
        \end{equation}
        and then 
        \begin{align}
            \circled{2} &= \frac{i}{4} \left\{ \round{1+ \frac{q\mu}{k}}^{i\omega_\text{log}} + \round{1 - \frac{q\mu}{k}}^{i\omega_\text{log}} - \round{1+ \frac{q\mu}{k}}^{-i\omega_\text{log}} - \round{1- \frac{q\mu}{k}}^{-i\omega_\text{log}} \right\}  \notag \\
            &= - \frac{1}{2} \left\{ \sin \left[ \omega_\text{log} \ln \round{1 + \frac{q\mu}{k}} \right] + \sin \left[ \omega_\text{log} \ln \round{1 - \frac{q\mu}{k}} \right] \right\} \,.
        \end{align}

        Putting everything back together in \cref{eq:operator_action_2}, we finally have
        \begin{align} \label{eq:operator_action_gen}
            &\sround{1 - \cosh \round{\mathbf{q} \cdot \nabla_{\mathbf{k}}} } P^\mathrm{w}_\text{log} \round{k} = \notag \\
            & \quad\, = \left\{ 1 - \frac{1}{2} \left[ \cos \round{\omega_\text{log} \ln \round{1 + \frac{q\mu}{k}}} + \cos \round{\omega_\text{log} \ln \round{1 - \frac{q\mu}{k}}} \right] \right\}  P^\mathrm{w}_\text{log} \round{k} \notag \\
            & \qquad \,\, - \left\{ \frac{1}{2} \left[ \sin \round{\omega_\text{log} \ln \round{1 + \frac{q\mu}{k}}} + \sin \round{\omega_\text{log} \ln \round{1 - \frac{q\mu}{k}}} \right] \right\} \frac{P^\mathrm{nw} \round{k}}{\omega_\text{log}} \frac{\dd \delta P_\zeta^\text{log} \round{k}}{\dd \ln k} \,.
        \end{align}
        Here we notice that a term with the derivative of the spectrum appeared. This is analogous to what we found in the case of scale dependent amplitudes when a term with a phase shift appeared in our computation of the damping. In this case it was possible to write it as derivative of the spectrum because, given the absence of a scale-dependent amplitude, all the scale dependence is concentrated within the oscillatory function. We decided to keep this notation to facilitate a direct comparison with Ref.~\cite{Beutler:2019ojk} and for an easier interpretation of this object. At this point we can compute the damping 
        factors with the formula in~\cref{eq:LO_power_packets} finding
        \begin{align}
            &\Sigma_\text{log} \left(k \right) \equiv \frac{1}{6\pi^2} \int_0^{k_\mathrm{S}} \dd q \; P_\mathrm{nw} \round{q} \biggl\{ 1 - \frac{3}{2 q^3 \round{36 + 49 \omega_\mathrm{log}^2 + 14 \omega_\mathrm{log}^4 + \omega_\mathrm{log}^6}} \notag \\
            & \quad \,\times \Bigl[ - \round{k + q} \cos\round{\omega_\mathrm{log} \log\round{1 + \frac{q}{k}}} \notag \\
            & \qquad \quad \,\,\, \times \left( 12 k^2 \round{-1 + \omega_\mathrm{log}^2} - 2 k q \round{-6 - 5 \omega_\mathrm{log}^2 + \omega_\mathrm{log}^4} - 3 q^2 \round{4 + 5 \omega_\mathrm{log}^2 + \omega_\mathrm{log}^4} \right) \notag \\
            & \qquad \,\, + \round{k - q} \cos\round{\omega_\mathrm{log} \log\round{1 - \frac{q}{k}}} \notag \\
            & \qquad \quad \,\,\, \times \left( 12 k^2 \round{-1 + \omega_\mathrm{log}^2} + 2 k q \round{-6 - 5 \omega_\mathrm{log}^2 + \omega_\mathrm{log}^4} - 3 q^2 \round{4 + 5 \omega_\mathrm{log}^2 + \omega_\mathrm{log}^4} \right) \notag \\
            & \qquad \,\, - \round{k + q} \omega_\mathrm{log} \sin\round{\omega_\mathrm{log} \log\round{1+\frac{q}{k}}} \notag \\
            & \qquad \quad \,\,\, \times \left( 2 k^2 \round{-11 + \omega_\mathrm{log}^2} + 10 k q \round{1 + \omega_\mathrm{log}^2} - q^2 \round{4 + 5 \omega_\mathrm{log}^2 + \omega_\mathrm{log}^4} \right) \notag \\
            & \qquad \,\, + \round{k - q} \omega_\mathrm{log} \sin\round{\omega_\mathrm{log} \log\round{1 - \frac{q}{k}}} \notag \\[-5pt]
            & \qquad \quad \,\,\, \times \left( 2 k^2 \round{-11 + \omega_\mathrm{log}^2} - 10 k q \round{1 + \omega_\mathrm{log}^2} - q^2 \round{4 + 5 \omega_\mathrm{log}^2 + \omega_\mathrm{log}^4} \right) \Bigr] \biggr\} \,,
        \end{align}
        and
        \begin{align}
            & \hat{\Sigma}_\text{log} \left(k \right) \equiv - \frac{1}{6\pi^2} \int_0^{k_\mathrm{S}} \dd q \; P_\mathrm{nw} \round{q}  \frac{3}{2 q^3 \round{36 + 49 \omega_\mathrm{log}^2 + 14 \omega_\mathrm{log}^4 + \omega_\mathrm{log}^6}} \notag \\
            & \quad \, \times \biggl\{ \round{k + q} \omega_\mathrm{log} \cos\sround{\omega_\mathrm{log} \log\round{1 + \frac{q}{k}}} \notag \\
            & \qquad\quad\,\,\, \times \left[ 2 k^2 \round{-11 + \omega_\mathrm{log}^2} + 10 k q \round{1 + \omega_\mathrm{log}^2} - q^2 \round{4 + 5 \omega_\mathrm{log}^2 + \omega_\mathrm{log}^4} \right] \notag \\
            & \qquad \,\, - \round{k - q} \omega_\mathrm{log}  \cos\sround{\omega_\mathrm{log} \log\round{1 - \frac{q}{k}}} \notag \\
            & \qquad\quad\,\,\, \times \left[ 2 k^2 \round{-11 + \omega_\mathrm{log}^2} - 10 k q \round{1 + \omega_\mathrm{log}^2} - q^2 \round{4 + 5 \omega_\mathrm{log}^2 + \omega_\mathrm{log}^4} \right] \notag \\
            & \qquad \,\, - \round{k + q} \sin\sround{\omega_\mathrm{log} \log\round{1 + \frac{q}{k}}} \notag \\
            & \qquad \quad \,\,\, \times \left[ 12 k^2 \round{-1 + \omega_\mathrm{log}^2} - 2 k q \round{-6 - 5 \omega_\mathrm{log}^2 + \omega_\mathrm{log}^4} - 3 q^2 \round{4 + 5 \omega_\mathrm{log}^2 + \omega_\mathrm{log}^4} \right] \notag \\
            & \qquad \,\, + \round{k - q} \sin\sround{\omega_\mathrm{log} \log\round{1 - \frac{q}{k}}} \notag \\[-5pt]
            & \qquad \quad \,\,\, \times \left[ 12 k^2 \round{-1 + \omega_\mathrm{log}^2} + 2 k q \round{-6 - 5 \omega_\mathrm{log}^2 + \omega_\mathrm{log}^4} - 3 q^2 \round{4 + 5 \omega_\mathrm{log}^2 + \omega_\mathrm{log}^4} \right] \biggr\} \,.
        \end{align}

        Finally let us briefly comment on how the appearance of the additional term in~\cref{eq:operator_action_gen} is due to a minor loss, in terms of powers of $k$, when taking the $2n$ gradients on the wiggle power spectrum. Indeed in Ref.~\cite{Vasudevan:2019ewf} just the derivatives of a factor $\round{\mathbf{q} \cdot \mathbf{k}}/k^2$, in contrast to the derivatives of a factor $\round{\mathbf{q} \cdot \mathbf{k}}/k$ in Ref.~\cite{Beutler:2019ojk}, are neglected. As a proof of this fact we can follow the steps in the derivation of Ref.~\cite{Beutler:2019ojk} factorizing the whole $\round{\mathbf{q} \cdot \mathbf{k}}/k^2$, and we will find a result analogous to the one of Ref.~\cite{Vasudevan:2019ewf}, that is~\cref{eq:operator_action}. The trick is, at each iteration, to rewrite the action of the derivative as follows 
        \begin{align}
            \partial_{k_i} \sround{ \sin \round{\omega_\text{log} \ln \frac{k}{k_*}} } &= \frac{1}{2i}  \sround{ \frac{i \omega_\text{log}}{k_*} \round{\frac{k}{k_*}}^{i\omega_\text{log} -1} - \frac{\round{- i \omega_\text{log}}}{k_*} \round{\frac{k}{k_*}}^{-i\omega_\text{log} -1} } \frac{k_i}{k} \notag \\
            &= \frac{1}{2i}  \sround{ i \omega_\text{log} \round{\frac{k}{k_*}}^{i\omega_\text{log}} - \round{- i \omega_\text{log}} \round{\frac{k}{k_*}}^{-i\omega_\text{log}} } \frac{k_i}{k^2} \,,
        \end{align}
        so that, in this approximation, the analogous of \cref{eq:nth_derivative} will be 
        \begin{align}
             &\partial_{k_{i_n}} \cdots \partial_{k_{i_{2n}}} \sin \round{\omega_\text{log} \ln \frac{k}{k_*}} = \notag \\
             &\quad \,= \frac{1}{2i} \frac{ k_{i_1} \cdots k_{i_{2n}}}{k^{4n}} \sround{ \round{i \omega_\text{log}}^{2n} \round{\frac{k}{k_*}}^{i \omega_\text{log}} - \round{-i\omega_\text{log}}^{2n} \round{\frac{k}{k_*}}^{-i \omega_\text{log}} } \,.
        \end{align}
        And finally, we can write 
        \begin{align}
            &\sround{1 - \cosh \round{\mathbf{q} \cdot \nabla_{\mathbf{k}}} } P^\mathrm{w}_\text{log} \round{k} = \notag \\
            & \quad \, = P^\mathrm{nw} \round{k} A_\text{log} \left\{ \vphantom{ \sround{ \round{i \omega_\text{log}}^{2n} \round{\frac{k}{k_*}}^{i \omega_\text{log}} - \round{-i\omega_\text{log}}^{2n} \round{\frac{k}{k_*}}^{-i \omega_\text{log}} } } \sin \round{ \omega_\text{log} \ln{\frac{k}{k_*}} } \noright \notag \\
            & \qquad \,\, \noleft - \frac{1}{2i} \sum_{n=0}^{\infty} \frac{1}{\round{2n}!} q_{i_1} \cdots q_{i_{2n}} \frac{ k_{i_1} \cdots k_{i_{2n}}}{k^{4n}} \sround{ \round{i \omega_\text{log}}^{2n} \round{\frac{k}{k_*}}^{i \omega_\text{log}} - \round{-i\omega_\text{log}}^{2n} \round{\frac{k}{k_*}}^{-i \omega_\text{log}} } \right\} \notag \\
            & \quad \, = P^\mathrm{nw} \round{k} A_\text{log} \left\{ \vphantom{ \sround{ \round{i \omega_\text{log}}^{2n} \round{\frac{k}{k_*}}^{i \omega_\text{log}} - \round{-i\omega_\text{log}}^{2n} \round{\frac{k}{k_*}}^{-i \omega_\text{log}} } } \sin \round{ \omega_\text{log} \ln{\frac{k}{k_*}} } \noright \notag \\
            & \qquad \,\, \noleft - \sum_{n=0}^{\infty} \frac{1}{\round{2n}!} \round{-1}^n \sround{\frac{\round{\mathbf{q} \cdot \mathbf{k}}}{k^2} \omega_\text{log} }^{2n} \times \frac{1}{2i} \sround{ \round{\frac{k}{k_*}}^{i \omega_\text{log}} - \round{\frac{k}{k_*}}^{-i \omega_\text{log}} } \right\} \notag \\
            & \quad \, = \sround{1 - \cos \round{\frac{\mathbf{q} \cdot \mathbf{k}}{k^2} \omega_\text{log} }} P^\mathrm{w}_\text{log} \round{k} \,,
        \end{align}
        which is exactly analogous to~\cref{eq:operator_action}.

        \subsection{Mixed term between baryon and primordial logarithmic oscillations}
            We can use a similar method to compute a more detailed version of~\cref{eq:S_mix_osc_1}. The $2n$-th derivative of the mixed oscillatory term can be calculated using induction. For example, for $n=1$, we have that\footnote{We carry out our computation only on the first term of the wiggle power spectrum. We will sum everything together at the end.}
            \begin{align}
                & \partial_{k_{i_1}} \partial_{k_{i_2}} \left[ \cos \round{\omega_\text{log} \ln \frac{k}{k_*} - \tilde{\omega}_\mathrm{BAO} k } \right] = \notag \\
                & \quad \, = \frac{1}{2} \frac{k_{i_1} k_{i_2}} {k^2} \left\{ \left[ \frac{i \omega_\text{log}}{k_*} \frac{\round{i \omega_\text{log}-1} }{k_*} \round{\frac{k}{k_*}}^{i \omega_\text{log} -2} + 2 \round{-i \tilde{\omega}_\mathrm{BAO}} \round{\frac{k}{k_*}}^{i \omega_\text{log} -1} \noright \noright \notag \\
                & \qquad\,\, \noleft \noleft + \round{-i \tilde{\omega}_\mathrm{BAO} }^2 \round{\frac{k}{k_*}}^{i \omega_\text{log}} \right] e^{-i \tilde{\omega}_\mathrm{BAO} k} +
                \begin{pmatrix}
                \omega_\text{log} \rightarrow -\omega_\text{log} \\
                \tilde{\omega}_\mathrm{BAO} \rightarrow - \tilde{\omega}_\mathrm{BAO} 
                \end{pmatrix} \right\}
                \,.
            \end{align}
            Now, assuming this to be true for any number of derivatives, we have
            \begin{align}
                & \partial_{k_{i_1}} \cdots \partial_{k_{i_{n}}} \left[ \cos \round{\omega_\text{log} \ln \frac{k}{k_*} - \tilde{\omega}_\mathrm{BAO} k } \right] = \notag \\
                & \quad \, = \frac{1}{2} \frac{k_{i_1} \cdots k_{i_n}}{k^n} \left\{ \sum_{m=0}^{n} \binom{n}{m} \frac{1}{k_*^{n-m}} \frac{\round{i \omega_\text{log}}! }{\round{i \omega_\text{log} -n +m }!}  \round{\frac{k}{k_*}}^{i \omega_\text{log} - n + m} \round{-i \tilde{\omega}_\mathrm{BAO}}^{m} \right\} e^{-i \tilde{\omega}_\mathrm{BAO} k} \notag \\
                & \qquad \,\, +
                \begin{pmatrix}
                \omega_\text{log} \rightarrow -\omega_\text{log} \\
                \tilde{\omega}_\mathrm{BAO} \rightarrow - \tilde{\omega}_\mathrm{BAO}
                \end{pmatrix} \,. \label{eq:inductive_n}
            \end{align}
            To check the proof, we add another derivative to the previous expression
            \begin{align}
                & \partial_{k_j} \partial_{k_{i_1}} \cdots \partial_{k_{i_{n}}} \left[ \cos \round{\omega_\text{log} \ln \frac{k}{k_*} - \tilde{\omega}_\mathrm{BAO} k } \right] = \notag \\
                & \quad \, =  \frac{1}{2} \frac{k_{i_1} \cdots k_{i_n} k_j}{k^{n+1}} \left\{ \sum_{m=0}^n \binom{n}{m} \frac{1}{k_*^{n-m+1}} \frac{\round{i \omega_\text{log}}!}{\round{i \omega_\text{log} -n-1+m}!} \round{\frac{k}{k_*}}^{i \omega_\text{log} - n-1 + m} \round{-i \tilde{\omega}_\mathrm{BAO}}^{m} \noright \notag \\
                & \qquad\quad \,\,\, + \noleft \sum_{m=0}^n \binom{n}{m} \frac{1}{k_*^{n-m}} \frac{\round{i \omega_\text{log}}!}{\round{i \omega_\text{log} -n+m}!} \round{\frac{k}{k_*}}^{i \omega_\text{log} - n + m} \round{-i \tilde{\omega}_\mathrm{BAO}}^{m+1} \right\} e^{-i \tilde{\omega}_\mathrm{BAO} k} \notag \\
                & \qquad \,\, +
                \begin{pmatrix}
                \omega_\text{log} \rightarrow -\omega_\text{log} \\
                \tilde{\omega}_\mathrm{BAO} \rightarrow - \tilde{\omega}_\mathrm{BAO}
                \end{pmatrix} \,.
            \end{align}
            For the first line, we have 
            \begin{align}
                \circled{1} &= \frac{1}{2} \frac{k_{i_1} \cdots k_{i_n} k_j}{k^{n+1}} \left\{ \binom{n}{0} \frac{1}{k_*^{n +1}} \frac{\round{i \omega_\text{log}}!}{\round{i \omega_\text{log} -n-1 }!} \round{\frac{k}{k_*}}^{i \omega_\text{log} - n-1 } \noright \notag \\ 
                & \quad \, \noleft + \sum_{m=1}^n \binom{n}{m} \frac{1}{k_*^{n-m+1}} \frac{\round{i \omega_\text{log}}!}{\round{i \omega_\text{log} -n-1+m}!} \round{\frac{k}{k_*}}^{i \omega_\text{log} - n-1 + m} \round{-i \tilde{\omega}_\mathrm{BAO}}^{m} \right\} e^{-i \tilde{\omega}_\text{BAO} k} \notag \\
                &= \frac{1}{2} \frac{k_{i_1} \cdots k_{i_n} k_j}{k^{n+1}} \left\{ \binom{n}{0} \frac{1}{k_*^{n +1}} \frac{\round{i \omega_\text{log}}!}{\round{i \omega_\text{log} -n-1 }!} \round{\frac{k}{k_*}}^{i \omega_\text{log} - n-1 } \noright \notag \\ 
                & \quad \, \noleft + \sum_{m=0}^n \binom{n-1}{m} \frac{1}{k_*^{n-m}} \frac{\round{i \omega_\text{log}}!}{\round{i \omega_\text{log} -n}!} \round{\frac{k}{k_*}}^{i \omega_\text{log} - n} \round{-i \tilde{\omega}_\mathrm{BAO}}^{m+1} \right\} e^{-i \tilde{\omega}_\mathrm{BAO} k} \,. \notag \\ 
            \end{align}
            For the second line, we have
            \begin{align}
                \circled{2} &= \frac{1}{2} \frac{k_{i_1} \cdots k_{i_n} k_j}{k^{n+1}} \left\{ \sum_{m=0}^{n-1} \binom{n}{m} \frac{1}{k_*^{n-m}} \frac{\round{i \omega_\text{log}}!}{\round{i \omega_\text{log} -n+m}!} \round{\frac{k}{k_*}}^{i \omega_\text{log} - n + m} \round{-i \tilde{\omega}_\mathrm{BAO}}^{m+1} \noright \notag \\
                & \quad \, \noleft + \binom{n}{n} \round{\frac{k}{k_*}}^{i \omega_\text{log}} \round{-i \tilde{\omega}_\mathrm{BAO}}^{n+1} \right\} e^{-i \tilde{\omega}_\mathrm{BAO} k} \,.
            \end{align}
            Adding the two terms we obtain\footnote{We recall that 
            \begin{equation}
                \binom{n+1}{m+1} = \binom{n}{m+1} + \binom{n}{m} \,.
            \end{equation}} 
            \begin{align}
                \circled{1} + \circled{2} &= \frac{1}{2} \frac{k_{i_1} \cdots k_{i_n} k_j}{k^{n+1}} \left\{ \binom{n}{0} \frac{1}{k_*^{n+1}} \frac{\round{i \omega_\text{log}}!}{\round{i \omega_\text{log} -n-1}!} \round{\frac{k}{k_*}}^{i \omega_\text{log} -n-1 } \noright \notag \\
                & \quad \, + \sum_{m=0}^{n-1} \binom{n+1}{m+1} \frac{1}{k_*^{n-m}} \frac{\round{i \omega_\text{log}}!}{\round{i \omega_\text{log} -n+m}!} \round{\frac{k}{k_*}}^{i \omega_\text{log} -n+m } \round{-i \tilde{\omega}_\mathrm{BAO}}^{m+1} \notag \\
                & \noleft \quad \, + \binom{n}{n} \round{\frac{k}{k_*}}^{i \omega_\text{log}} \round{-i \tilde{\omega}_\mathrm{BAO}}^{n+1} \right\} e^{-i \omega_text{BAO}} \notag \\
                &= \frac{1}{2} \frac{k_{i_1} \cdots k_{i_n} k_j}{k^{n+1}} \left\{ \binom{n}{0} \frac{1}{k_*^{n+1}} \frac{\round{i \omega_\text{log}}!}{\round{i \omega_\text{log} -n-1}!} \round{\frac{k}{k_*}}^{i \omega_\text{log} -n-1 } \noright \notag \\
                & \quad \, + \sum_{m=1}^{n} \binom{n+1}{m} \frac{1}{k_*^{n+1-m}} \frac{\round{i \omega_\text{log}}!}{\round{i \omega_\text{log} -n-1+m}!} \round{\frac{k}{k_*}}^{i \omega_\text{log} -n-1+m } \round{-i \tilde{\omega}_\mathrm{BAO}}^{m} \notag \\
                & \noleft \quad \, + \binom{n}{n} \round{\frac{k}{k_*}}^{i \omega_\text{log}} \round{-i \tilde{\omega}_\mathrm{BAO}}^{n+1} \right\} e^{-i \tilde{\omega}_\mathrm{BAO} k} \,.
            \end{align}
            Now, since 
            \begin{equation}
                \binom{n}{0} = \binom{n+1}{0} = 1 \; , \qquad \binom{n}{n} = \binom{n+1}{n+1} = 1 \,,
            \end{equation}
            we have that 
            \begin{align}
                \circled{1} + \circled{2} &= \frac{1}{2} \frac{k_{i_1} \cdots k_{i_n} k_j}{k^{n+1}} \left\{ \binom{n+1}{0} \frac{1}{k_*^{n+1}} \frac{\round{i \omega_\text{log}}!}{\round{i \omega_\text{log} -n-1}!} \round{\frac{k}{k_*}}^{i \omega_\text{log} -n-1 } \noright \notag \\
                & \quad \, + \sum_{m=1}^{n} \binom{n+1}{m} \frac{1}{k_*^{n+1-m}} \frac{\round{i \omega_\text{log}}!}{\round{i \omega_\text{log} -n-1+m}!} \round{\frac{k}{k_*}}^{i \omega_\text{log} -n-1+m } \round{-i \tilde{\omega}_\mathrm{BAO}}^{m} \notag \\
                & \quad \, + \noleft \binom{n+1}{n+1} \round{\frac{k}{k_*}}^{i \omega_\text{log}} \round{-i \tilde{\omega}_\mathrm{BAO}}^{n+1} \right\} e^{-i \tilde{\omega}_\mathrm{BAO} k} \notag \\
                &= \frac{1}{2} \frac{k_{i_1} \cdots k_{i_n} k_j}{k^{n+1}} e^{-i \tilde{\omega}_\mathrm{BAO} k} \notag \\
                & \quad \, \times \left\{ \sum_{m=0}^{n+1} \binom{n+1}{m} \frac{1}{k_*^{n+1-m}} \frac{\round{i \omega_\text{log}}!}{\round{i \omega_\text{log} -n-1+m}!} \round{\frac{k}{k_*}}^{i \omega_\text{log} -n-1+m } \round{-i \tilde{\omega}_\text{BAO}}^{m}  \right\}  \,,
            \end{align}
            which is exactly~\cref{eq:inductive_n} at the $n+1$ iteration, fact that completes our inductive proof. 
            
            As we have done before, now we can rewrite the action of the operator appearing in~\cref{eq:operator_action} now as\footnote{Here the two additional terms in the third line are due to the presence of the second cosine term in the wiggle mixed power spectrum, neglected during the computation of this appendix as mentioned in the previous footnote.}
            \begin{align}
                &\sround{1 - \cosh \round{\mathbf{q} \cdot \nabla_{\mathbf{k}}} } P_\text{BAO}^\mathrm{w} \left(k\right) \delta P^\text{log}_\zeta \left(k\right) = \notag \\
                & \quad\, = \sround{ 1 - \sum_{n=0}^{\infty} \frac{1}{\round{2n}!} \round{\mathbf{q} \cdot \nabla_{\mathbf{k}} }^{2n} } P_\text{BAO}^\mathrm{w} \left(k\right) \delta P^\text{log}_\zeta \left(k\right) \notag \\ 
                & \quad\, = P_L^\mathrm{nw}  A_\mathrm{BAO} \round{k} \left(k\right) A_\text{log} \left\{ \sin \round{\tilde{\omega}_\mathrm{BAO} k} \sin \round{ \omega_\mathrm{log} \ln \frac{k}{k_*} } \noright \notag \\
                & \qquad\,\, -\frac{e^{i \tilde{\omega}_\mathrm{BAO} k}}{2} \sum_{n=0}^{\infty} \frac{1}{\round{2n}!} \frac{(i \omega_\mathrm{log})!}{(-2n + i\omega_\mathrm{log})!} \round{\frac{q \mu}{k_*}}^{2n} {}_1F_1(-2n, 1 - 2n + i\omega_\mathrm{log}, i k \tilde{\omega}_\mathrm{BAO}) \left( \frac{k}{k_*} \right)^{-2n + i\omega_\mathrm{log}} \notag \\ 
                & \qquad \,\, + \noleft 
                \begin{pmatrix}
                \omega_\text{log} \rightarrow -\omega_\text{log} \\
                \tilde{\omega}_\mathrm{BAO} \rightarrow - \tilde{\omega}_\mathrm{BAO}
                \end{pmatrix} - \begin{pmatrix}
                \omega_\text{log} \rightarrow \omega_\text{log} \\
                \tilde{\omega}_\mathrm{BAO} \rightarrow - \tilde{\omega}_\mathrm{BAO}
                \end{pmatrix}
                - \begin{pmatrix}
                \omega_\text{log} \rightarrow -\omega_\text{log} \\
                \tilde{\omega}_\mathrm{BAO} \rightarrow \tilde{\omega}_\mathrm{BAO}
                \end{pmatrix}
                \right\} \notag \\
            \end{align}
            where ${}_pF_q$ is the hypergeometric function, defined as 
            \begin{equation}
                {}_pF_q(a_1, a_2, \ldots, a_p; b_1, b_2, \ldots, b_q; z) = \sum_{n=0}^{\infty} \frac{(a_1)_n (a_2)_n \cdots (a_p)_n}{(b_1)_n (b_2)_n \cdots (b_q)_n} \frac{z^n}{n!} \,,
            \end{equation}  
            with $\round{a}_n$ representing the Pochhammer symbol 
            \begin{equation}
                \round{a}_n = a \round{a + 1} \round{a + 2} \cdots \round{a + n - 1} = \frac{\Gamma \round{a+n}}{\Gamma \round{a}} \,,
            \end{equation}
            with an analogous expression for $\round{b}_n$. Unfortunately this expression does not admit any closed representation and thus we are not able to give a single formula for the damping in this case which, to be computed require the use of specific numeric methods. However, given the small magnitude of this correction observed in the strong approximation, we do not pursue a precise evaluation in this weak approximation, which is beyond the purposes of this work.

\bibliographystyle{JHEP}
\bibliography{bibliography}

\providecommand{\href}[2]{#2}\begingroup\raggedright\begin{thebibliography}{100}

\bibitem{Starobinsky:1980te}
A.A.~{Starobinsky}, \emph{{A new type of isotropic cosmological models without singularity}}, \href{https://doi.org/10.1016/0370-2693(80)90670-X}{\emph{Physics Letters B} {\bfseries 91} (1980) 99}.

\bibitem{Guth:1980zm}
A.H.~{Guth}, \emph{{Inflationary universe: A possible solution to the horizon and flatness problems}}, \href{https://doi.org/10.1103/PhysRevD.23.347}{\emph{\prd} {\bfseries 23} (1981) 347}.

\bibitem{Linde:1981mu}
A.D.~{Linde}, \emph{{A new inflationary universe scenario: A possible solution of the horizon, flatness, homogeneity, isotropy and primordial monopole problems}}, \href{https://doi.org/10.1016/0370-2693(82)91219-9}{\emph{Physics Letters B} {\bfseries 108} (1982) 389}.

\bibitem{Linde:1983gd}
A.D.~{Linde}, \emph{{Chaotic inflation}}, \href{https://doi.org/10.1016/0370-2693(83)90837-7}{\emph{Physics Letters B} {\bfseries 129} (1983) 177}.

\bibitem{Mukhanov:1981xt}
V.F.~{Mukhanov} and G.V.~{Chibisov}, \emph{{Quantum fluctuations and a nonsingular universe}}, {\emph{Soviet Journal of Experimental and Theoretical Physics Letters} {\bfseries 33} (1981) 532}.

\bibitem{Starobinsky:1992ts}
A.A.~{Starobinsky}, \emph{{Spectrum of adiabatic perturbations in the universe when there are singularities in the inflationary potential.}}, {\emph{Soviet Journal of Experimental and Theoretical Physics Letters} {\bfseries 55} (1992) 489}.

\bibitem{Adams:2001vc}
J.~{Adams}, B.~{Cresswell} and R.~{Easther}, \emph{{Inflationary perturbations from a potential with a step}}, \href{https://doi.org/10.1103/PhysRevD.64.123514}{\emph{\prd} {\bfseries 64} (2001) 123514} [\href{https://arxiv.org/abs/astro-ph/0102236}{{\ttfamily astro-ph/0102236}}].

\bibitem{Chen:2006xjb}
X.~{Chen}, R.~{Easther} and E.A.~{Lim}, \emph{{Large non-Gaussianities in single-field inflation}}, \href{https://doi.org/10.1088/1475-7516/2007/06/023}{\emph{\jcap} {\bfseries 2007} (2007) 023} [\href{https://arxiv.org/abs/astro-ph/0611645}{{\ttfamily astro-ph/0611645}}].

\bibitem{Achucarro:2010da}
A.~{Ach{\'u}carro}, J.-O.~{Gong}, S.~{Hardeman}, G.A.~{Palma} and S.P.~{Patil}, \emph{{Features of heavy physics in the CMB power spectrum}}, \href{https://doi.org/10.1088/1475-7516/2011/01/030}{\emph{\jcap} {\bfseries 2011} (2011) 030} [\href{https://arxiv.org/abs/1010.3693}{{\ttfamily 1010.3693}}].

\bibitem{Chen:2011zf}
X.~{Chen}, \emph{{Primordial Features as Evidence for Inflation}}, \href{https://doi.org/10.1088/1475-7516/2012/01/038}{\emph{\jcap} {\bfseries 2012} (2012) 038} [\href{https://arxiv.org/abs/1104.1323}{{\ttfamily 1104.1323}}].

\bibitem{Braglia:2020fms}
M.~{Braglia}, D.K.~{Hazra}, L.~{Sriramkumar} and F.~{Finelli}, \emph{{Generating primordial features at large scales in two field models of inflation}}, \href{https://doi.org/10.1088/1475-7516/2020/08/025}{\emph{\jcap} {\bfseries 2020} (2020) 025} [\href{https://arxiv.org/abs/2004.00672}{{\ttfamily 2004.00672}}].

\bibitem{Chen:2008wn}
X.~{Chen}, R.~{Easther} and E.A.~{Lim}, \emph{{Generation and characterization of large non-Gaussianities in single field inflation}}, \href{https://doi.org/10.1088/1475-7516/2008/04/010}{\emph{\jcap} {\bfseries 2008} (2008) 010} [\href{https://arxiv.org/abs/0801.3295}{{\ttfamily 0801.3295}}].

\bibitem{Flauger:2009ab}
R.~{Flauger}, L.~{McAllister}, E.~{Pajer}, A.~{Westphal} and G.~{Xu}, \emph{{Oscillations in the CMB from axion monodromy inflation}}, \href{https://doi.org/10.1088/1475-7516/2010/06/009}{\emph{\jcap} {\bfseries 2010} (2010) 009} [\href{https://arxiv.org/abs/0907.2916}{{\ttfamily 0907.2916}}].

\bibitem{Flauger:2010ja}
R.~{Flauger} and E.~{Pajer}, \emph{{Resonant non-gaussianity}}, \href{https://doi.org/10.1088/1475-7516/2011/01/017}{\emph{\jcap} {\bfseries 2011} (2011) 017} [\href{https://arxiv.org/abs/1002.0833}{{\ttfamily 1002.0833}}].

\bibitem{Chen:2010bka}
X.~{Chen}, \emph{{Folded resonant non-Gaussianity in general single field inflation}}, \href{https://doi.org/10.1088/1475-7516/2010/12/003}{\emph{\jcap} {\bfseries 2010} (2010) 003} [\href{https://arxiv.org/abs/1008.2485}{{\ttfamily 1008.2485}}].

\bibitem{Martin:2000xs}
J.~{Martin} and R.H.~{Brandenberger}, \emph{{Trans-Planckian problem of inflationary cosmology}}, \href{https://doi.org/10.1103/PhysRevD.63.123501}{\emph{\prd} {\bfseries 63} (2001) 123501} [\href{https://arxiv.org/abs/hep-th/0005209}{{\ttfamily hep-th/0005209}}].

\bibitem{Easther:2002xe}
R.~{Easther}, B.R.~{Greene}, W.H.~{Kinney} and G.~{Shiu}, \emph{{Generic estimate of trans-Planckian modifications to the primordial power spectrum in inflation}}, \href{https://doi.org/10.1103/PhysRevD.66.023518}{\emph{\prd} {\bfseries 66} (2002) 023518} [\href{https://arxiv.org/abs/hep-th/0204129}{{\ttfamily hep-th/0204129}}].

\bibitem{Armendariz-Picon:2003knj}
C.~{Armend{\'a}riz-Pic{\'o}n} and E.A.~{Lim}, \emph{{Vacuum choices and the predictions of inflation}}, \href{https://doi.org/10.1088/1475-7516/2003/12/006}{\emph{\jcap} {\bfseries 2003} (2003) 006} [\href{https://arxiv.org/abs/hep-th/0303103}{{\ttfamily hep-th/0303103}}].

\bibitem{Chen:2014joa}
X.~{Chen} and M.H.~{Namjoo}, \emph{{Standard Clock in primordial density perturbations and cosmic microwave background}}, \href{https://doi.org/10.1016/j.physletb.2014.11.002}{\emph{Physics Letters B} {\bfseries 739} (2014) 285} [\href{https://arxiv.org/abs/1404.1536}{{\ttfamily 1404.1536}}].

\bibitem{Chen:2018cgg}
X.~{Chen}, A.~{Loeb} and Z.-Z.~{Xianyu}, \emph{{Unique Fingerprints of Alternatives to Inflation in the Primordial Power Spectrum}}, \href{https://doi.org/10.1103/PhysRevLett.122.121301}{\emph{\prl} {\bfseries 122} (2019) 121301} [\href{https://arxiv.org/abs/1809.02603}{{\ttfamily 1809.02603}}].

\bibitem{Chluba:2015bqa}
J.~{Chluba}, J.~{Hamann} and S.P.~{Patil}, \emph{{Features and new physical scales in primordial observables: Theory and observation}}, \href{https://doi.org/10.1142/S0218271815300232}{\emph{International Journal of Modern Physics D} {\bfseries 24} (2015) 1530023} [\href{https://arxiv.org/abs/1505.01834}{{\ttfamily 1505.01834}}].

\bibitem{Slosar:2019gvt}
A.~{Slosar}, X.~{Chen}, C.~{Dvorkin}, D.~{Meerburg}, B.~{Wallisch}, D.~{Green} et~al., \emph{{Scratches from the Past: Inflationary Archaeology through Features in the Power Spectrum of Primordial Fluctuations}}, \href{https://doi.org/10.48550/arXiv.1903.09883}{\emph{\baas} {\bfseries 51} (2019) 98} [\href{https://arxiv.org/abs/1903.09883}{{\ttfamily 1903.09883}}].

\bibitem{2022arXiv220308128A}
A.~{Ach{\'u}carro}, M.~{Biagetti}, M.~{Braglia}, G.~{Cabass}, R.~{Caldwell}, E.~{Castorina} et~al., \emph{{Inflation: Theory and Observations}}, \href{https://doi.org/10.48550/arXiv.2203.08128}{\emph{arXiv e-prints} (2022) arXiv:2203.08128} [\href{https://arxiv.org/abs/2203.08128}{{\ttfamily 2203.08128}}].

\bibitem{Akrami:2018odb}
{Planck Collaboration: Y. Akrami}, F.~{Arroja}, M.~{Ashdown}, J.~{Aumont}, C.~{Baccigalupi}, M.~{Ballardini} et~al., \emph{{Planck 2018 results. X. Constraints on inflation}}, \href{https://doi.org/10.1051/0004-6361/201833887}{\emph{\aap} {\bfseries 641} (2020) A10} [\href{https://arxiv.org/abs/1807.06211}{{\ttfamily 1807.06211}}].

\bibitem{Handley:2019fll}
W.J.~{Handley}, A.N.~{Lasenby}, H.V.~{Peiris} and M.P.~{Hobson}, \emph{{Bayesian inflationary reconstructions from Planck 2018 data}}, \href{https://doi.org/10.1103/PhysRevD.100.103511}{\emph{\prd} {\bfseries 100} (2019) 103511} [\href{https://arxiv.org/abs/1908.00906}{{\ttfamily 1908.00906}}].

\bibitem{Beutler:2019ojk}
F.~{Beutler}, M.~{Biagetti}, D.~{Green}, A.~{Slosar} and B.~{Wallisch}, \emph{{Primordial features from linear to nonlinear scales}}, \href{https://doi.org/10.1103/PhysRevResearch.1.033209}{\emph{Physical Review Research} {\bfseries 1} (2019) 033209} [\href{https://arxiv.org/abs/1906.08758}{{\ttfamily 1906.08758}}].

\bibitem{Ballardini:2022wzu}
M.~{Ballardini}, F.~{Finelli}, F.~{Marulli}, L.~{Moscardini} and A.~{Veropalumbo}, \emph{{New constraints on primordial features from the galaxy two-point correlation function}}, \href{https://doi.org/10.1103/PhysRevD.107.043532}{\emph{\prd} {\bfseries 107} (2023) 043532} [\href{https://arxiv.org/abs/2202.08819}{{\ttfamily 2202.08819}}].

\bibitem{Mergulhao:2023ukp}
T.~{Mergulh{\~a}o}, F.~{Beutler} and J.A.~{Peacock}, \emph{{Primordial feature constraints from BOSS + eBOSS}}, \href{https://doi.org/10.1088/1475-7516/2023/08/012}{\emph{\jcap} {\bfseries 2023} (2023) 012} [\href{https://arxiv.org/abs/2303.13946}{{\ttfamily 2303.13946}}].

\bibitem{Wang:2000js}
Y.~{Wang} and G.J.~{Mathews}, \emph{{Model-independent Primordial Power Spectrum from MAXIMA, BOOMERANG, and DASI Data}}, \href{https://doi.org/10.1086/340492}{\emph{\apj} {\bfseries 573} (2002) 1} [\href{https://arxiv.org/abs/astro-ph/0011351}{{\ttfamily astro-ph/0011351}}].

\bibitem{Peiris:2003ff}
H.V.~{Peiris}, E.~{Komatsu}, L.~{Verde}, D.N.~{Spergel}, C.L.~{Bennett}, M.~{Halpern} et~al., \emph{{First-Year Wilkinson Microwave Anisotropy Probe (WMAP) Observations: Implications For Inflation}}, \href{https://doi.org/10.1086/377228}{\emph{\apjs} {\bfseries 148} (2003) 213} [\href{https://arxiv.org/abs/astro-ph/0302225}{{\ttfamily astro-ph/0302225}}].

\bibitem{Mukherjee:2003ag}
P.~{Mukherjee} and Y.~{Wang}, \emph{{Model-independent Reconstruction of the Primordial Power Spectrum from Wilkinson Microwave Anistropy Probe Data}}, \href{https://doi.org/10.1086/379161}{\emph{\apj} {\bfseries 599} (2003) 1} [\href{https://arxiv.org/abs/astro-ph/0303211}{{\ttfamily astro-ph/0303211}}].

\bibitem{Covi:2006ci}
L.~{Covi}, J.~{Hamann}, A.~{Melchiorri}, A.~{Slosar} and I.~{Sorbera}, \emph{{Inflation and WMAP three year data: Features are still present}}, \href{https://doi.org/10.1103/PhysRevD.74.083509}{\emph{\prd} {\bfseries 74} (2006) 083509} [\href{https://arxiv.org/abs/astro-ph/0606452}{{\ttfamily astro-ph/0606452}}].

\bibitem{Hamann:2007pa}
J.~{Hamann}, L.~{Covi}, A.~{Melchiorri} and A.~{Slosar}, \emph{{New constraints on oscillations in the primordial spectrum of inflationary perturbations}}, \href{https://doi.org/10.1103/PhysRevD.76.023503}{\emph{\prd} {\bfseries 76} (2007) 023503} [\href{https://arxiv.org/abs/astro-ph/0701380}{{\ttfamily astro-ph/0701380}}].

\bibitem{Meerburg:2011gd}
P.D.~{Meerburg}, R.A.M.J.~{Wijers} and J.P.~{van der Schaar}, \emph{{WMAP7 constraints on oscillations in the primordial power spectrum}}, \href{https://doi.org/10.1111/j.1365-2966.2011.20311.x}{\emph{\mnras} {\bfseries 421} (2012) 369} [\href{https://arxiv.org/abs/1109.5264}{{\ttfamily 1109.5264}}].

\bibitem{Planck:2013jfk}
{Planck Collaboration: P.~A.~R. Ade}, N.~{Aghanim}, C.~{Armitage-Caplan}, M.~{Arnaud}, M.~{Ashdown}, F.~{Atrio-Barandela} et~al., \emph{{Planck 2013 results. XXII. Constraints on inflation}}, \href{https://doi.org/10.1051/0004-6361/201321569}{\emph{\aap} {\bfseries 571} (2014) A22} [\href{https://arxiv.org/abs/1303.5082}{{\ttfamily 1303.5082}}].

\bibitem{Meerburg:2013dla}
P.D.~{Meerburg}, D.N.~{Spergel} and B.D.~{Wandelt}, \emph{{Searching for oscillations in the primordial power spectrum. II. Constraints from Planck data}}, \href{https://doi.org/10.1103/PhysRevD.89.063537}{\emph{\prd} {\bfseries 89} (2014) 063537} [\href{https://arxiv.org/abs/1308.3705}{{\ttfamily 1308.3705}}].

\bibitem{Benetti:2013cja}
M.~{Benetti}, \emph{{Updating constraints on inflationary features in the primordial power spectrum with the Planck data}}, \href{https://doi.org/10.1103/PhysRevD.88.087302}{\emph{\prd} {\bfseries 88} (2013) 087302} [\href{https://arxiv.org/abs/1308.6406}{{\ttfamily 1308.6406}}].

\bibitem{Miranda:2013wxa}
V.~{Miranda} and W.~{Hu}, \emph{{Inflationary steps in the Planck data}}, \href{https://doi.org/10.1103/PhysRevD.89.083529}{\emph{\prd} {\bfseries 89} (2014) 083529} [\href{https://arxiv.org/abs/1312.0946}{{\ttfamily 1312.0946}}].

\bibitem{Easther:2013kla}
R.~{Easther} and R.~{Flauger}, \emph{{Planck constraints on monodromy inflation}}, \href{https://doi.org/10.1088/1475-7516/2014/02/037}{\emph{\jcap} {\bfseries 2014} (2014) 037} [\href{https://arxiv.org/abs/1308.3736}{{\ttfamily 1308.3736}}].

\bibitem{Achucarro:2014msa}
A.~{Ach{\'u}carro}, V.~{Atal}, B.~{Hu}, P.~{Ortiz} and J.~{Torrado}, \emph{{Inflation with moderately sharp features in the speed of sound: Generalized slow roll and in-in formalism for power spectrum and bispectrum}}, \href{https://doi.org/10.1103/PhysRevD.90.023511}{\emph{\prd} {\bfseries 90} (2014) 023511} [\href{https://arxiv.org/abs/1404.7522}{{\ttfamily 1404.7522}}].

\bibitem{Hazra:2014goa}
D.K.~{Hazra}, A.~{Shafieloo}, G.F.~{Smoot} and A.A.~{Starobinsky}, \emph{{Wiggly whipped inflation}}, \href{https://doi.org/10.1088/1475-7516/2014/08/048}{\emph{\jcap} {\bfseries 2014} (2014) 048} [\href{https://arxiv.org/abs/1405.2012}{{\ttfamily 1405.2012}}].

\bibitem{Hazra:2014jwa}
D.K.~{Hazra}, A.~{Shafieloo} and T.~{Souradeep}, \emph{{Primordial power spectrum from Planck}}, \href{https://doi.org/10.1088/1475-7516/2014/11/011}{\emph{\jcap} {\bfseries 2014} (2014) 011} [\href{https://arxiv.org/abs/1406.4827}{{\ttfamily 1406.4827}}].

\bibitem{Hu:2014hra}
B.~{Hu} and J.~{Torrado}, \emph{{Searching for primordial localized features with CMB and LSS spectra}}, \href{https://doi.org/10.1103/PhysRevD.91.064039}{\emph{\prd} {\bfseries 91} (2015) 064039} [\href{https://arxiv.org/abs/1410.4804}{{\ttfamily 1410.4804}}].

\bibitem{Ade:2015lrj}
{Planck Collaboration: P.~A.~R. Ade}, N.~{Aghanim}, M.~{Arnaud}, F.~{Arroja}, M.~{Ashdown}, J.~{Aumont} et~al., \emph{{Planck 2015 results. XX. Constraints on inflation}}, \href{https://doi.org/10.1051/0004-6361/201525898}{\emph{\aap} {\bfseries 594} (2016) A20} [\href{https://arxiv.org/abs/1502.02114}{{\ttfamily 1502.02114}}].

\bibitem{Gruppuso:2015zia}
A.~{Gruppuso} and A.~{Sagnotti}, \emph{{Observational hints of a pre-inflationary scale?}}, \href{https://doi.org/10.1142/S0218271815440083}{\emph{International Journal of Modern Physics D} {\bfseries 24} (2015) 1544008} [\href{https://arxiv.org/abs/1506.08093}{{\ttfamily 1506.08093}}].

\bibitem{Gruppuso:2015xqa}
A.~{Gruppuso}, N.~{Kitazawa}, N.~{Mandolesi}, P.~{Natoli} and A.~{Sagnotti}, \emph{{Pre-inflationary relics in the CMB?}}, \href{https://doi.org/10.1016/j.dark.2015.12.001}{\emph{Physics of the Dark Universe} {\bfseries 11} (2016) 68} [\href{https://arxiv.org/abs/1508.00411}{{\ttfamily 1508.00411}}].

\bibitem{Hazra:2016fkm}
D.K.~{Hazra}, A.~{Shafieloo}, G.F.~{Smoot} and A.A.~{Starobinsky}, \emph{{Primordial features and Planck polarization}}, \href{https://doi.org/10.1088/1475-7516/2016/09/009}{\emph{\jcap} {\bfseries 2016} (2016) 009} [\href{https://arxiv.org/abs/1605.02106}{{\ttfamily 1605.02106}}].

\bibitem{Torrado:2016sls}
J.~{Torrado}, B.~{Hu} and A.~{Ach{\'u}carro}, \emph{{Robust predictions for an oscillatory bispectrum in Planck 2015 data from transient reductions in the speed of sound of the inflaton}}, \href{https://doi.org/10.1103/PhysRevD.96.083515}{\emph{\prd} {\bfseries 96} (2017) 083515} [\href{https://arxiv.org/abs/1611.10350}{{\ttfamily 1611.10350}}].

\bibitem{Zeng:2018ufm}
C.~{Zeng}, E.D.~{Kovetz}, X.~{Chen}, Y.~{Gong}, J.B.~{Mu{\~n}oz} and M.~{Kamionkowski}, \emph{{Searching for oscillations in the primordial power spectrum with CMB and LSS data}}, \href{https://doi.org/10.1103/PhysRevD.99.043517}{\emph{\prd} {\bfseries 99} (2019) 043517} [\href{https://arxiv.org/abs/1812.05105}{{\ttfamily 1812.05105}}].

\bibitem{Canas-Herrera:2020mme}
G.~{Ca{\~n}as-Herrera}, J.~{Torrado} and A.~{Ach{\'u}carro}, \emph{{Bayesian reconstruction of the inflaton's speed of sound using CMB data}}, \href{https://doi.org/10.1103/PhysRevD.103.123531}{\emph{\prd} {\bfseries 103} (2021) 123531} [\href{https://arxiv.org/abs/2012.04640}{{\ttfamily 2012.04640}}].

\bibitem{Braglia:2021ckn}
M.~{Braglia}, X.~{Chen} and D.K.~{Hazra}, \emph{{Comparing multi-field primordial feature models with the Planck data}}, \href{https://doi.org/10.1088/1475-7516/2021/06/005}{\emph{\jcap} {\bfseries 2021} (2021) 005} [\href{https://arxiv.org/abs/2103.03025}{{\ttfamily 2103.03025}}].

\bibitem{Braglia:2021sun}
M.~{Braglia}, X.~{Chen} and D.K.~{Hazra}, \emph{{Uncovering the history of cosmic inflation from anomalies in cosmic microwave background spectra}}, \href{https://doi.org/10.1140/epjc/s10052-022-10461-3}{\emph{European Physical Journal C} {\bfseries 82} (2022) 498} [\href{https://arxiv.org/abs/2106.07546}{{\ttfamily 2106.07546}}].

\bibitem{Braglia:2021rej}
M.~{Braglia}, X.~{Chen} and D.K.~{Hazra}, \emph{{Primordial standard clock models and CMB residual anomalies}}, \href{https://doi.org/10.1103/PhysRevD.105.103523}{\emph{\prd} {\bfseries 105} (2022) 103523} [\href{https://arxiv.org/abs/2108.10110}{{\ttfamily 2108.10110}}].

\bibitem{Naik:2022mxn}
S.S.~{Naik}, K.~{Furuuchi} and P.~{Chingangbam}, \emph{{Particle production during inflation: a Bayesian analysis with CMB data from Planck 2018}}, \href{https://doi.org/10.1088/1475-7516/2022/07/016}{\emph{\jcap} {\bfseries 2022} (2022) 016} [\href{https://arxiv.org/abs/2202.05862}{{\ttfamily 2202.05862}}].

\bibitem{Hamann:2021eyw}
J.~{Hamann} and J.~{Wons}, \emph{{Optimising inflationary features the Bayesian way}}, \href{https://doi.org/10.1088/1475-7516/2022/03/036}{\emph{\jcap} {\bfseries 2022} (2022) 036} [\href{https://arxiv.org/abs/2112.08571}{{\ttfamily 2112.08571}}].

\bibitem{Antony:2022ert}
A.~{Antony}, F.~{Finelli}, D.K.~{Hazra} and A.~{Shafieloo}, \emph{{Discordances in Cosmology and the Violation of Slow-Roll Inflationary Dynamics}}, \href{https://doi.org/10.1103/PhysRevLett.130.111001}{\emph{\prl} {\bfseries 130} (2023) 111001} [\href{https://arxiv.org/abs/2202.14028}{{\ttfamily 2202.14028}}].

\bibitem{Antony:2024vrx}
A.~{Antony}, F.~{Finelli}, D.K.~{Hazra}, D.~{Paoletti} and A.~{Shafieloo}, \emph{{A search for super-imposed oscillations to the primordial power spectrum in Planck and SPT-3G 2018 data}}, \href{https://doi.org/10.48550/arXiv.2403.19575}{\emph{arXiv e-prints} (2024) arXiv:2403.19575} [\href{https://arxiv.org/abs/2403.19575}{{\ttfamily 2403.19575}}].

\bibitem{Planck:2013wtn}
{Planck Collaboration: P.~A.~R. Ade}, N.~{Aghanim}, C.~{Armitage-Caplan}, M.~{Arnaud}, M.~{Ashdown}, F.~{Atrio-Barandela} et~al., \emph{{Planck 2013 results. XXIV. Constraints on primordial non-Gaussianity}}, \href{https://doi.org/10.1051/0004-6361/201321554}{\emph{\aap} {\bfseries 571} (2014) A24} [\href{https://arxiv.org/abs/1303.5084}{{\ttfamily 1303.5084}}].

\bibitem{Fergusson:2014tza}
J.R.~{Fergusson}, H.F.~{Gruetjen}, E.P.S.~{Shellard} and B.~{Wallisch}, \emph{{Polyspectra searches for sharp oscillatory features in cosmic microwave sky data}}, \href{https://doi.org/10.1103/PhysRevD.91.123506}{\emph{\prd} {\bfseries 91} (2015) 123506} [\href{https://arxiv.org/abs/1412.6152}{{\ttfamily 1412.6152}}].

\bibitem{Planck:2015zfm}
{Planck Collaboration: P.~A.~R. Ade}, N.~{Aghanim}, M.~{Arnaud}, F.~{Arroja}, M.~{Ashdown}, J.~{Aumont} et~al., \emph{{Planck 2015 results. XVII. Constraints on primordial non-Gaussianity}}, \href{https://doi.org/10.1051/0004-6361/201525836}{\emph{\aap} {\bfseries 594} (2016) A17} [\href{https://arxiv.org/abs/1502.01592}{{\ttfamily 1502.01592}}].

\bibitem{Meerburg:2015owa}
P.D.~{Meerburg}, M.~{M{\"u}nchmeyer} and B.~{Wandelt}, \emph{{Joint resonant CMB power spectrum and bispectrum estimation}}, \href{https://doi.org/10.1103/PhysRevD.93.043536}{\emph{\prd} {\bfseries 93} (2016) 043536} [\href{https://arxiv.org/abs/1510.01756}{{\ttfamily 1510.01756}}].

\bibitem{Planck:2019izv}
{Planck Collaboration: Y. Akrami}, F.~{Arroja}, M.~{Ashdown}, J.~{Aumont}, C.~{Baccigalupi}, M.~{Ballardini} et~al., \emph{{Planck 2018 results. IX. Constraints on primordial non-Gaussianity}}, \href{https://doi.org/10.1051/0004-6361/201935891}{\emph{\aap} {\bfseries 641} (2020) A9} [\href{https://arxiv.org/abs/1905.05697}{{\ttfamily 1905.05697}}].

\bibitem{Wang:1998gb}
Y.~{Wang}, D.N.~{Spergel} and M.A.~{Strauss}, \emph{{Cosmology in the Next Millennium: Combining Microwave Anisotropy Probe and Sloan Digital Sky Survey Data to Constrain Inflationary Models}}, \href{https://doi.org/10.1086/306558}{\emph{\apj} {\bfseries 510} (1999) 20} [\href{https://arxiv.org/abs/astro-ph/9802231}{{\ttfamily astro-ph/9802231}}].

\bibitem{Zhan:2005rz}
H.~{Zhan}, L.~{Knox}, J.A.~{Tyson} and V.~{Margoniner}, \emph{{Exploring Large-Scale Structure with Billions of Galaxies}}, \href{https://doi.org/10.1086/500077}{\emph{\apj} {\bfseries 640} (2006) 8} [\href{https://arxiv.org/abs/astro-ph/0508119}{{\ttfamily astro-ph/0508119}}].

\bibitem{Huang:2012mr}
Z.~{Huang}, L.~{Verde} and F.~{Vernizzi}, \emph{{Constraining inflation with future galaxy redshift surveys}}, \href{https://doi.org/10.1088/1475-7516/2012/04/005}{\emph{\jcap} {\bfseries 2012} (2012) 005} [\href{https://arxiv.org/abs/1201.5955}{{\ttfamily 1201.5955}}].

\bibitem{Chen:2016vvw}
X.~{Chen}, C.~{Dvorkin}, Z.~{Huang}, M.H.~{Namjoo} and L.~{Verde}, \emph{{The future of primordial features with large-scale structure surveys}}, \href{https://doi.org/10.1088/1475-7516/2016/11/014}{\emph{\jcap} {\bfseries 2016} (2016) 014} [\href{https://arxiv.org/abs/1605.09365}{{\ttfamily 1605.09365}}].

\bibitem{Chen:2016zuu}
X.~{Chen}, P.D.~{Meerburg} and M.~{M{\"u}nchmeyer}, \emph{{The future of primordial features with 21 cm tomography}}, \href{https://doi.org/10.1088/1475-7516/2016/09/023}{\emph{\jcap} {\bfseries 2016} (2016) 023} [\href{https://arxiv.org/abs/1605.09364}{{\ttfamily 1605.09364}}].

\bibitem{Ballardini:2016hpi}
M.~{Ballardini}, F.~{Finelli}, C.~{Fedeli} and L.~{Moscardini}, \emph{{Probing primordial features with future galaxy surveys}}, \href{https://doi.org/10.1088/1475-7516/2016/10/041}{\emph{\jcap} {\bfseries 2016} (2016) 041} [\href{https://arxiv.org/abs/1606.03747}{{\ttfamily 1606.03747}}].

\bibitem{Xu:2016kwz}
Y.~{Xu}, J.~{Hamann} and X.~{Chen}, \emph{{Precise measurements of inflationary features with 21 cm observations}}, \href{https://doi.org/10.1103/PhysRevD.94.123518}{\emph{\prd} {\bfseries 94} (2016) 123518} [\href{https://arxiv.org/abs/1607.00817}{{\ttfamily 1607.00817}}].

\bibitem{Fard:2017oex}
M.~{Ansari Fard} and S.~{Baghram}, \emph{{Late time sky as a probe of steps and oscillations in primordial Universe}}, \href{https://doi.org/10.1088/1475-7516/2018/01/051}{\emph{\jcap} {\bfseries 2018} (2018) 051} [\href{https://arxiv.org/abs/1709.05323}{{\ttfamily 1709.05323}}].

\bibitem{Palma:2017wxu}
G.A.~{Palma}, D.~{Sapone} and S.~{Sypsas}, \emph{{Constraints on inflation with LSS surveys: features in the primordial power spectrum}}, \href{https://doi.org/10.1088/1475-7516/2018/06/004}{\emph{\jcap} {\bfseries 2018} (2018) 004} [\href{https://arxiv.org/abs/1710.02570}{{\ttfamily 1710.02570}}].

\bibitem{Ballardini:2017qwq}
M.~{Ballardini}, F.~{Finelli}, R.~{Maartens} and L.~{Moscardini}, \emph{{Probing primordial features with next-generation photometric and radio surveys}}, \href{https://doi.org/10.1088/1475-7516/2018/04/044}{\emph{\jcap} {\bfseries 2018} (2018) 044} [\href{https://arxiv.org/abs/1712.07425}{{\ttfamily 1712.07425}}].

\bibitem{Ballardini:2018noo}
M.~{Ballardini}, \emph{{Probing primordial features with the primary CMB}}, \href{https://doi.org/10.1016/j.dark.2018.11.006}{\emph{Physics of the Dark Universe} {\bfseries 23} (2019) 100245} [\href{https://arxiv.org/abs/1807.05521}{{\ttfamily 1807.05521}}].

\bibitem{Ballardini:2019tuc}
M.~{Ballardini}, R.~{Murgia}, M.~{Baldi}, F.~{Finelli} and M.~{Viel}, \emph{{Non-linear damping of superimposed primordial oscillations on the matter power spectrum in galaxy surveys}}, \href{https://doi.org/10.1088/1475-7516/2020/04/030}{\emph{\jcap} {\bfseries 2020} (2020) 030} [\href{https://arxiv.org/abs/1912.12499}{{\ttfamily 1912.12499}}].

\bibitem{Debono:2020emh}
I.~{Debono}, D.K.~{Hazra}, A.~{Shafieloo}, G.F.~{Smoot} and A.A.~{Starobinsky}, \emph{{Constraints on features in the inflationary potential from future Euclid data}}, \href{https://doi.org/10.1093/mnras/staa1765}{\emph{\mnras} {\bfseries 496} (2020) 3448} [\href{https://arxiv.org/abs/2003.05262}{{\ttfamily 2003.05262}}].

\bibitem{Li:2021jvz}
Y.~{Li}, H.-M.~{Zhu} and B.~{Li}, \emph{{Non-linear reconstruction of features in the primordial power spectrum from large-scale structure}}, \href{https://doi.org/10.1093/mnras/stac1544}{\emph{\mnras} {\bfseries 514} (2022) 4363} [\href{https://arxiv.org/abs/2102.09007}{{\ttfamily 2102.09007}}].

\bibitem{Euclid:2023shr}
{Euclid Collaboration: M. Ballardini}, Y.~{Akrami}, F.~{Finelli}, D.~{Karagiannis}, B.~{Li}, Y.~{Li} et~al., \emph{{Euclid: The search for primordial features}}, \href{https://doi.org/10.1051/0004-6361/202348162}{\emph{\aap} {\bfseries 683} (2024) A220} [\href{https://arxiv.org/abs/2309.17287}{{\ttfamily 2309.17287}}].

\bibitem{Crocce:2007dt}
M.~{Crocce} and R.~{Scoccimarro}, \emph{{Nonlinear evolution of baryon acoustic oscillations}}, \href{https://doi.org/10.1103/PhysRevD.77.023533}{\emph{\prd} {\bfseries 77} (2008) 023533} [\href{https://arxiv.org/abs/0704.2783}{{\ttfamily 0704.2783}}].

\bibitem{Scoccimarro2011}
R.~{Scoccimarro}, L.~{Hui}, M.~{Manera} and K.C.~{Chan}, \emph{{Large-scale bias and efficient generation of initial conditions for nonlocal primordial non-Gaussianity}}, \href{https://doi.org/10.1103/PhysRevD.85.083002}{\emph{\prd} {\bfseries 85} (2012) 083002} [\href{https://arxiv.org/abs/1108.5512}{{\ttfamily 1108.5512}}].

\bibitem{Creminelli:2013poa}
P.~{Creminelli}, J.~{Gleyzes}, M.~{Simonovi{\'c}} and F.~{Vernizzi}, \emph{{Single-field consistency relations of large scale structure part II: resummation and redshift space}}, \href{https://doi.org/10.1088/1475-7516/2014/02/051}{\emph{\jcap} {\bfseries 2014} (2014) 051} [\href{https://arxiv.org/abs/1311.0290}{{\ttfamily 1311.0290}}].

\bibitem{Baldauf:2015xfa}
T.~{Baldauf}, M.~{Mirbabayi}, M.~{Simonovi{\'{c}}} and M.~{Zaldarriaga}, \emph{{Equivalence principle and the baryon acoustic peak}}, \href{https://doi.org/10.1103/PhysRevD.92.043514}{\emph{\prd} {\bfseries 92} (2015) 043514} [\href{https://arxiv.org/abs/1504.04366}{{\ttfamily 1504.04366}}].

\bibitem{Blas:2015qsi}
D.~{Blas}, M.~{Garny}, M.M.~{Ivanov} and S.~{Sibiryakov}, \emph{{Time-sliced perturbation theory for large scale structure I: general formalism}}, \href{https://doi.org/10.1088/1475-7516/2016/07/052}{\emph{\jcap} {\bfseries 2016} (2016) 052} [\href{https://arxiv.org/abs/1512.05807}{{\ttfamily 1512.05807}}].

\bibitem{Senatore:2017pbn}
L.~{Senatore} and G.~{Trevisan}, \emph{{On the IR-resummation in the EFTofLSS}}, \href{https://doi.org/10.1088/1475-7516/2018/05/019}{\emph{\jcap} {\bfseries 2018} (2018) 019} [\href{https://arxiv.org/abs/1710.02178}{{\ttfamily 1710.02178}}].

\bibitem{Blas:2016sfa}
D.~{Blas}, M.~{Garny}, M.M.~{Ivanov} and S.~{Sibiryakov}, \emph{{Time-sliced perturbation theory II: baryon acoustic oscillations and infrared resummation}}, \href{https://doi.org/10.1088/1475-7516/2016/07/028}{\emph{\jcap} {\bfseries 2016} (2016) 028} [\href{https://arxiv.org/abs/1605.02149}{{\ttfamily 1605.02149}}].

\bibitem{Vasudevan:2019ewf}
A.~{Vasudevan}, M.M.~{Ivanov}, S.~{Sibiryakov} and J.~{Lesgourgues}, \emph{{Time-sliced perturbation theory with primordial non-Gaussianity and effects of large bulk flows on inflationary oscillating features}}, \href{https://doi.org/10.1088/1475-7516/2019/09/037}{\emph{\jcap} {\bfseries 2019} (2019) 037} [\href{https://arxiv.org/abs/1906.08697}{{\ttfamily 1906.08697}}].

\bibitem{Vlah:2015zda}
Z.~{Vlah}, U.~{Seljak}, M.~{Yat Chu} and Y.~{Feng}, \emph{{Perturbation theory, effective field theory, and oscillations in the power spectrum}}, \href{https://doi.org/10.1088/1475-7516/2016/03/057}{\emph{\jcap} {\bfseries 2016} (2016) 057} [\href{https://arxiv.org/abs/1509.02120}{{\ttfamily 1509.02120}}].

\bibitem{Chen:2020ckc}
S.-F.~{Chen}, Z.~{Vlah} and M.~{White}, \emph{{Modeling features in the redshift-space halo power spectrum with perturbation theory}}, \href{https://doi.org/10.1088/1475-7516/2020/11/035}{\emph{\jcap} {\bfseries 2020} (2020) 035} [\href{https://arxiv.org/abs/2007.00704}{{\ttfamily 2007.00704}}].

\bibitem{Ballardini:2022vzh}
M.~{Ballardini} and F.~{Finelli}, \emph{{On the primordial origin of the smoothing excess in the Planck temperature power spectrum in light of LSS data}}, \href{https://doi.org/10.1088/1475-7516/2022/10/083}{\emph{\jcap} {\bfseries 2022} (2022) 083} [\href{https://arxiv.org/abs/2207.14410}{{\ttfamily 2207.14410}}].

\bibitem{2024arXiv240600103C}
S.-F.~{Chen}, Z.~{Vlah} and M.~{White}, \emph{{The Bispectrum in Lagrangian Perturbation Theory}}, \href{https://doi.org/10.48550/arXiv.2406.00103}{\emph{arXiv e-prints} (2024) arXiv:2406.00103} [\href{https://arxiv.org/abs/2406.00103}{{\ttfamily 2406.00103}}].

\bibitem{Ivanov:2018gjr}
M.M.~{Ivanov} and S.~{Sibiryakov}, \emph{{Infrared resummation for biased tracers in redshift space}}, \href{https://doi.org/10.1088/1475-7516/2018/07/053}{\emph{\jcap} {\bfseries 2018} (2018) 053} [\href{https://arxiv.org/abs/1804.05080}{{\ttfamily 1804.05080}}].

\bibitem{Ivanov:2019pdj}
M.M.~{Ivanov}, M.~{Simonovi{\'c}} and M.~{Zaldarriaga}, \emph{{Cosmological parameters from the BOSS galaxy power spectrum}}, \href{https://doi.org/10.1088/1475-7516/2020/05/042}{\emph{\jcap} {\bfseries 2020} (2020) 042} [\href{https://arxiv.org/abs/1909.05277}{{\ttfamily 1909.05277}}].

\bibitem{Ivanov:2019hqk}
M.M.~{Ivanov}, M.~{Simonovi{\'c}} and M.~{Zaldarriaga}, \emph{{Cosmological parameters and neutrino masses from the final P l a n c k and full-shape BOSS data}}, \href{https://doi.org/10.1103/PhysRevD.101.083504}{\emph{\prd} {\bfseries 101} (2020) 083504} [\href{https://arxiv.org/abs/1912.08208}{{\ttfamily 1912.08208}}].

\bibitem{Nishimichi:2020tvu}
T.~{Nishimichi}, G.~{D'Amico}, M.M.~{Ivanov}, L.~{Senatore}, M.~{Simonovi{\'c}}, M.~{Takada} et~al., \emph{{Blinded challenge for precision cosmology with large-scale structure: Results from effective field theory for the redshift-space galaxy power spectrum}}, \href{https://doi.org/10.1103/PhysRevD.102.123541}{\emph{\prd} {\bfseries 102} (2020) 123541} [\href{https://arxiv.org/abs/2003.08277}{{\ttfamily 2003.08277}}].

\bibitem{Ivanov:2020ril}
M.M.~{Ivanov}, E.~{McDonough}, J.C.~{Hill}, M.~{Simonovi{\'c}}, M.W.~{Toomey}, S.~{Alexander} et~al., \emph{{Constraining early dark energy with large-scale structure}}, \href{https://doi.org/10.1103/PhysRevD.102.103502}{\emph{\prd} {\bfseries 102} (2020) 103502} [\href{https://arxiv.org/abs/2006.11235}{{\ttfamily 2006.11235}}].

\bibitem{Chudaykin:2020ghx}
A.~{Chudaykin}, K.~{Dolgikh} and M.M.~{Ivanov}, \emph{{Constraints on the curvature of the Universe and dynamical dark energy from the full-shape and BAO data}}, \href{https://doi.org/10.1103/PhysRevD.103.023507}{\emph{\prd} {\bfseries 103} (2021) 023507} [\href{https://arxiv.org/abs/2009.10106}{{\ttfamily 2009.10106}}].

\bibitem{DESI:2024mwx}
A.G.~{Adame}, J.~{Aguilar}, S.~{Ahlen}, S.~{Alam}, D.M.~{Alexander}, M.~{Alvarez} et~al., \emph{{DESI 2024 VI: cosmological constraints from the measurements of baryon acoustic oscillations}}, \href{https://doi.org/10.1088/1475-7516/2025/02/021}{\emph{\jcap} {\bfseries 2025} (2025) 021} [\href{https://arxiv.org/abs/2404.03002}{{\ttfamily 2404.03002}}].

\bibitem{Euclid:2024yrr}
{Euclid Collaboration: Y. Mellier}, {Abdurro'uf}, J.A.~{Acevedo Barroso}, A.~{Ach{\'u}carro}, J.~{Adamek}, R.~{Adam} et~al., \emph{{Euclid: I. Overview of the Euclid mission}}, \href{https://doi.org/10.1051/0004-6361/202450810}{\emph{\aap} {\bfseries 697} (2025) A1} [\href{https://arxiv.org/abs/2405.13491}{{\ttfamily 2405.13491}}].

\bibitem{Carroll:2013oxa}
S.M.~{Carroll}, S.~{Leichenauer} and J.~{Pollack}, \emph{{Consistent effective theory of long-wavelength cosmological perturbations}}, \href{https://doi.org/10.1103/PhysRevD.90.023518}{\emph{\prd} {\bfseries 90} (2014) 023518} [\href{https://arxiv.org/abs/1310.2920}{{\ttfamily 1310.2920}}].

\bibitem{Valageas:2003gm}
P.~{Valageas}, \emph{{A new approach to gravitational clustering: A path-integral formalism and large-N expansions}}, \href{https://doi.org/10.1051/0004-6361:20040125}{\emph{\aap} {\bfseries 421} (2004) 23} [\href{https://arxiv.org/abs/astro-ph/0307008}{{\ttfamily astro-ph/0307008}}].

\bibitem{Bean:2008na}
R.~{Bean}, X.~{Chen}, G.~{Hailu}, S.H.H.~{Tye} and J.~{Xu}, \emph{{Duality cascade in brane inflation}}, \href{https://doi.org/10.1088/1475-7516/2008/03/026}{\emph{\jcap} {\bfseries 2008} (2008) 026} [\href{https://arxiv.org/abs/0802.0491}{{\ttfamily 0802.0491}}].

\bibitem{Miranda:2012rm}
V.~{Miranda}, W.~{Hu} and P.~{Adshead}, \emph{{Warp features in DBI inflation}}, \href{https://doi.org/10.1103/PhysRevD.86.063529}{\emph{\prd} {\bfseries 86} (2012) 063529} [\href{https://arxiv.org/abs/1207.2186}{{\ttfamily 1207.2186}}].

\bibitem{Braglia:2022ftm}
M.~{Braglia}, X.~{Chen}, D.~{Kumar Hazra} and L.~{Pinol}, \emph{{Back to the features: assessing the discriminating power of future CMB missions on inflationary models}}, \href{https://doi.org/10.1088/1475-7516/2023/03/014}{\emph{\jcap} {\bfseries 2023} (2023) 014} [\href{https://arxiv.org/abs/2210.07028}{{\ttfamily 2210.07028}}].

\bibitem{McAllister:2008hb}
L.~{McAllister}, E.~{Silverstein} and A.~{Westphal}, \emph{{Gravity waves and linear inflation from axion monodromy}}, \href{https://doi.org/10.1103/PhysRevD.82.046003}{\emph{\prd} {\bfseries 82} (2010) 046003} [\href{https://arxiv.org/abs/0808.0706}{{\ttfamily 0808.0706}}].

\bibitem{Silverstein:2008sg}
E.~{Silverstein} and A.~{Westphal}, \emph{{Monodromy in the CMB: Gravity waves and string inflation}}, \href{https://doi.org/10.1103/PhysRevD.78.106003}{\emph{\prd} {\bfseries 78} (2008) 106003} [\href{https://arxiv.org/abs/0803.3085}{{\ttfamily 0803.3085}}].

\bibitem{Seo:2007ns}
H.-J.~{Seo} and D.J.~{Eisenstein}, \emph{{Improved Forecasts for the Baryon Acoustic Oscillations and Cosmological Distance Scale}}, \href{https://doi.org/10.1086/519549}{\emph{\apj} {\bfseries 665} (2007) 14} [\href{https://arxiv.org/abs/astro-ph/0701079}{{\ttfamily astro-ph/0701079}}].

\bibitem{Planck:2018vyg}
{Planck Collaboration: N. Aghanim}, Y.~{Akrami}, M.~{Ashdown}, J.~{Aumont}, C.~{Baccigalupi}, M.~{Ballardini} et~al., \emph{{Planck 2018 results. VI. Cosmological parameters}}, \href{https://doi.org/10.1051/0004-6361/201833910}{\emph{\aap} {\bfseries 641} (2020) A6} [\href{https://arxiv.org/abs/1807.06209}{{\ttfamily 1807.06209}}].

\bibitem{Silk:1967kq}
J.~{Silk}, \emph{{Cosmic Black-Body Radiation and Galaxy Formation}}, \href{https://doi.org/10.1086/149449}{\emph{\apj} {\bfseries 151} (1968) 459}.

\bibitem{Hu:1995en}
W.~{Hu} and N.~{Sugiyama}, \emph{{Small-Scale Cosmological Perturbations: an Analytic Approach}}, \href{https://doi.org/10.1086/177989}{\emph{\apj} {\bfseries 471} (1996) 542} [\href{https://arxiv.org/abs/astro-ph/9510117}{{\ttfamily astro-ph/9510117}}].

\bibitem{Eisenstein:1997ik}
D.J.~{Eisenstein} and W.~{Hu}, \emph{{Baryonic Features in the Matter Transfer Function}}, \href{https://doi.org/10.1086/305424}{\emph{\apj} {\bfseries 496} (1998) 605} [\href{https://arxiv.org/abs/astro-ph/9709112}{{\ttfamily astro-ph/9709112}}].

\bibitem{Tassev:2013pn}
S.~Tassev, M.~Zaldarriaga and D.~Eisenstein, \emph{{Solving Large Scale Structure in Ten Easy Steps with COLA}}, \href{https://doi.org/10.1088/1475-7516/2013/06/036}{\emph{JCAP} {\bfseries 06} (2013) 036} [\href{https://arxiv.org/abs/1301.0322}{{\ttfamily 1301.0322}}].

\bibitem{Tassev:2015mia}
S.~{Tassev}, D.J.~{Eisenstein}, B.D.~{Wandelt} and M.~{Zaldarriaga}, \emph{{sCOLA: The N-body COLA Method Extended to the Spatial Domain}}, \href{https://doi.org/10.48550/arXiv.1502.07751}{\emph{arXiv e-prints} (2015) arXiv:1502.07751} [\href{https://arxiv.org/abs/1502.07751}{{\ttfamily 1502.07751}}].

\bibitem{Winther:2017jof}
H.A.~{Winther}, K.~{Koyama}, M.~{Manera}, B.S.~{Wright} and G.-B.~{Zhao}, \emph{{COLA with scale-dependent growth: applications to screened modified gravity models}}, \href{https://doi.org/10.1088/1475-7516/2017/08/006}{\emph{\jcap} {\bfseries 2017} (2017) 006} [\href{https://arxiv.org/abs/1703.00879}{{\ttfamily 1703.00879}}].

\bibitem{Wright:2017dkw}
B.S.~{Wright}, H.A.~{Winther} and K.~{Koyama}, \emph{{COLA with massive neutrinos}}, \href{https://doi.org/10.1088/1475-7516/2017/10/054}{\emph{\jcap} {\bfseries 2017} (2017) 054} [\href{https://arxiv.org/abs/1705.08165}{{\ttfamily 1705.08165}}].

\bibitem{Howlett:2015hfa}
C.~{Howlett}, M.~{Manera} and W.J.~{Percival}, \emph{{L-PICOLA: A parallel code for fast dark matter simulation}}, \href{https://doi.org/10.1016/j.ascom.2015.07.003}{\emph{Astronomy and Computing} {\bfseries 12} (2015) 109} [\href{https://arxiv.org/abs/1506.03737}{{\ttfamily 1506.03737}}].

\bibitem{Viel:2010bn}
M.~{Viel}, M.G.~{Haehnelt} and V.~{Springel}, \emph{{The effect of neutrinos on the matter distribution as probed by the intergalactic medium}}, \href{https://doi.org/10.1088/1475-7516/2010/06/015}{\emph{\jcap} {\bfseries 2010} (2010) 015} [\href{https://arxiv.org/abs/1003.2422}{{\ttfamily 1003.2422}}].

\bibitem{Villaescusa-Navarro:2018bpd}
F.~{Villaescusa-Navarro}, S.~{Naess}, S.~{Genel}, A.~{Pontzen}, B.~{Wandelt}, L.~{Anderson} et~al., \emph{{Statistical Properties of Paired Fixed Fields}}, \href{https://doi.org/10.3847/1538-4357/aae52b}{\emph{\apj} {\bfseries 867} (2018) 137} [\href{https://arxiv.org/abs/1806.01871}{{\ttfamily 1806.01871}}].

\bibitem{2016JCAP...09..015M}
J.E.~{McEwen}, X.~{Fang}, C.M.~{Hirata} and J.A.~{Blazek}, \emph{{FAST-PT: a novel algorithm to calculate convolution integrals in cosmological perturbation theory}}, \href{https://doi.org/10.1088/1475-7516/2016/09/015}{\emph{\jcap} {\bfseries 2016} (2016) 015} [\href{https://arxiv.org/abs/1603.04826}{{\ttfamily 1603.04826}}].

\bibitem{2017JCAP...02..030F}
X.~{Fang}, J.A.~{Blazek}, J.E.~{McEwen} and C.M.~{Hirata}, \emph{{FAST-PT II: an algorithm to calculate convolution integrals of general tensor quantities in cosmological perturbation theory}}, \href{https://doi.org/10.1088/1475-7516/2017/02/030}{\emph{\jcap} {\bfseries 2017} (2017) 030} [\href{https://arxiv.org/abs/1609.05978}{{\ttfamily 1609.05978}}].

\bibitem{2017JCAP...11..007B}
D.~{Baumann}, D.~{Green} and M.~{Zaldarriaga}, \emph{{Phases of New Physics in the BAO Spectrum}}, \href{https://doi.org/10.1088/1475-7516/2017/11/007}{\emph{\jcap} {\bfseries 2017} (2017) 007} [\href{https://arxiv.org/abs/1703.00894}{{\ttfamily 1703.00894}}].

\bibitem{Cyr-Racine:2011bjz}
F.-Y.~{Cyr-Racine} and F.~{Schmidt}, \emph{{Oscillating bispectra and galaxy clustering: A novel probe of inflationary physics with large-scale structure}}, \href{https://doi.org/10.1103/PhysRevD.84.083505}{\emph{\prd} {\bfseries 84} (2011) 083505} [\href{https://arxiv.org/abs/1106.2806}{{\ttfamily 1106.2806}}].

\bibitem{2014PhRvD..89f3540G}
J.-O.~{Gong}, K.~{Schalm} and G.~{Shiu}, \emph{{Correlating correlation functions of primordial perturbations}}, \href{https://doi.org/10.1103/PhysRevD.89.063540}{\emph{\prd} {\bfseries 89} (2014) 063540} [\href{https://arxiv.org/abs/1401.4402}{{\ttfamily 1401.4402}}].

\bibitem{2015JCAP...10..062M}
S.~{Mooij}, G.A.~{Palma}, G.~{Panotopoulos} and A.~{Soto}, \emph{{Consistency relations for sharp features in the primordial spectra}}, \href{https://doi.org/10.1088/1475-7516/2015/10/062}{\emph{\jcap} {\bfseries 2015} (2015) 062} [\href{https://arxiv.org/abs/1507.08481}{{\ttfamily 1507.08481}}].

\bibitem{Cabass:2018roz}
G.~{Cabass}, E.~{Pajer} and F.~{Schmidt}, \emph{{Imprints of Oscillatory Bispectra on Galaxy Clustering}}, \href{https://doi.org/10.1088/1475-7516/2018/09/003}{\emph{\jcap} {\bfseries 2018} (2018) 003} [\href{https://arxiv.org/abs/1804.07295}{{\ttfamily 1804.07295}}].

\end{thebibliography}\endgroup

\end{document}